%% file: ms.tex
\documentclass[12pt,preprint]{aastex}

\def\deg{\degr}
\def\dex#1{10$^{#1}$}
\def\tdex#1{$\times$10$^{#1}$}
\def\kms{km\,s$^{-1}$}
\def\cmm#1{cm$^{-#1}$}

\def\mA{m\AA}

\def\Msunpyr{M$_\odot$\,yr$^{-1}$}
\def\vlsr{$v_{\rm LSR}$}
\def\nav{$N_a(v)$}
\def\simlt{\mbox{$^{<}_{\sim}$}}

\def\Ht{H$_2$}
\def\Copernicus{{\it Copernicus}}
\def\IUE{{\it IUE}}
\def\HST{{\it HST}}
\def\GHRS{{GHRS}}
\def\STIS{{STIS}}
\def\FUSE{{\it FUSE}}
\def\COS{{COS}}

\def\ch#1{\colhead{#1}}
\def\e{$\pm$}
\def\Lya{Ly$\alpha$}
\def\Lyb{Ly$\beta$}
\def\Lyg{Ly$\gamma$}
\def\Lyk{Ly$\kappa$}
\def\OVIa{\OVI\,$\lambda$1031.926}
\def\OVIb{\OVI\,$\lambda$1037.617}
\def\CIVa{\CIV\,$\lambda$1548.204}
\def\CIVb{\CIV\,$\lambda$1550.78}
\def\NVa{\NV\,$\lambda$1238.821}
\def\NVb{\NV\,$\lambda$1242.804}
\def\SiIVa{\SiIV\,$\lambda$1393.760}
\def\SiIVb{\SiIV\,$\lambda$1402.772}
\def\SiIIIa{\SiIII\,$\lambda$1206.500}
\def\FeIIIa{\FeIII\,$\lambda$1122.524}
\def\CIIIa{\CIII\,$\lambda$977.020}
\def\CIIa{\CII\,$\lambda$1334.532}
\def\CIIb{\CII\,$\lambda$1036.337} 

\def\SIIa{\SII\,$\lambda$1250.584}

\def\SIId{\SII\,$\lambda$1250.584, 1253.811}
\def\SIIIa{\SIII\,$\lambda$1012.501}
\def\FeIIIa{\FeIII\,$\lambda$1122.524}
\def\ArIa{\ArI\,$\lambda$1048.220}

\def\CIVd{\CIV\,$\lambda$$\lambda$1548, 1550}

\def\absl{$|$$l$$|$}
\def\absb{$|$$b$$|$}
\def\Nsinb{$N$\,sin$|$$b$$|$}
\def\Nisinb#1{$N$(#1)\,sin$|$$b$$|$}

\def\logNisinb#1{log\,$N$(#1)\,sin$|$$b$$|$}

\def\Sobs{2}

\def\Smeas{3}
\def\Sresults{4}
\def\Svdiff{4.1}
\def\Sextent{4.2}
\def\Swidth{4.3}
\def\Srotate{4.4}
\def\Scoldens{4.5}
\def\Ssky{4.6}

\def\Sratiohist{4.8}

\def\Smodel{5}

\def\Sdisc{6}

\def\VALUE{\relax}

\def\HI{\ion{H}{1}}

\def\CI{\ion{C}{1}}
\def\CII{\ion{C}{2}}
\def\CIII{\ion{C}{3}}
\def\CIV{\ion{C}{4}}
\def\NI{\ion{N}{1}}

\def\NIII{\ion{N}{3}}
\def\NIV{\ion{N}{4}}
\def\NV{\ion{N}{5}}
\def\OI{\ion{O}{1}}

\def\OIII{\ion{O}{3}}

\def\OV{\ion{O}{5}}
\def\OVI{\ion{O}{6}}

\def\NeVIII{\ion{Ne}{8}}

\def\AlIII{\ion{Al}{3}}

\def\SiII{\ion{Si}{2}}
\def\SiIII{\ion{Si}{3}}
\def\SiIV{\ion{Si}{4}}

\def\SII{\ion{S}{2}}
\def\SIII{\ion{S}{3}}

\def\ArI{\ion{Ar}{1}}

\def\FeII{\ion{Fe}{2}}
\def\FeIII{\ion{Fe}{3}}

\def\Tobs{1}
\def\Tres{2}
\def\Fmapid{1}
\def\Fspec{2}
\def\Fextent{3}
\def\Fwidthhist{4}
\def\Fvelo{5}
\def\Fcoldens{6}
\def\Fskycollongtot{7}
\def\Fskycollatitot{8}
\def\FmapOVIvellincol{9}
\def\Fratiohist{10}
\def\Fratiomodel{11}
\def\FAcoolcurve{A1}
\def\FAzone{A2}
\def\FAfractions{A3}
\def\FAcoldens{A4}

\begin{document}
\title{Characterizing Transition Temperature Gas in the Galactic Corona}

\author{%
Bart P.\ Wakker\altaffilmark{1},
Blair D.\ Savage\altaffilmark{1},
Andrew J. Fox\altaffilmark{2,3},
Robert Benjamin\altaffilmark{4},
\altaffiltext{1}{Department of Astronomy, University of Wisconsin, Madison, WI 53706; wakker@astro.wisc.edu, savage@astro.wisc.edu}
\altaffiltext{2}{European Southern Observatory, Alonso de C\'ordova 3107, Casilla 19001, Vitacura, Santiago, Chile}
\altaffiltext{3}{Institute of Astronomy, University of Cambridge, Madingley Road, Cambridge, CB3 0HA, UK; afox@eso.org}
\altaffiltext{4}{Department of Physics, University of Wisconsin, Whitewater, 800 West Main Street, Whitewater WI 53190; benjamir@wisp.physics.wisc.edu}
}


\begin{abstract}
We present a study of the properties of the transition temperature
($T$$\sim$\dex5~K) gas in the Milky Way corona, based on measurements of \OVI,
\NV, \CIV, \SiIV\ and \FeIII\ absorption lines seen in the far ultraviolet
spectra of 58 sightlines to extragalactic targets, obtained with {\it
Far-Ultraviolet Spectroscopic Explorer} (FUSE) and {\it Space Telescope
Imaging Spectrograph}. In many sightlines the Galactic absorption profiles show
multiple components, which are analyzed separately. We find that the
highly-ionized atoms are distributed irregularly in a layer with a scaleheight
of about 3~kpc, which rotates along with the gas in the disk, without an obvious
gradient in the rotation velocity away from the Galactic plane. Within this
layer the gas has randomly oriented velocities with a dispersion of 40--60~\kms.
On average the integrated column densities are log\,$N$(\OVI)=14.3,
log\,$N$(\NV)=13.5, log\,$N$(\CIV)=14.2, log\,$N$(\SiIV)=13.6 and
log\,$N$(\FeIII)=14.2, with a dispersion of just 0.2~dex in each case. In
sightlines around the Galactic Center and Galactic North Pole all column
densities are enhanced by a factor $\sim$2, while at intermediate latitudes in
the southern sky there is a deficit in $N$(\OVI) of about a factor 2, but no
deficit for the other ions. We compare the column densities and ionic ratios to
a series of theoretical predictions: collisional ionization equilibrium, shock
ionization, conductive interfaces, turbulent mixing, thick disk supernovae,
static non-equilibrium ionization (NIE) radiative cooling and an NIE radiative
cooling model in which the gas flows through the cooling zone. None of these
models can fully reproduce the data, but it is clear that non-equilibrium
ionization radiative cooling is important in generating the transition
temperature gas.
\end{abstract}



\section{Introduction}
\par Understanding the processes in Milky Way Disk, Thick Disk and Halo gas that
generate gas at temperatures $T$$>$\dex5~K is a crucial element in learning
about the lifecycle of gas in galaxies. Such gas can be produced by supernovae,
winds and infall, and its presence traces the circumstances of galactic
evolution. Spitzer (1956) originally proposed the existence of a ``hot
interstellar Galactic Corona'', with $T$$\sim$\dex6~K, to account for the
presence of neutral clouds about 1~kpc above the Galactic plane (M\"unch 1952;
M\"unch \& Zirin 1961), the idea being that these clouds could only persist if
they were confined by the pressure of an external medium. After the detection of
the diffuse galactic soft X-ray background (Bunner et al.\ 1973, Williamson et
al.\ 1974) and of \OVI\ absorption (Rogerson et al.\ 1973; York 1974; Jenkins \&
Meloy 1974), Shapiro \& Field (1976) proposed that this hot gas would produce a
``Galactic Fountain'' -- a process in which the ISM is heated by supernova
explosions, expands away from the Galactic plane, tries to settle
hydrostatically, but as it cools clouds with $T$$\sim$\dex4~K rain back on the
disk. This is a natural way to produce moving clouds of cooler gas embedded in a
hot medium. The precise predictions of the process depend on many factors, most
importantly the ratio of free-fall time to cooling time (Bregman 1980; Kahn
1981; Houck \& Bregman 1990). Other ways to produce hot coronal gas have also
been proposed. E.g., a Galactic wind originating in the center (see Everett et
al.\ 2008 and references therein) or heat generated by accreting clouds
(considered by Henley et al.\ 2010).
\par In his original paper, Spitzer (1956) pointed out that the
(1s$^2$2s)$^2$$S_{1/2}$ $\rightarrow$ (1s$^2$2p)$^2$$P_{1/2,3/2}$ electronic
transitions of the lithium-like ions O$^{+5}$, N$^{+4}$ and C$^{+3}$ (and the
similar Si-ion Si$^{+3}$) produce resonance doublet absorption lines in the UV
(at $\lambda$$\sim$1030, 1240, 1550 and 1400\,\AA, respectively). These ions
have ionization potentials for production of 113.9, 77.5, 47.9 and 33.5~eV and
trace gas in the 1--3\tdex5~K temperature range, which originates in processes
that convert \dex6~K into \dex4~K gas and vice versa. At the temperatures
sampled by these ions, the ISM is unstable -- left to itself it will rapidly
cool to below \dex4~K. The presence of these ions therefore implies the
existence of \dex6~K or hotter gas that has cooled down or of a process that is
heating cooler gas.
\par Here two notes about nomenclature are in order. First, Spitzer used
``Corona'' for the region containing hot gas out to a few kpc from the Galactic
plane. ``Halo'' has often been used instead, referring to the region of space
outside the Galactic disk. As the scaleheights of \OVI, NV\ and \CIV\ are found
to be only $\sim$3~kpc, Savage et al.\ (2003) refer to the region where the
\dex5~K gas occurs as a ``Thick Disk'', as opposed to a much more extended
($>$50~kpc) Corona that is implied by the detection of \OVI\ absorption
associated with the Magellanic Stream (Sembach et al.\ 2003). We will follow
Savage et al.\ (2003) in using this terminology. Second, the term ``hot gas''
has often been used to describe gas with $T$$>$\dex5~K, using ``warm gas'' if
the temperature is $\sim$5000~K to a few \dex4~K. However, since gas at
temperatures $\sim$\dex5~K has very different properties than gas at
$T$$\sim$\dex5~K (not the least of which is that is it cools rapidly), We will
instead follow the convention of Savage \& Wakker (2009), who proposed to
instead use the term ``transition temperature gas'' for this quickly-evolving
phase, reserving ``hot gas'' for the X-ray emitting phase with $T$$>$\dex6~K.
\par The desire to detect the highly-ionized interstellar atoms was one of
Spitzer's main motivators to push for the creation of a UV satellite in space,
resulting in the launch of the Copernicus satellite in 1972. Other UV
spectroscopic instruments followed, including the International Ultraviolet
Explorer (\IUE), the Goddard High Resolution Spectrograph (\GHRS), the Space
Telescope Imaging Spectrograph (\STIS), the Far Ultraviolet Spectroscopic
Explorer (\FUSE) and the Cosmic Origins Spectrograph (\COS). These instruments
can detect the highly-ionized ions in the Milky Way disk and corona in
absorption against both stars and extragalactic targets. Many such studies have
been done: York (1974, 1977), Jenkins \& Meloy (1974), Jenkins (1978a, 1978b),
Cowie et al.\ (1979, 1981), Savage \& de Boer (1979, 1981), Savage \& Massa
(1987), Sembach et al.\ (1990, 1991, 1994a, 1994b, 1995, 1999, 2001), Savage et
al.\ (1990, 1995, 1997a, 1997b, 2001a, 2001b, 2003, 2005), Sembach \& Savage
(1992, 1994), Tripp et al.\ (1993), Savage \& Sembach (1994), Sembach (1994),
Jenkins et al.\ (1998, 2000), Oegerle et al.\ (2000, 2005), Sterling et al.\
(2002), Howk et al.\ (2002, 2003), Fox et al.\ (2003), Zsarg\'o et al.\ (2003),
Wakker et al.\ (2003), Knauth et al.\ (2003), Lehner et al.\ (2003), Indebetouw
\& Shull (2004b), Ganguly et al.\ (2005), Savage \& Lehner (2006), Keeney et
al.\ (2006), Bowen et al.\ (2008), Savage \& Wakker (2009), Lehner et al.\
(2011). These studies show that in the disk the average density of the ionized
hydrogen plasma in which the highly-ionized species occur is about
3\tdex{-4}\,\cmm3, while the scale height of the ions is 2.6$\pm$0.6~kpc for
\OVI, 3.6$^{+1.0}_{-0.8}$~kpc for \CIV, and 3.2$^{+1.0}_{-0.6}$~kpc for \SiIV).
The velocity spread of the gas is too low to explain the scaleheight as due to
turbulent pressure. The distribution is highly irregular. Finally, the
ionization conditions show that the gas is far out of equilibrium.
\par Starting in the early 1990s, the \GHRS\ and \STIS\ instruments on the
Hubble Space Telescope (\HST) have observed 85 extragalactic targets, usually
for the purpose of studying intergalactic absorption in the sightlines. Toward
each of these sightlines, absorption by Galactic \NV, \CIV\ and/or \SiIV\ is
seen. Combined with the $\sim$150 sightlines toward which \OVI\ can be measured
using \FUSE, this dataset presents an opportunity to learn more about the
properties of the transition temperature gas in the Milky Way, allowing us to
address issues such as its kinematics relative to differential galactic
rotation, its vertical velocity spread, the total amount, its distribution on
the sky and relation to other structures, and the physical processes that
produce it.
\par In this paper we provide a table of extragalactic targets observed with
\HST\ spectrographs between 1992 and 2007, i.e., until the breakdown of the
\STIS\ spectrograph (Section~\Sobs). At the time this paper was finished (mid
2011), the \COS\ spectrograph (launched in fall 2009) already produced many new
spectra covering \NV, \CIV\ and \SiIV\ absorption. However, as of September 2011
only a handful are public. and only two sightlines would be added to our sample
of sightlines for which high S/N ratio data are available for all four
highly-ionized species. Future studies will allow detailed studies of \NV, \CIV\
and \SiIV\ in a few hundred sightlines (although \OVI\ will be missing for most
of these new sightlines). For each of the targets observed by \HST\ before 2007,
we include \FUSE\ observations if they are available. We then determine the
absorption that we think is associated with the thick disk of the Galaxy
(Section~\Smeas). In a large fraction of the sightlines, absorption is also
visible at high velocities ($\vert$\vlsr$\vert$$>$100~\kms; see e.g., Sembach et
al.\ 2003; Fox et al.\ 2006). We do not include those components in this paper.
The data also include absorption lines of ions with low ionization potential
(\CI, \CII, \NI, \OI, \SiII, \SII, \ArI, \FeII). We will not address these ions
here. A few abundant ions with intermediate ionization potentials (13.6--35~eV)
also produce strong UV absorption lines (\CIII, \NIII, \OIII, \AlIII, \SiIII,
\SIII\ and \FeIII). Of this set, the Galactic \CIIIa\ and \SiIIIa\ absorption
are always saturated, the Galactic \OIII\ and most of the \NIII\ lines are in
the EUV ($\lambda$$<$912~\AA) and thus hidden by the Galactic \HI\ absorption,
\NIII\,$\lambda$989\ blends with a strong \SiII\ line. \AlIII\,$\lambda$1854 was
observed by \IUE\ toward many stars (see e.g.\ Savage \& Wakker 2009), but lies
outside the wavelength range of most existing \HST\ spectra. However, \SIIIa\
and \FeIIIa\ are visible in \FUSE\ spectra (although the \SIII\ line is usually
contaminated by Galactic H$_2$ absorption). In this paper, we will include
\SIII\ and \FeIII\ in our analyses, allowing a comparison of the properties of
the warm ionized medium with the more highly-ionized transition temperature
gas.
\par We list the measured column densities and other parameters of \OVI, \NV,
\CIV, \SiIV, \FeIII\ and \SIII\ for the Thick Disk component of the high-ion
absorption lines in Section~\Smeas. Section~\Sresults\ presents a discussion of
the properties of the absorption, such as their distribution across the sky, the
kinematical relationships between the different ions, and the column density
ratios. We address the physical processes that cause these ions to appear in
Section~\Smodel, where we present a discussion of the predictions for ionic
column density ratios implied by the various models that have been proposed.
Compared to earlier discussions (Spitzer 1990; Savage et al.\ 1997b; Fox et al.\
2004; Indebetouw \& Shull 2004a) we look more precisely at the numerical
predictions for ionic ratios, and we consider several new models. Finally, we
discuss what we have learned in Section~\Sdisc.

\section{Observations}
\subsection{Sources of data}
\par In Table~\Tobs\ we give observational information for the high-resolution
($R$$>$10000) \GHRS, \STIS-G140M or \STIS-E140M spectra that were taken before
2007 and for which the signal-to-noise ratio per 20~\kms\ near either the \NV\
or \CIV\ line is larger than 7. This S/N limit is used because at lower S/N
ratios the Galactic absorption lines become difficult to measure and the error
in the ratios derived between ionic column densities is larger than a factor 2.

\input table1.tex

\par Of the set of observations, 9 targets were only observed using the \GHRS,
18 were only observed using \STIS-G140M and 31 were observed using \STIS-E140M.
Of the 58 targets used in our analysis, 43 also have \FUSE\ data with S/N near
\OVI\ greater than 7. There are 55 targets having data with S/N$>$7 for \NV, 33
have such data covering \CIV, while for 26 targets the S/N ratio near all three
lines is $>$7. These 26 form the main sample for the current study.
\par Figure~\Fmapid\ shows the distribution of the analyzed targets on the sky.
The integrated Galactic \HI\ column density is also shown, with a contour at a
\HI\ column density of 7\tdex{20}\,\cmm2. This shows that the extragalactic
UV-bright targets that can be observed with space spectrographs only sample gas
at high galactic latitudes, and that these targets are spread fairly evenly
across the sky, with the exception of a concentration of targets in the ``hole''
near $l$=120\deg, $b$=45\deg, where the extinction is below average, allowing
more targets to shine through the Galactic dust. This is also the region of sky
where the large high-velocity cloud complex~C can be found (Hulsbosch \& Wakker
1988), targets behind which have been disproportionally selected for
observations with \HST\ spectrographs.

\begin{figure} 
\plotfiddle{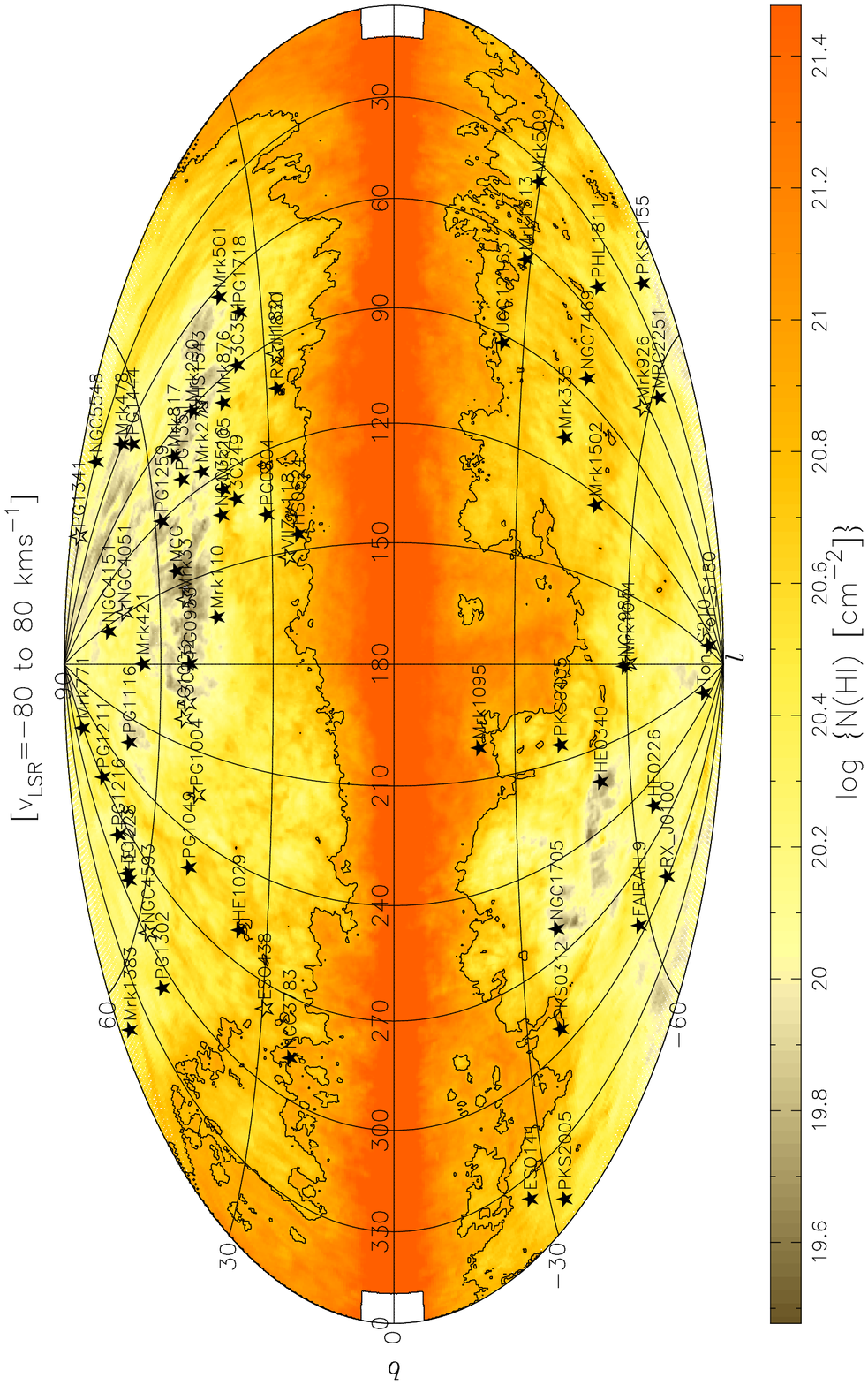}{0in}{270}{280}{450}{0}{0}
\figurenum{1}
\caption{%
Distribution of targets with \NV\ and/or \CIV\ spectra, on top of a map of the
logarithm of the column density of the 21-cm \HI\ sky, integrated between $-$80
and 80~\kms, with a contour overlaid for $N$(\HI)=7\tdex{20}\,\cmm2. The data
are shown in an Aitoff projection of Galactic coordinates centered on the
Anti-Center direction. The \HI\ data are from the Leiden-Argentina-Bonn (LAB)
survey (Kalberla et al.\ 2005).
}\end{figure}

\par Table~\Tobs\ and Figure~\Fmapid\ include 12 targets for which the status
column in the table has the value ``Skip''. This is the case when the spectra
have a sufficiently high S/N ratio, but there are problems with the
measurements. E.g., toward H\,1821+643 strong absorption is seen that is
associated with the planetary nebula K1-16. For NGC\,4151, both the \STIS-E140M
and \FUSE\ spectrum have high S/N ($\sim$14), and Galactic absorption is clearly
seen in all highly-ionized species, but because of the low redshift of
NGC\,4151, it is blended with lines associated with the background galaxy,
making the Galactic lines impossible to measure. For 3C\,232 several \STIS-G140M
grating settings cover \NV, \CIV\ and \SiIV, but the absorptions in the
individual lines of the \CIV\ and \SiIV\ doublet do not seem to match up. Toward
PG\,1004+130 the Galactic \OVI\ appears to be broad, as do \SiIIIa\ and \CIIb.
However, a series of intrinsic \OIII\ lines probably contaminates the Galactic
\OVI. Eight other sightlines with high S/N near 1238\,\AA\ are excluded from the
final sample (ESO\,438$-$G09, HS\,1543+5921, Mrk\,33, Mrk\,926, Mrk\,1044,
NGC\,4593, PG\,1001+291 and PG\,1341+258). These have low S/N or no data for
\OVI, and either the two lines of the \NV\ and/or \CIV\ doublet do not match or
there are intrinsic AGN lines contaminating the ISM lines. We still include
these sightlines in Table~\Tobs, so that we can show their spectra in
Figure~\Fspec, allowing the reader to check on the problems. However, they are
not used in the analyses.


\subsection{Calibration}
\par For \HST\ data, the calibrated fits files in the MAST archive were
retrieved. This is the only step needed, except for observations with the
\STIS-G140M grating and central wavelength 1222~\AA, where a 1-pixel shift seems
necessary (see Wakker \& Savage 2009). That conclusion is based on fourteen
sightlines with good data and relatively simple ISM absorption lines. For these
sightlines the \SIId\ lines can be fit both in a G140M spectrum centered on
1222~\AA\ and in one centered on 1272~\AA, and sometimes also in an E140M
spectrum. To align the lines in the 1222~\AA-centered spectrum with those in the
other spectra, an average redward shift of 12~\kms\ is needed, which corresponds
to one pixel.
\par Finally, for observations with the \STIS-E140M echelle, the MAST fits files
give the data for each of the 42 orders separately. These orders were combined
into a single spectrum by interpolating the photon counts and errors onto a
common grid, adding the photon counts and converting back to a flux.
\par The processing of the \FUSE\ data was described in detail by Wakker et al.\
(2003) and Wakker (2006), and therefore only a summary is given here. First, the
spectra were calibrated using v2.1 or v2.4 of the \FUSE\ calibration pipeline. 
To correct for residual wavelength shifts, the central velocities of the Milky 
Way interstellar lines were determined for each detector segment
(LiF1A/1B/2A/2B, SiC2A/2B) of each individual observation. The \FUSE\ segments
were then aligned with the LSR interstellar velocities implied by the
\STIS-E140M spectra, or, if no E140M data were available, with the LSR velocity
of the strongest component in the 21-cm \HI\ spectrum. For targets with a
\STIS-E140M spectrum, the interstellar reference velocity was determined by
fitting all Milky Way lines in that spectrum; the \STIS\ wavelength calibration
is accurate  to about 1~\kms\ (Kim-Quijano et al.\ 2007). For sightlines with
S/N$>$10 near   1031~\AA, the resulting shifts were given by Wakker (2006).
Using these shifts, LiF1A and LiF2B data are added together to produce the final
spectrum for each target. The final velocity scale references the velocities to
the Local Standard of Rest (LSR).


\subsection{Presentation of spectra}
\par We present the Galactic high-ionization absorption lines for six sightlines
in Figure~\Fspec. These sightlines illustrate each of the items mentioned in the
figure caption. The entire set of spectra is available in the electronic version
of this paper. The plots include the available data for the \OVI, \NV, \CIV\ and
\SiIV\ doublets, as well as their apparent column density profiles. Note that we
regridded the spectra to 10~\kms\ velocity bins for this figure. This does not
affect the \FUSE\ or \STIS-G140M data, but does slightly degrade the \STIS-E140M
spectra (enhancing their S/N ratio). The apparent column density is defined as:
$$ N_a(v) = {m_e c\over \pi e^2}\,f\,\lambda\ \ln{C(v)\over F(v)}, $$ with
$F(v)$ the observed profile, $C(v)$ the fitted continuum and $f$ the oscillator
strength. In addition, to allow comparing the kinematics of the high-ionization
lines with those of lower ionization gas, we show a representative strong and
weak line of an intermediate- and a low-ionization species, as described in the
figure caption.
\par We also show profiles of the ratios between different ions. These are
derived from the apparent column density profiles and their errors. We display
the ionic ratio only in the range of velocities where it is larger than the
error in this ratio, and only for sightlines and line-combinations where both
doublet lines of the various ions were detected at a significant level.
\par We show the actual spectra in Fig.~\Fspec, i.e.\ they are not normalized
relative to the continuum. However, for each absorption line we chose a vertical
scale that places the continuum at about 80\% of the maximum. For the apparent
column density and ratio profiles, we use the same vertical scale for each ion
throughout, to allow easy comparisons of the relative strength of the lines
between different sightlines.
\par A sightline of note is NGC\,4151 (spectra available in the electronic
version of this paper). Here the \OVI\ lines appears offset from all other
absorption lines, which {\it do} line up with each other, even showing a
component between $-$100 and $-$60~\kms\ in \CIV, \SiIII\ and \CII. There is no
problem with the velocity calibration, as the Galactic H$_2$ absorption lines
near \OVIa\ are fitted perfectly well. The most likely explanation is that the
\OVI\ absorption between \vlsr=25 and 60~\kms\ is related to the strong P-Cygni
profile produced by the outflowing gas in NGC\,4151 itself. We include the
measurements for most ions in this sightline, but we exclude the \OVI\ lines.
\par We note that in all cases the profiles of the high-ionization species cover
a similar range of velocities. In all but a few cases (PG\,1116+215,
PKS\,2155$-$304; see electronic version) the shapes and velocity extent of the
intermediate-ionization lines (i.e.\ \SiIII\ and \FeIII) follow that of the
high-ionization lines. On the other hand, the strong low-ionization absorption
lines tend to be narrower than the high-ionization absorption lines. The weak
low-ionization lines (usually exemplified by \SIIa) often show clear component
structure, which is usually, but not always reflected in the high-ionization
absorption (contrast e.g.\ ESO\,141$-$G55, Mrk\,205, PG\,1302$-$102 to 3C\,273.0,
Mrk\,509, PG\,1116+215 and PKS\,2005$-$489). Sometimes there is a narrow
low-ionization component near $v$$\sim$0~\kms, with no absorption at higher
velocities, where there is strong high-ionization absorption (e.g.\ toward
HE\,0226$-$4110 and Mrk\,509).

\begin{figure} 
\plotfiddle{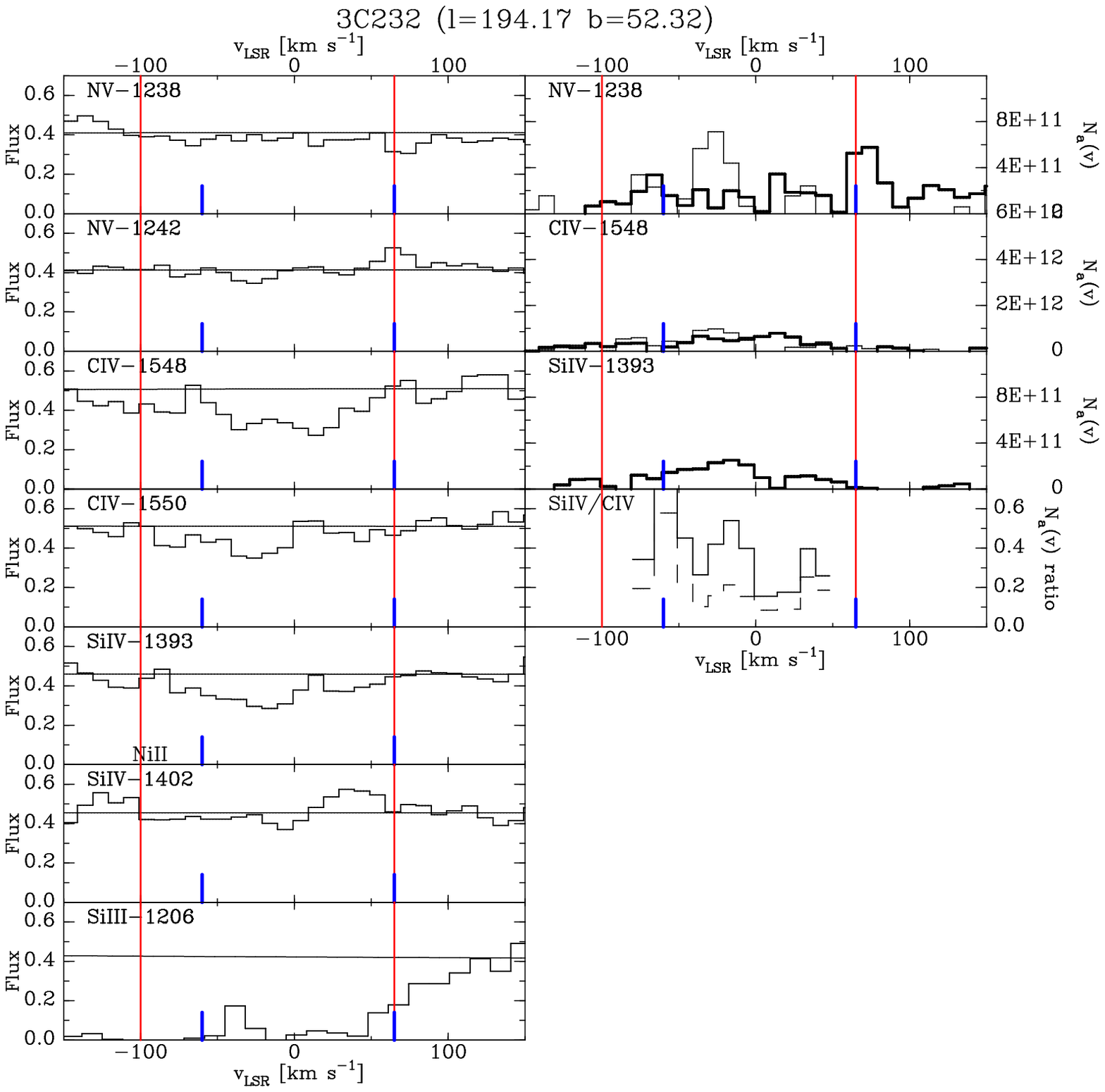}{0in}{0}{350}{300}{0}{0} 
\figurenum{2a}
\caption{%
The panels show the available data for the four high-ionization lines, as well
as for a strong and a weak intermediate-ionization and a strong and weak
low-ionization line (when data is available). The panels on the left give the
spectra for the high-ionization lines, as shown by the labels. The top set of
panels on the right shows the apparent optical depth profiles for the
high-ionization species, with the stronger absorption line shown as thick lines.
No apparent optical depth profiles are shown for lines that were deemed
unmeasurable, even though a panel is always shown if data were present. Below
the optical depth profiles are panels showing the ratios between two ions, but
only in spectral regions where the S/N in the ratio is $>$3 in a 10~\kms\
velocity bin. Strong intermediate ionization is represented by \SiIII\ (or by
\CIII\ when \SiIII\ is not available). Weak intermediate ionization is given by
\FeIII. The strong low-ionization line is either \CIIa\ or \CIIb, while the weak
low-ionization line is \SIIa\ or \ArIa. The intermediate- and low-ionization
lines are always given in the bottom one or two panels. Line identifications are
given at the top left of each panel. The continuum fit (including a model for
H$_2$ absorption) is also shown. In a few cases (e.g.\ for \OVIb\ toward
Mrk\,509), the continuum looks much higher than the data, because it was fit
through pixels far outside the displayed range of velocities. All spectra were
rebinned to 10~\kms\ pixels. Other absorption lines in each spectral window are
identified by labels near the bottom of each panel. The short vertical blue
lines show the range of velocities within 60~\kms\ of those consistent with
standard Galactic differential rotation (see Section~\Sextent\ for the
justification of this number). The red vertical lines indicate the velocity
limits of integration for the different Galactic components. Note that a figure
is available for each sightline in the on-line version of this paper.
}\end{figure}

\begin{figure} 
\plotfiddle{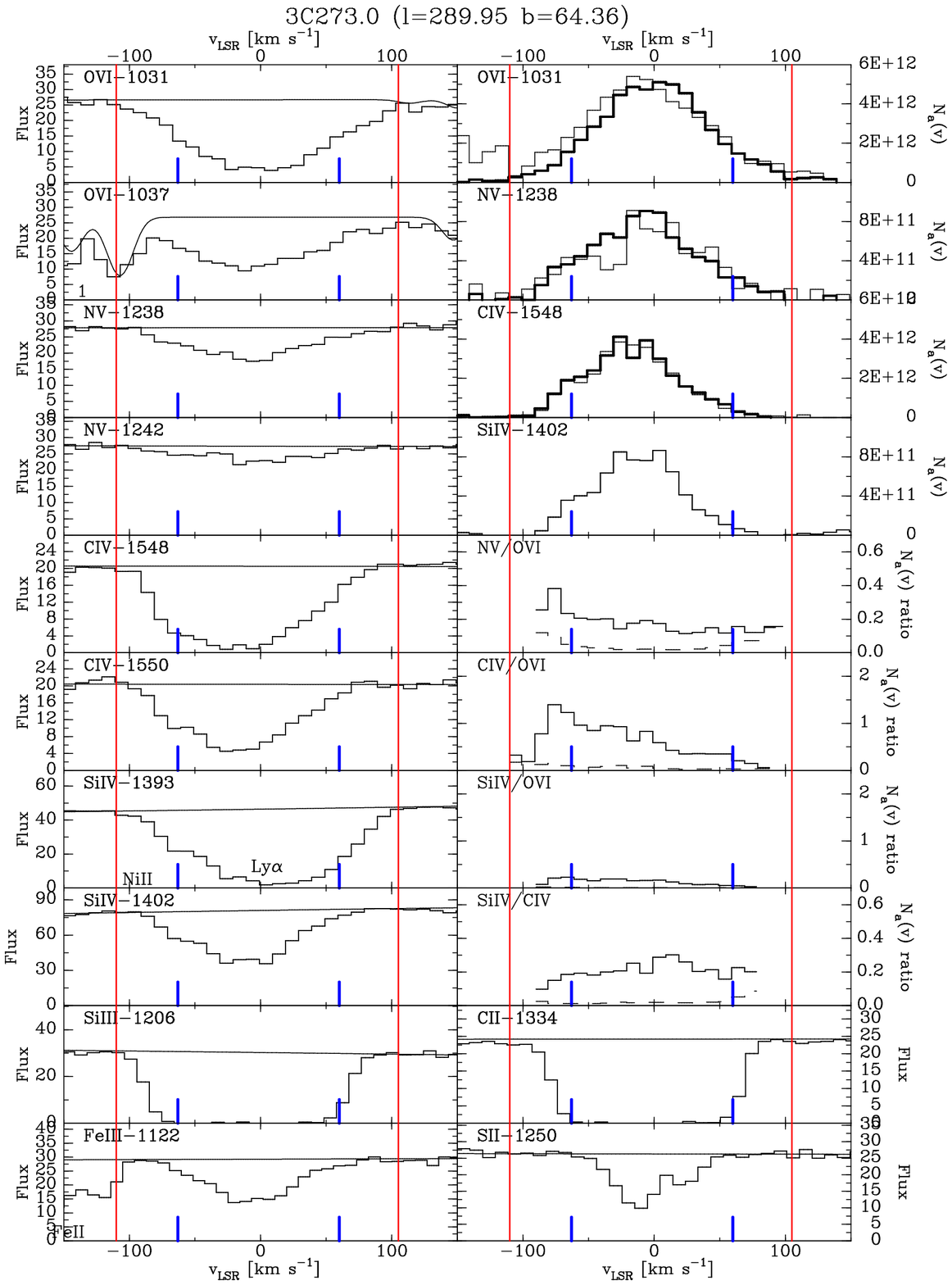}{0in}{0}{450}{600}{0}{0} 
\end{figure}
\begin{figure} 
\plotfiddle{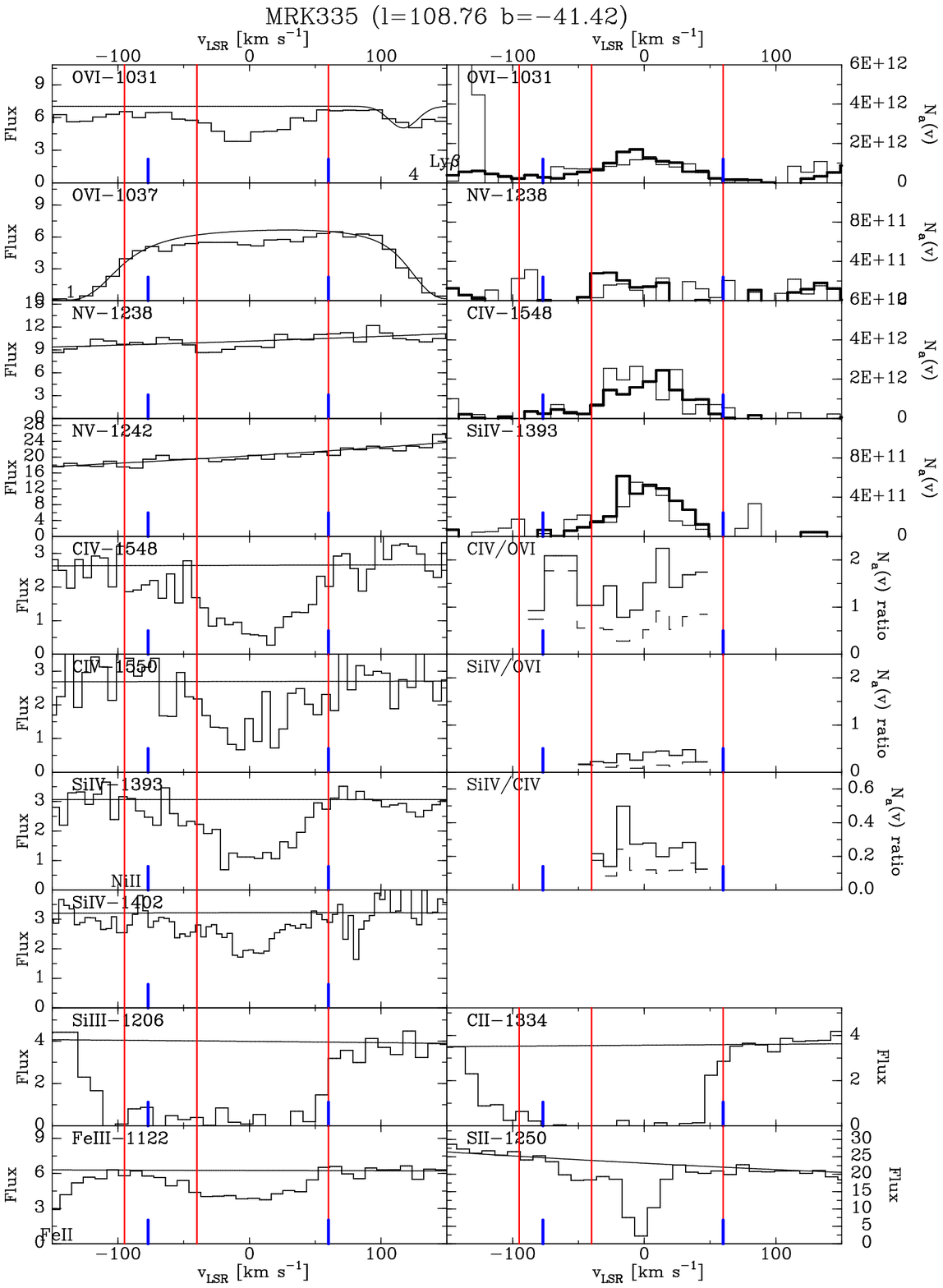}{0in}{0}{450}{600}{0}{0} 
\end{figure}
\begin{figure} 
\plotfiddle{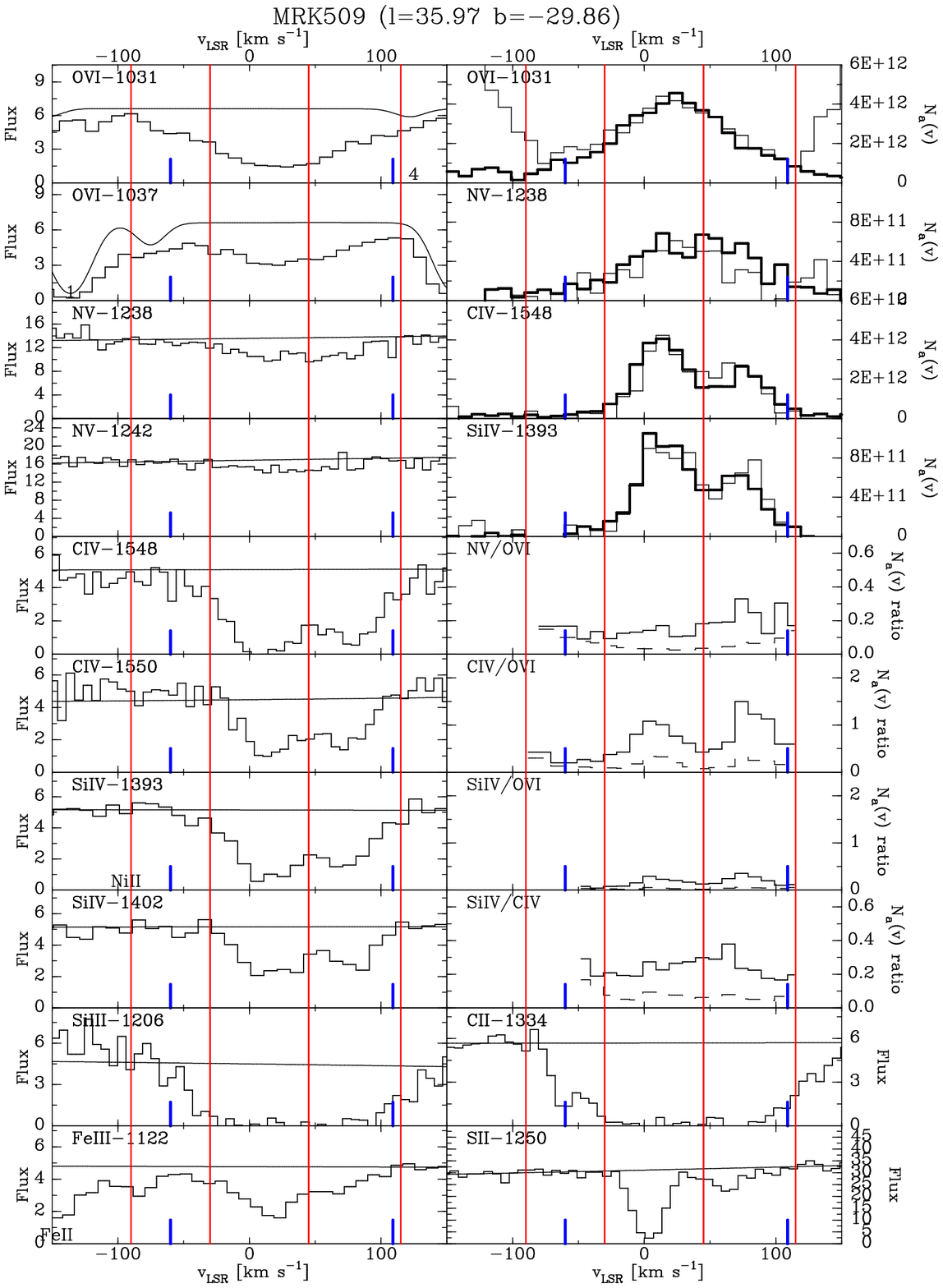}{0in}{0}{450}{600}{0}{0} 
\end{figure}
\begin{figure} 
\plotfiddle{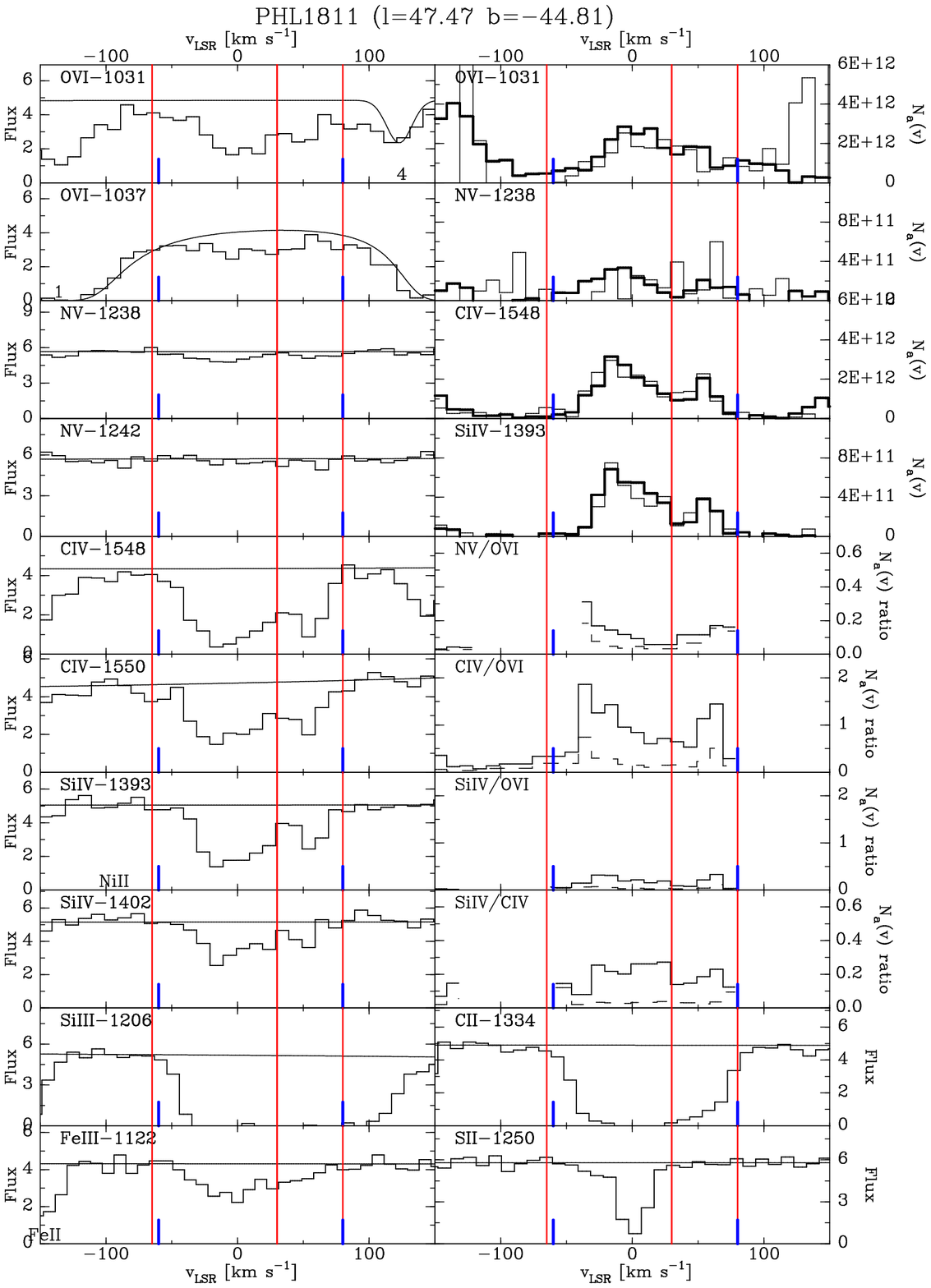}{0in}{0}{450}{600}{0}{0} 
\end{figure}
\begin{figure} 
\plotfiddle{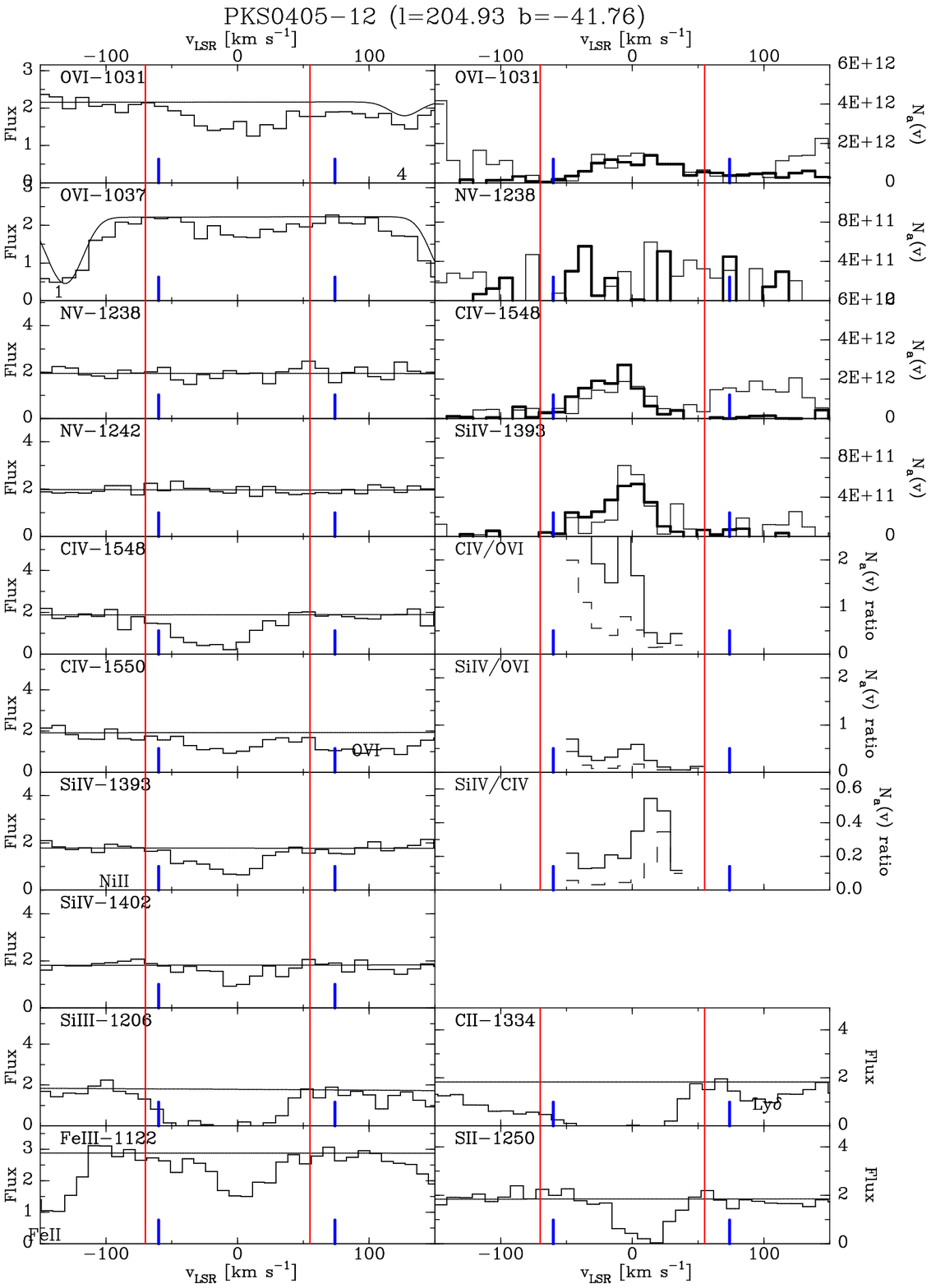}{0in}{0}{450}{600}{0}{0} 
\end{figure}



\section{Measurements}
\par The Galactic high-ionization lines tend not to have simple gaussian
profiles, but rather they typically appear to be a mixture of several broad,
blended components. Since the typical optical depth of the lines is on the order
of 1--2, and since we can only measure two lines for each ion, it is not
advisable to estimate column densities by converting an equivalent width
measurement into a column density or applying a curve-of-growth method. We also
do not resort to fitting components to the line profiles as the signal-to-noise
ratio is often low ($\sim$10--15), the profiles are usually asymmetric, and the
component structure is not always clear. Profile fits would give formally
correct results, and would allow us to calculate the total column density
accurately, but the fitted central velocities and widths of the gaussian
components would more often than not be misleading, as a good fit would require
several components, which usually cannot be defined unambiguously.
\par Because of these considerations, we instead integrated the apparent column
density (\nav) profiles (Savage \& Sembach 1991). For resolved and unsaturated
lines this represents the true column density distribution of the ion, broadened
by the instrumental profile. If the \nav\ profiles of the two members of each
doublet match, then one can conclude that there is no unresolved saturation. In
practice we find that this is a good assumption for the Galactic high-ionization
lines, as the resulting column densities always match to within the errors (with
one exception, see below).
\par To derive the column density profiles, we first fitted a local continuum to
each of the absorption lines. To do this, we started by selecting line-free
regions in a window about $\pm$1000~\kms\ wide around each line and then fitted
a first, second or third order Legendre polynomial to these points, following
the method of Sembach \& Savage (1992). In addition, a model for the molecular
hydrogen absorption is applied, using the H$_2$ column densities found by Wakker
(2006). Any associated H$_2$ optical depth is removed from the apparent optical
depth profile. This is mostly important for the \OVIa\ line, while for \OVIb\
and \SIIIa\ the H$_2$ contamination is often too large to be removed reliably.
We only list the results for those two lines if the contamination is
sufficiently minor. In some sightlines we list the measurements of
H$_2$-decontaminated \OVI\ lines, but do not use them for the analyses.
\par We visually determined the velocity extent of the absorption associated
with the Milky Way Thick Disk. In many sightlines absorption occurs at
velocities more negative than $-$80~\kms\ or more positive than 80~\kms. This
high-velocity absorption was analyzed by Sembach et al.\ (2003). It is often
associated with known 21-cm \HI\ high-velocity clouds, which trace the accretion
of low-metallicity material and tidal streams (Wakker et al.\ 1999, 2007;
Richter 2006). In this paper we concentrate on the high-ionization absorption
associated with the gas at velocities below about 80~\kms, which originates in
the Disk and lower Galactic halo (i.e.\ the Thick Disk). An analysis of the
high-velocity gas is deferred to the future.
\par Next, we looked at the structure of all absorption lines toward each target
and used the visual appearance of each profile and the comparison between the
profiles for different ions (high-, intermediate and low-ionization species) to
decide whether one can discern individual components and in which velocity
ranges these occur. The structure in the low-ionization lines is taken into
consideration for the determinations, but it does not drive the resulting
interpretation for the high-ionization line component structure. We also noted
cases where the line of the highly-ionized species is blended with other
galactic or intergalactic absorption lines.
\par We then integrated the \nav\ profile, both over the full velocity range
over which we think absorption originates in the Thick Disk, and over the
different selected velocity ranges for each individual component. For each \nav\
integration we determined a statistical and a systematic error, following the
methods described by Wakker et al.\ (2003). The statistical error combines the
Poisson noise, detector background and continuum placement, while the systematic
error takes into account the placement of integration limits and fixed-pattern
noise. We flagged the cases where we judged it possible to derive reasonable
values for line ratios.
\par We also calculated a central velocity ($v_0$) and FWHM ($W$, full width at
half maximum, $W$=$\sqrt{4\,\ln2}\,b$, where $b$ is the Doppler parameter)
linewidth from the first and second moments of the \nav\ profile. For the
analyses we also correct the derived FWHMs for instrumental broadening (see
notes to Table~\Tres\ for details).
$$  N      = \int N_a(v) dv, $$
$$  v_0    = [\int v N_a(v) dv ] / [\int N_a(v) dv], $$
$$  W      = \sqrt{(4{\rm ln}2)}\left(\ [\int (v - v_0)^2 N_a(v) dv ] / [\int N_a(v) dv]\ \right)^2. $$
We applied our best judgments to determine whether the resulting values are
likely to represent a real width, as opposed to just being set by the center and
width of the integration range. Only in about half of the sightlines were the
individual components separated clearly enough to derive individual velocities
and widths. However, for the full Thick Disk integration we can derive a column
density weighted average velocity and width in almost all cases.
\par We checked all doublet measurements for consistency. The two measurements
for each half of each doublet were deemed to be compatible if the column density
difference was less than the sum of the statistical and systematic errors for
each individual component. For example, for the $-$110 to +105~\kms\ integration
toward 3C\,273.0, the \NV\ lines give column densities of
13.94$\pm$0.02$\pm$0.01 and 13.90$\pm$0.03$\pm$0.01. As 13.94 and 13.90 are
closer together than 0.02+0.01+0.03+0.01=0.07 dex, these two values are deemed
compatible. We justify adding the errors linearly instead of in quadrature by
noting that just by chance the measurement for a particular line will be more
than 1$\sigma$ above (or below) the real value in 16\% of the cases. In 16\% of
those cases the measurement of the other line of the doublet will be more than
1$\sigma$ off in the other direction. Thus, with 258 cases where we measure both
members of a doublet, we expect 6.6 cases where the column densities implied by
the two member of the doublet differ by more than the sum of the errors. We find
seven such cases; in four of these the excess the difference is just 0.01 dex.
In only one case do the two lines of the doublet suggest the presence of
saturation. This is for the \CIV\ lines in the direction toward ESO\,141$-$G55,
which are very deep, resulting in a factor 2 difference in the column densities
derived from the two members of the doublet. Since our check shows that in all
but one case the absorption lines are unlikely to be saturated, we estimated
final values for the ionic column densities by averaging the measurements toward
the two lines of the doublet, if both measurements were deemed to be useful.
There is one exception: for the saturated \CIV\ line toward ESO\,141$-$G55, we
instead used the rule-of-thumb derived by Sembach \& Savage (1991) that
$\log\,N$(ion)= 2\,log$N$(weak) $-$ $\log\,N$(strong).
\par The resulting measurements are presented in Table~\Tres. This table shows
for each target, and for each component, the measured column densities in each
high-ionization line, as well as the central velocities, widths and ratios for
those cases where the result was deemed to be reasonably reliable. Note that the
widths that we list are corrected for instrumental broadening (see notes for
details).
\par We compared our values to those listed by Savage et al.\ (1997b) in their
earlier study of highly-ionized atoms toward extragalactic targets and find that
for the most part our measurements of \NV\ and \CIV\ column densities are the
same within the errors. There are two exceptions: (a) H\,1821+643, for which
Savage et al.\ (1997b) had not realized that there is strong contamination
associated with the nearby planetary nebula K1-16, and (b) PKS\,2155$-$304, for
which we concluded that the \NVa\ line is contaminated by an intergalactic \Lya\
line because the two lines of the doublet do not match, and the absorption that
might be Galactic \NV\ extends to much more negative velocities than is the case
for the other high-ionization lines.
\par We also checked our measurements against the numbers presented by
Indebetouw \& Shull (2004b). To make their measurements, they integrated all of
the high-ionization absorption, mixing together high-velocity clouds and Milky
Way Thick Disk components. In a few cases they did not remove blending
intergalactic absorption and they did not remove the absorption associated with
the planetary nebula K1-16 near the H\,1821+643 sightline. Furthermore, their
velocity limits are often inconsistent with our judgments. Also, in many cases
the column densities they give for the two lines in a \NV\ or \CIV\ doublet are
often rather different, sometimes by as much as 0.3~dex. In addition, the column
density errors in their table are much larger than what we found (typically they
are listed as 0.2--0.6~dex). Finally, in sightlines where there are no
high-velocity clouds or other problems the column densities listed by Indebetouw
\& Shull (2004b) still usually differ from our results. Due to all these issues,
we did not make a detailed comparison of our results to the measurements given
by Indebetouw \& Shull (2004b).

\input table2.tex


\section{Results}
\par In this section we analyze several different aspects of the highly ionized
gas, including the kinematic properties of the absorption lines (extent,
component width, relation to Galactic rotation), the distribution of column
densities, the distribution of velocities and column densities across the sky,
and the distribution of column density ratios. These analyses show that the gas
in which these ions occur participates in differential galactic rotation, but
has a large velocity dispersion. There is no evidence for a gradient in the
rotation velocity with height above the Galactic plane. The highly-ionized ions
are present across the sky, in similar amounts in every direction, but with a
patchy distribution. There are some well-delineated regions where the average
properties are noticeably different (factor two average enhancement around the
Galactic Center and near the Galactic North Pole, \OVI\ showing a factor two
deficit at intermediate southern latitude). The column densities and column
density ratios in individual parcels of transition temperature gas have a narrow
distribution, suggesting a uniformity of its properties.

\subsection{Velocity differences between ions}
\par We compared the centroids of the absorption profiles of the different ions
(\OVI, \NV, \CIV, \SiIV, \FeIII) to each other and found that they correlate
very well. In particular, for the 9 components where both the \OVI\ and \NV\
absorption lines are well-defined and simple, the centroids have an average
offset of 0$\pm$8~\kms, i.e.\ the rms difference is well within the resolution
of the \FUSE\ \OVI\ spectra, as well as within the alignment errors associated
with the wavelength calibration of the \FUSE\ data. For the 24 components with
good data for both \CIV\ and \SiIV\ the difference in their centroids is
0$\pm$3~\kms. The centroids of \OVI\ components are about 4$\pm$8~\kms\ more
negative than those of \CIV, \SiIV\ and \FeIII, which is not a significant
difference. The \NV\ and \CIV\ lines appear to be well-aligned (average
difference $-$2$\pm$4~\kms\ for 7 components), and all of the cases where both
of these lines are seen are derived from \STIS-E140M spectra. Thus, the
preponderance of the evidence suggests that there are {\it no} velocity offsets
between the centroids of the absorption lines of the different highly-ionized
species.



\subsection{Velocity extent of the Galactic absorption}
\par In each direction, there is a range of velocities that is compatible with
differential galactic rotation. Following Wakker (2004), we estimate this range
using a model in which gas rotates at 220~\kms\ around the center of the Milky
Way for galactocentric radii 0.5 to 26~kpc in a layer with thickness 3~kpc
inside the 8.0~kpc solar radius, increasing to 6~kpc at 26~kpc from the Galactic
Center. At radii $<$0.5~kpc the velocity of rotation is assumed to decrease
linearly to 0~\kms\ at $R$=0. Since most of our targets are at high galactic
latitude, this usually produces a narrow range from 0 to about plus or minus
10~\kms, except for directions below $|$$b$$|$=30\deg, where a range of 0 to
about plus or minus 40~\kms\ is found.
\par Figure~\Fextent\ shows the observed extents of the absorption profiles
graphically. The vertical lines give the range of velocities over which Galactic
absorption is seen, while the short horizontal lines are drawn at the median
difference of $\pm$60~\kms\ more positive or negative than those the simple
model predicts. As this figure shows, the observed range of velocities over
which absorption is seen is much larger than that predicted by differential
galactic rotation. Absorption typically extends to velocities that are 45 to
100~\kms\ more negative than expected, with an extreme of 125~\kms\ toward
PG\,0953+414. At high latitudes ($b$$>$45\deg) the typical profile also extends
to velocities that are 30 to 80~\kms\ more positive. Looking separately at
positive and negative velocities for four sky quadrants
($b$=$-$90\deg:$-$45\deg, $b$=$-$45\deg:0\deg, $b$=0\deg:45\deg\ and
$b$=45\deg:90\deg\ (10, 10, 9 and 16 sightlines, respectively), the median and
the average deviation from the maximum negative or positive velocity predicted
by the simple model of rotation is about 60~\kms, {\it except} for negative
velocities near the north polar cap, where the median and average are 90~\kms.
That is, for sightlines near the north galactic pole there appears to be an
excess of absorption at velocities more negative than $-$60~\kms.
\par The velocity range over which the Milky Way transition-temperature gas
absorption extends is likely proportional to the outflow velocity of the hot
gas. A value of about 60~\kms\ is in line with the predictions of Houck \&
Bregman (1990) for the formation velocities of neutral (i.e.\ cooled) clouds in
a low-temperature (5\tdex5~K) Galactic Fountain, with a density of
\dex{-3}\,\cmm3\ and a (two-sided) flow rate of 0.4~\Msunpyr. Statistically
speaking, however, the absorption profiles have a similar extent at positive and
negative velocities, independent of Galactic longitude or latitude. Thus, on
average, vertical motions are as large as motions parallel to the plane. This
suggests that the material is highly turbulent. The only evidence for a
deviation from this pattern is that there appears to be a slight preference
for infall in the North Galactic Cap region. This may just be a temporary
local occurrence in the solar neighborhood, however.

\begin{figure} 
\plotfiddle{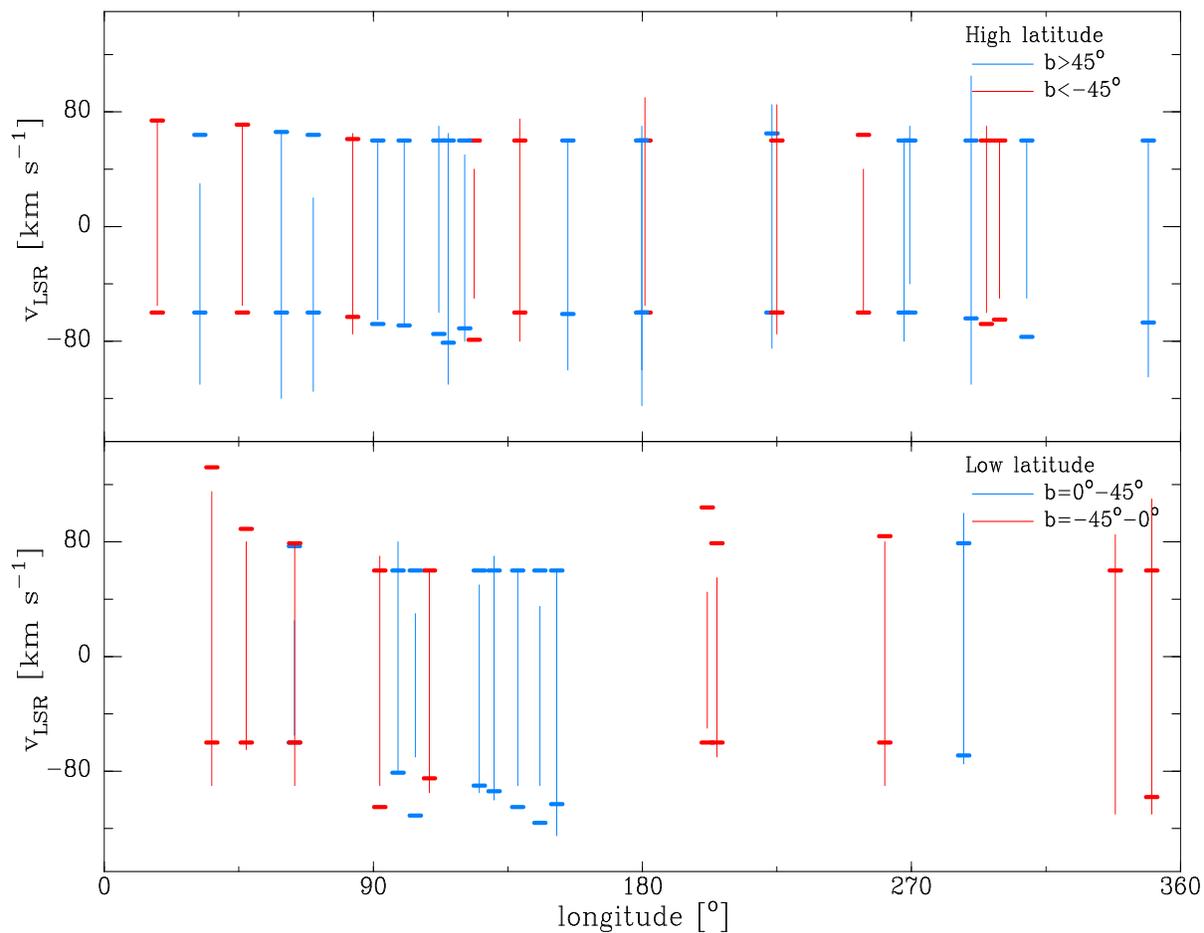}{0cm}{0}{450}{350}{10}{0}
\figurenum{3}
\caption{%
Comparison of the extent of the Galactic high-ionization absorption with the
range of velocities allowed by differential galactic rotation, as function of
galactic longitude. The thick horizontal bars are placed at velocities
$\pm$60~\,\kms\ larger/smaller than the maximum/minimum velocity expected from
rotation. The vertical lines indicate the observed range of velocities over
which high-ionization absorption is seen. The top panel shows high-latitude
sightlines, the bottom panel low-latitude sightlines, with northern sightlines
in blue and southern sightlines in red.
}\end{figure}


\subsection{Line width distribution}
\par In Figure~\Fwidthhist\ we look at the distribution of linewidths, rather
than at the total extent of the absorption. These widths are found by
calculating the second moment ($b$) of the \nav\ distribution, converting it
into the full-width-at-half-maximum (FWHM=1.665\,$b$) and correcting for
instrumental broadening. The figure has two panels for each ion. In the top
panel we use the full absorption profile, while in the bottom panel we use the
FWHM for individual components (see Section~\Smeas). Note that in about half of
the sightlines only a single component can be discerned. Only components where
the measured width was deemed reliable are included. This means that in some
sightlines the width of the full profile can be measured, but the widths of the
two or three components that make up the profile are individually unreliable.
Thus, the total number of measurements of the full line width is larger than the
number of well-measured components.
\par For many sightlines it is possible to separately calculate the widths of
both members of the \OVI, \NV, \CIV\ and \SiIV\ doublets. In every case the two
widths are identical to within the errors. Therefore, we use only the width of
the strong line for the histograms in Figure~\Fwidthhist\ (except when only the
weak line was measurable). Widths are not calculated if the Galactic profile
blends with high-velocity absorption and it is not clear where to place the
boundary (for Mrk\,205, Mrk\,279, Mrk\,817, Mrk\,876, and NGC\,3783). All cases
where we deemed it possible to estimate a width are listed in Table~\Tres\ and
used in Figure~\Fwidthhist.
\par We do not explicitly show a figure that directly compares the widths of the
ions in individual absorption components, but these correlate very well. That
is, if FWHM(\OVI) is large, then FWHM(\NV) is large. The correlation coefficient
between the two widths is 0.97. FWHM(\CIV) and FWHM(\SiIV) also correlate
closely (correlation coefficient 0.94). The correlation between FWHM(\OVI) or
FWHM(\NV) and FWHM(\CIV) or FWHM(\SiIV) is not quite as good, but still strong
($\sim$0.92). Thus, \OVI\ goes closely together with \NV, but somewhat less
strongly with \CIV\ and \SiIV, even though \NV, \CIV\ and \SiIV\ are determined
from \STIS-E140M spectra (with 6.5~\kms\ resolution), while \OVI\ is seen in
\FUSE\ spectra (20~\kms\ resolution). Thus, the differences in linewidths are
not likely to be an artifact of the higher resolution of the \STIS-E140M
spectra.

\begin{figure} 
\plotfiddle{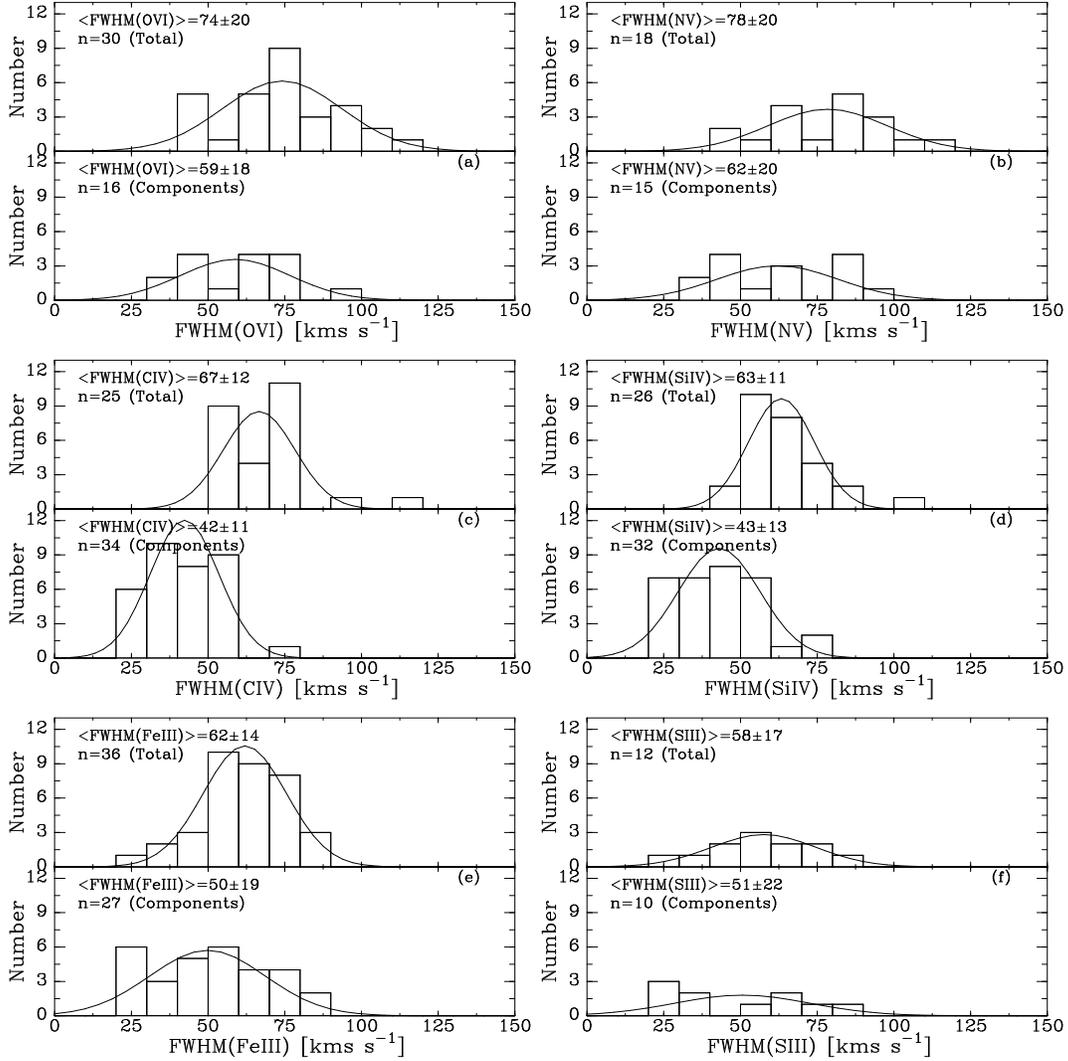}{0in}{0}{400}{400}{30}{0}
\figurenum{4}
\caption{%
Distribution of linewidths for the six ions. There are two panels for each ion,
with the top panel in each pair showing the distributions of the FWHM
(calculated from the second moment) for the integrated Galactic absorption, and
the bottom panel showing the values for individual components. The FWHM of the
full Galactic absorption profile is a general indicator for the velocity spread
of the high-ionization absorption associated with the Milky Way disk and lower
corona. The FWHM of the individual components indicates the velocity spread in
single structures. The average and spread of each distribution are given in the
label, as are the number of sightlines/components that are used.
}\end{figure}

\par As Fig.~\Fwidthhist\ shows, the FWHM of the full Galactic absorption ranges
from about 50~\kms\ to about 120~\kms, with an average value in the range
60--80~\kms, dependent on the ion. The widths of the \OVI\ and \NV\ absorption
lines appear similar, while those of \CIV, \SiIV\ and \FeIII\ appear to be
slightly narrower. We applied a $t$-test to see whether these differences are
statistically significant. This test is most useful when applied to the
distributions of the widths of the individual components. Then the difference in
the average linewidth of \OVI\ and \NV\ (\VALUE 59$\pm$18~\kms\ and \VALUE
62$\pm$20~\kms) is not significant (the probability that they do not differ is
55\%). This is also the case for the difference between the average widths of
\CIV\ and \SiIV\ (\VALUE 42$\pm$11~\kms\ and \VALUE 43$\pm$13~\kms, probability
that they do not differ is 80\%). However, the difference between each pair of
ions {\it is} significant, with probabilities that the widths do not differ of
$<$0.5\% when comparing \OVI\ or \NV\ to \CIV\ or \SiIV. The average width of
the \FeIII\ line (\VALUE 50$\pm$19~\kms) is marginally larger than that of \CIV\
(probability it does not differ 5\%), and insignificantly smaller than that of
\OVI\ (probability it does not differ 2\%). Thus, we cannot really tell whether
\FeIII\ belongs in the \CIV/\SiIV\ group or in the \OVI/\NV\ group. Because of
its lower ionization potential, the former is more likely, however. For \SIII\
the number of points is too small to say anything useful.
\par Thus, the evidence points to a situation in which the \OVI\ and \NV\ lines
have similar widths, as do the \CIV, and \SiIV\ lines, with the latter two ions
having narrower lines on average. The general correlation of the linewidths and
the alignment in velocity between all ions (see Section~\Svdiff) shows that the
different ions must originate in the same general structures, but also that they
cannot be completely co-spatial. Further, since even at a temperature of \dex5~K
the thermal width of the lightest of our elements (carbon) is only 20~\kms, most
of the observed linewidth cannot be due to thermal broadening. This implies that
there is either a high level of turbulence or there are strong velocity
gradients in the gas.
\par A difference in line widths between \OVI/\NV\ vs \CIV/\SiIV\ was previously
found by Fox et al.\ (2003), who analyzed high-resolution (3.5~\kms) \STIS-E140H
data of the halo star HD\,116852. They saw both a narrow and a broad \CIV\ and
\SiIV\ line, but \OVI\ and \NV\ did not show the narrow absorption. They
interpreted this as indicating that there are regions in the Thick Disk that are
sufficiently warm to contain \CIV, but not warm enough to have \OVI. In an
analysis of 38 high-resolution \STIS-E140H spectra of stars, Lehner et al.\
(2011) find the same effect in several other sightlines, with the average FWHM
of the narrow \CIV\ and \SiIV\ components about $\sim$20~\kms. They further find
only broad absorption components for \OVI\ and \NV\ (average FWHM 65$\pm$22 and
50$\pm$25~\kms, respectively), The S/N ratio of our data is usually not nearly
as high as it is for these stellar spectra, making it more difficult to separate
narrow and broad \CIV/\SiIV\ components. Further, all of our sightlines pass
through several kpc of gas, whereas the sightlines to the stars typically are
only 1--3~kpc long. This will produce more blending between structures at
different distances/velocities for the AGN sightlines. Nevertheless, we find
that the distributions of the widths of the \OVI\ and \NV\ lines in sightlines
through the Thick Disk are very similar to those in sightlines through the disk.
We also find that the \CIV/\SiIV\ lines are narrower, but we do not find the
very narrow components seen in the disk. This is not necessarily a result of the
poorer resolution of our spectra (6.5~\kms\ vs 3.5~\kms), since lines that are
20~\kms\ wide should still have been resolved. Thus, the presence of narrow
\CIV/\SiIV\ absorption components may indicate a real difference between low-
and high-latitude sightlines. It may also be due to the longer pathlength
through the disk for the AGN sightlines. Or it may be caused by the limitations
of the lower S/N ratio in the extra-galactic sightlines.

\subsection{Signature of Galactic rotation}
\par In order to determine whether the velocity of the absorbing gas reflects
Galactic rotation, we use the kinematical model described in Sect.~\Sextent, but
added an exponential density distribution having scaleheight 2.6~kpc. We then
find the velocity and density along the sightline in steps of 0.2~kpc, convert
this density to an optical depth and add a gaussian with FWHM 60~\kms. Summing
over the path produces a predicted apparent column density profile, from which a
centroid can be calculated. We look at two versions of this model, one in which
the coronal gas co-rotates with the disk at all heights above the plane and one
in which there is a decrease in the rotation velocity of 10~\kms\ per kpc.  The
latter is inspired by the results of e.g., Fraternali et al.\ (2004), Heald et
al.\ (2006, 2007) and Oosterloo et al.\ (2007), who find that in several nearby
spiral galaxies H$\alpha$ and \HI\ emission show gradients of
$\sim$15--25~\kms\,kpc$^{-1}$ at heights more than 1~kpc above the plane. Since
in reality the distribution of the transition temperature gas is highly
irregular, the only justified use of this model is to give an expectation value
for the centroid of the absorption profile, but it cannot be used to model the
individual absorption profiles.

\begin{figure} 
\plotfiddle{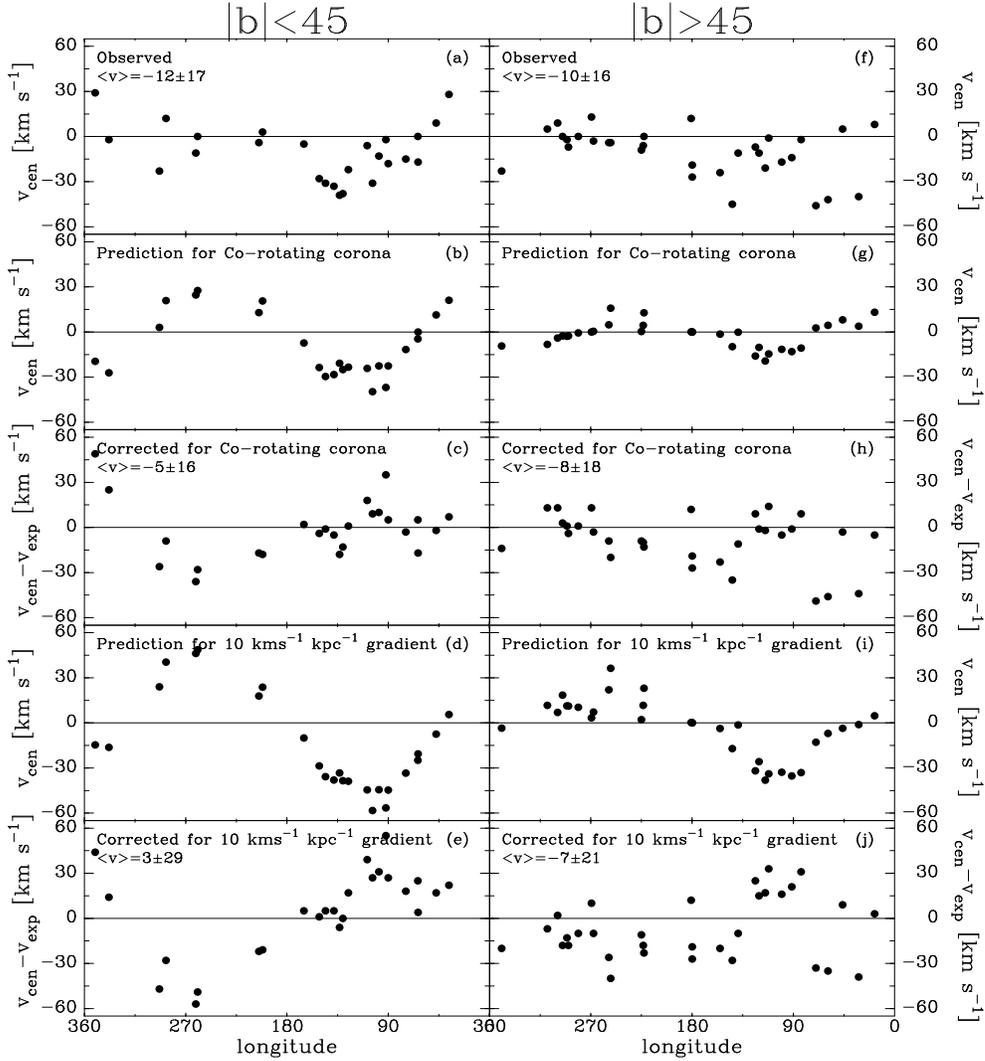}{0in}{0}{370}{400}{30}{0}
\figurenum{5}
\caption{%
Comparison of observed and expected centroid velocities of the profiles of
absorption by highly-ionized atoms, as function of longitude. Panels on the left
are for sightlines at low latitudes (\absb$<$45\deg), panels on the right are
for high latitudes. The top panels give the observations for \OVI. The second
and third panel give the predictions and difference between observed and
predicted values for a co-rotating corona, while the bottom two panels give
this for a corona having a 10~\kms\ decrease in velocity per kpc away from the
Galactic mid plane.
}\end{figure}

\par Using our model, we find that galactic rotation does have a subtle, but
measurable effect on the centroids of the profiles. This can be seen in
Figure~\Fvelo, which shows the observations, and the predictions for a
co-rotating corona as well as those for a corona with a vertical velocity
gradient. The differences between the observations and these two models are also
shown. The panels on the left give this for low latitude sightlines, while those
on the right are valid for sightlines at high latitudes. For the high latitude
sightlines we expect to see little to no systematic variation with longitude in
the centroid velocity, which is indeed observed to be the case. At low
latitudes, the co-rotating corona predicts centroids to be about $-$40~\kms\
near $l$=90\deg, and about +30~\kms\ near $l$=270\deg, while the model with a
10~\kms\,kpc${-1}$ vertical velocity gradient predicts values of about $-$60 and
+60~\kms, respectively. Increasing the gradient to 20~\kms\,kpc$^{-1}$ increased
this range to about $\pm$70~\kms. The observations appear more in line with the
co-rotating corona version of the model. Excluding the two outliers at
$l$$>$300~\kms\ (ESO\,141$-$G55 and PKS\,2005$-$489), the average difference
between predicted and observed centroids is \VALUE 5~\kms, with an rms spread of
\VALUE 16~\kms\ (see panel (c)). From Figure~\Fvelo\ we conclude that Galactic
rotation does leave an imprint on the velocities of the highly-ionized atoms.
However, the effect is subtle, leading to centroids differing from 0~\kms\ by no
more than 40~\kms.
\par The assumed vertical distribution of the average density of the transition
temperature has a large influence on the resulting predictions for the centroid
velocities. If we were to assume that the density of the transition temperature
gas is constant between 0 and several ($\sim$5) kpc above the Galactic plane, we
would predict very little difference between a co-rotating corona and one with a
vertical gradient, but the centroids would have values up to $\pm$50~\kms\
(i.e.\ much larger than what is observed). If we instead assume a scaleheight of
0.5~kpc, the predicted centroids would range between $\pm$10~\kms\ for a
co-rotating corona and between $\pm$30~\kms\ for one having a 10~\kms\,kpc${-1}$
vertical gradient. The predictions for the latter case are about as good as the
ones for our standard model.
\par Thus, it is possible to exclude a model in which the transition temperature
extends to very large heights above the plane. However, it is not really
possible to differentiate between a model with a co-rotating corona and a scale
height of 2.6~kpc and a corona with a vertical velocity gradient and a scale
height of 0.5~kpc. On the other hand, using the fact that the directly derived
scale height is about 2.6~kpc, we can conclude that the transition temperature
gas appears to be co-rotating up to heights of a few kpc.


\subsection{Column density distribution}
\par Figure~\Fcoldens\ presents the distributions of the column densities for
the different ions. For each ion we show the total Galactic column density in
the top panel and the column densities of the individual components in the
bottom panel. Upper limits are indicated by the hatched areas. In the bottom
panel of each pair, the dotted vertical line shows the column density detection
limit corresponding to a line with 20~\mA\ equivalent width and FWHM 40~\kms.
Such an equivalent width is the detection limit when integrating over a 75~\kms\
window in data with an S/N ratio of about 20. These limits are 13.2 dex for
\OVI, 13.0 dex for \NV, 12.7 dex for \CIV, 12.4 dex for \SiIV, 13.4 dex for
\FeIII\ and 13.7 dex for \SIII. Clearly, except for \NV, the typical detection
limit is 0.6 dex to 1.1 dex below the average component column density, making
these lines always easily detectable. For \NV\ the detection limit is usually
only about 0.3 dex below the typical value, which is mostly caused by the fact
that the interstellar abundance of nitrogen is about 0.8 dex less than that of
oxygen and carbon. Therefore, although \NV\ is detected in 43 Galactic
components, for 46 components only an upper limit is possible. In contrast, an
upper limit is found for \OVI, \CIV\ and \SiIV\ in just 3, 3 and 4 components,
respectively.
\par Savage et al.\ (2003) derived an average Galactic \OVI\ column density of
14.32$\pm$0.19, which is basically the same as the value of 14.31$\pm$0.20 found
here. They also looked at the average value of \logNisinb{\OVI}, which is
14.21$\pm$0.23, or about 0.11 dex lower. We do not show the distributions of
\logNisinb{\OVI} separately, as they are very similar to the ones shown in
Figure~\Fcoldens, but shifted to column densities that are $\sim$0.15~dex lower.
The spread in the column densities of highly-ionized atoms is such that for 95\%
of the sightlines the integrated column density is within a factor $\sim$2.5
from the average. I.e., not only is the transition temperature gas widespread,
{\it on average} it is fairly evenly distributed.
\par The histograms in Fig.~\Fcoldens\ show that along paths through the
Galactic Thick Disk, \OVI, \CIV\ and \SiIV\ are always found. Only \NV\ is
sometimes missing, which is most likely due to the fact that the line is weaker,
which in turn in due to the fact that the cosmic abundance of nitrogen is lower
than that of oxygen and carbon. To calculate the average and dispersion of the
distribution of $N$(\NV), we only use the detections, and exclude the upper
limits. We justify this on the basis of the fact that generally the ratio
\NV/\OVI\ has a narrow range (see Sect.~\Sratiohist\ below) and in sightlines
where \NV\ is not detected the implied limit to the \NV/\OVI\ ratio is within
the range of detected values or above it. That is, given $N$(\OVI) and \NV/\OVI,
the non-detection of \NV\ is generally not significant. There are only five (out
of forty) components for which the measured upper limit on the ratio \NV/\OVI\
is slightly more than 2$\sigma$ below the average ratio; i.e., based on
$N$(\OVI), a detection of \NV\ would have been expected.

\begin{figure} 
\plotfiddle{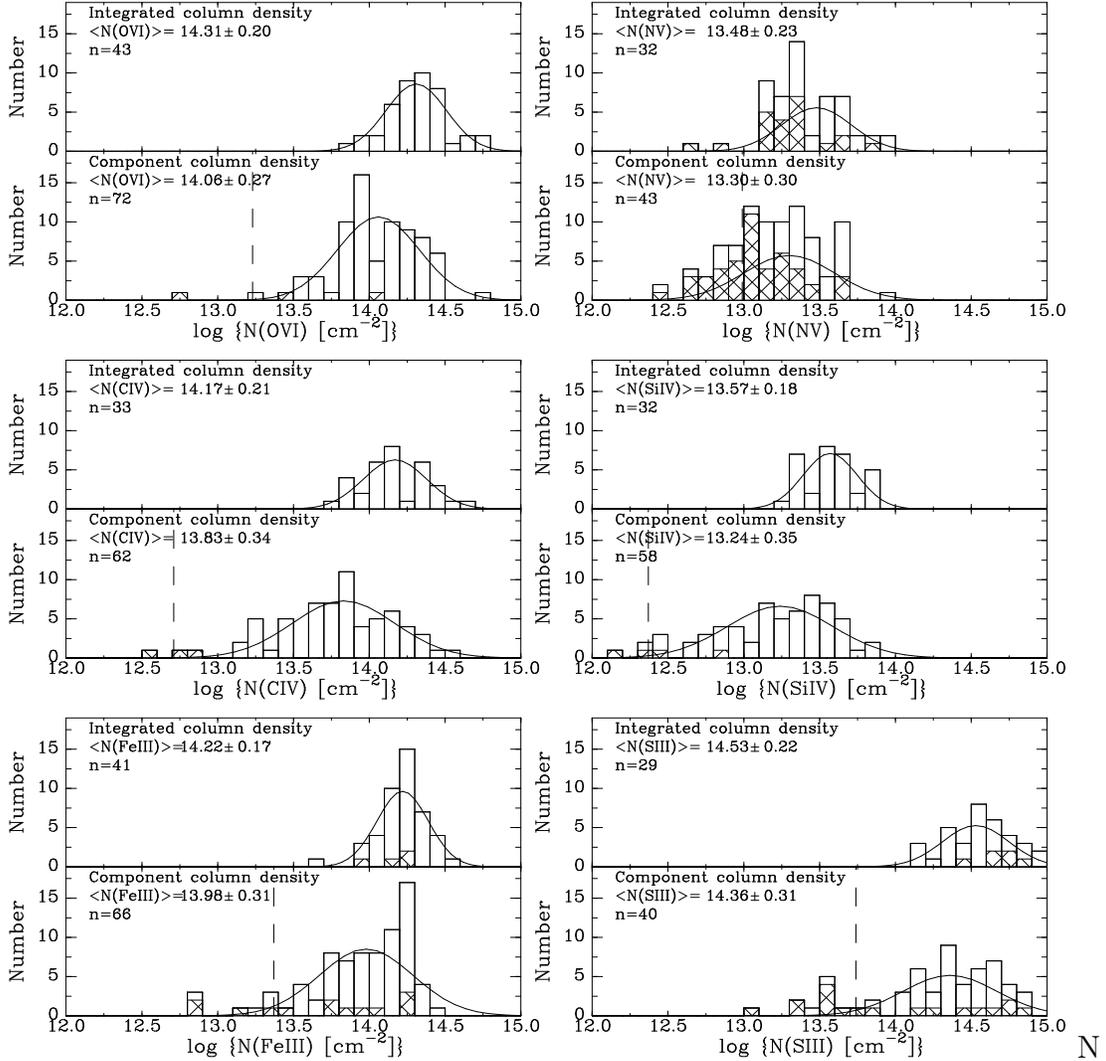}{0in}{0}{400}{400}{30}{0}
N\figurenum{6}
\caption{%
Distributions of column densities for the four high- and two
intermediate-ionization species. The top panel of each pair is for the
integrated column densities, the bottom panel for the individual components. The
hatched parts of the histograms correspond to upper limits. The dotted vertical
lines in the bottom panels show the detection limit for a 20\,\mA\ line with
FWHM 40~\kms, while the curves are gaussians using the parameters listed in the
label (i.e., number, average and dispersion). Limits were excluded in these
calculations (see text for a discussion of why this is justified in the case of
\NV).
}\end{figure}


\subsection{Sky distribution}
\par Savage et al.\ (2003) discovered that the \OVI\ column density is
systematically enhanced in the northern sky relative to the southern sky, but
they did not see a systematic dependence of $N$(\OVI) on longitude. To be
precise, they found an average for \logNisinb{OVI} of 14.26 for $b$$>$0\deg\ and
14.09 for $b$$<$0\deg. We can now refine this conclusion.
\par In Figures~\Fskycollongtot\ and \Fskycollatitot\ we show the logarithmic
column densities of the six ions as a function of Galactic longitude. In
Figure~\Fskycollatitot\ we further include curves showing the expected change of
column density with latitude for a plane-parallel layer (i.e.,
log\Nisinb{90\deg}).
\par When looking at Figures~\Fskycollongtot\ and \Fskycollatitot\ (at first
ignoring the colors), it appears that sightlines with high column density occur
relatively more often at longitudes within 30\deg\ of $l$=0\deg, with about a
factor of two difference between directions near $l$=0\deg\ vs directions near
$l$=180\deg. This effect is not seen in the column densities of \FeIII\ or
\SIII. Further, sightlines at $b$$>$45\deg\ tend to lie above the blue
log\Nisinb{90\deg} line. When exploring this further, we found that column
densities are generally enhanced near the Galactic Center and near the Galactic
north pole. Further, \OVI\ is deficient in parts of the southern sky. These
regions are delineated in Figure~\FmapOVIvellincol. We note that there are about
100 additional sightlines toward which $N$(\OVI) can be measured. These
sightlines include the ones that were in the original \FUSE\ \OVI\ survey
(Wakker et al.\ 2003) as well as sightlines observed between 2003 and 2007. We
have not yet measured each of these with the same care as the sightlines in this
paper, but a preliminary assessment shows that for these additional sightlines
\OVI\ column densities are also enhanced in the region of excess and depressed
in the region of deficit defined by the sightlines measured in this paper. In
Figures~\Fskycollongtot\ and \Fskycollatitot\ we show the points in the region
with enhanced column densities in red. It is clear from these figures that the
excess column density seen in \OVI\ in the region around the Galactic Center is
also seen in the other ions. Comparing the average log\Nsinb\ for the region
with an excess to those in the ``normal'' sky gives differences in the logarithm
of the polar column densities of \VALUE 0.20, 0.36, 0.31, 0.26, 0.21 and
0.29~dex for \OVI, \NV, \CIV, \SiIV, \FeIII\ and \SIII, respectively. That
column densities are enhanced over the Galactic Center suggests the presence of
enhanced activity. It is more difficult to explain the 0.17~dex deficit seen in
$N$(\OVI) in the southern sky, especially since it is not reflected in any of
the other ions.
\par The total hydrogen column corresponding to each ion can be estimated from
log\,$<$\Nsinb$>-\log\,Af$, with log$<$\Nsinb$>$ the logarithm of the average
polar column density (see Figure~\Fskycollatitot\ for the values), $A$ the
element abundance and $f$ the ionization fraction. The maximum ionization
fraction of the ions is on the order of 20\% at temperatures near \dex5~K (see
e.g., Gnat \& Sternberg 2007). We take $f$=10\%, and use solar abundances of
$\log\,A$=$-$3.31 for oxygen, $-$3.57 for carbon, $-$4.17 for nitrogen and
$-$4.49 for silicon (Asplund et al.\ 2009). Then, the average polar column
densities (logarithmic values 14.17, 13.22, and 13.94 for \OVI, \NV, and \CIV,
respectively) imply log\,$N$(H,eqv)$\sim$18.4 for the transition temperature
gas. This number can be compared to the total neutral hydrogen column density
(log\,$N$(\HI)$\sim$20.28) and total ionized hydrogen column density
(log\,$N$(H$^+$)$\sim$20.0) (Savage \& Wakker 2009).
\par We note that Savage et al.\ (2003) compared the distribution of \OVI\
column densities to that of the radio-continuum loops and X-ray emission. They
found no obvious correlation. We rechecked these comparisons and we still see no
clear relation between $N$(\OVI) and other known structures. We also compared
the pattern of the column densities and velocities to the distribution on the
sky of the high- and intermediate-velocity \HI\ clouds (see Wakker 2004 for sky
maps of these objects). In the north, a large part of the sky at $l$=100\deg\ to
$l$-240\deg, $b$$>$30\deg\ is covered by negative-velocity gas ($v$$\sim$$-$40
to $-$90~\kms), in three structures called the ``Intermediate-Velocity Arch'',
the ``Intermediate-Velocity Spur'' and the ``Low-Latitude Intermediate-Velocity
Arch'' (as named by Kuntz \& Danly 1996). Although the \OVI\ absorption in this
part of the sky tends to have negative centroids, \OVI\ absorption with negative
centroids occurs both on and away from these \HI\ structures, and there is no
clear correlation. Furthermore, the region of sky where the excess high-ion
column density is seen covers area both overlapping with and away from the
intermediate-velocity gas. However, some connection between the two phenomena
would be logical, since the intermediate-velocity clouds (IVCs) are about 1~kpc
above the Galactic plane (Wakker 2001), which is similar to the scale height of
the transition temperature gas. The connection clearly does not manifest itself
as a simple correlation between the column densities of the neutral and
transition temperature gas, but must be more complex. Understanding the
connection will require many more sightlines toward which the highly-ionized
species are detected, and/or maps of the emission from these ions.

\begin{figure} 
\plotfiddle{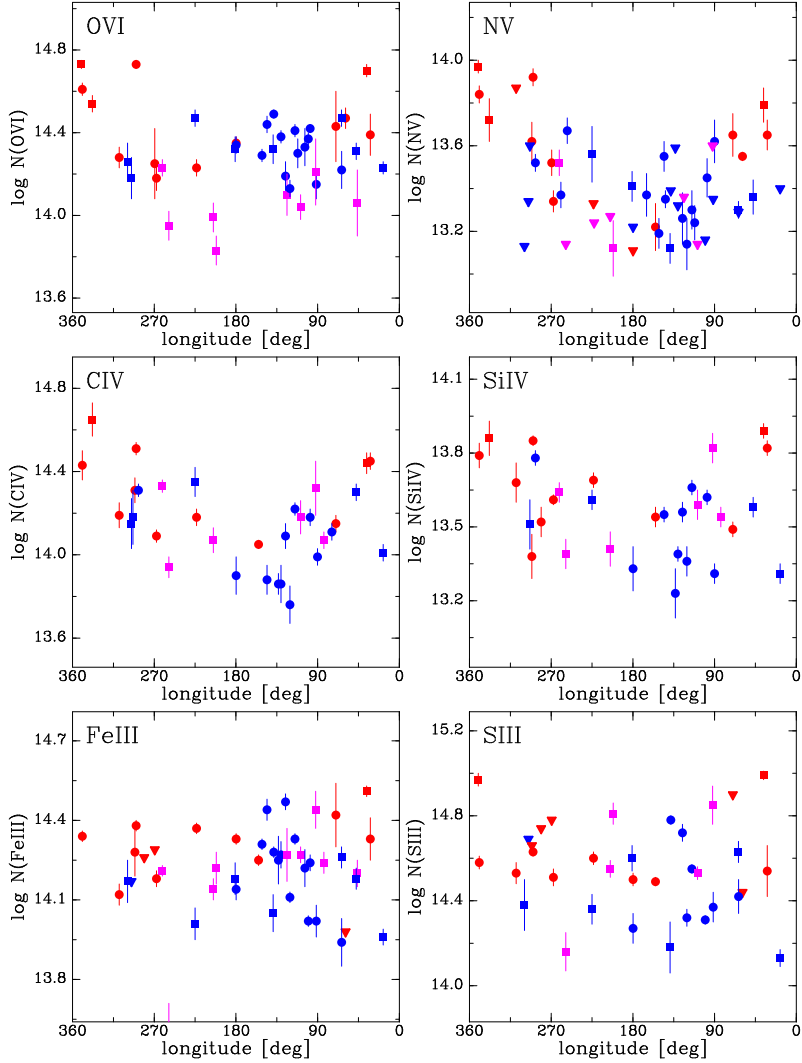}{0in}{0}{300}{400}{40}{0}
\figurenum{7}
\caption{%
Scatter plot of ionic column densities against Galactic longitude. Circles are
for sightlines with $b$$>$0\deg, squares are for sightlines with $b$$<$0\deg.
Triangles correspond to upper limits. Red points are for sightlines where
$N$(\OVI) is systematically enhanced (an area of sky delineated as
\absl$<$40\deg, \absb$<$50\deg\ combined with \absl$<$80\deg, $b$$>$40\deg).
Purple points are for sightlines showing a systematic deficit in $N$(\OVI)
(approximately $l$=70 to 280\deg, $b$=$-$60 to $-$10\deg) and blue points cover
the rest of the sky.
}\end{figure}

\begin{figure} 
\plotfiddle{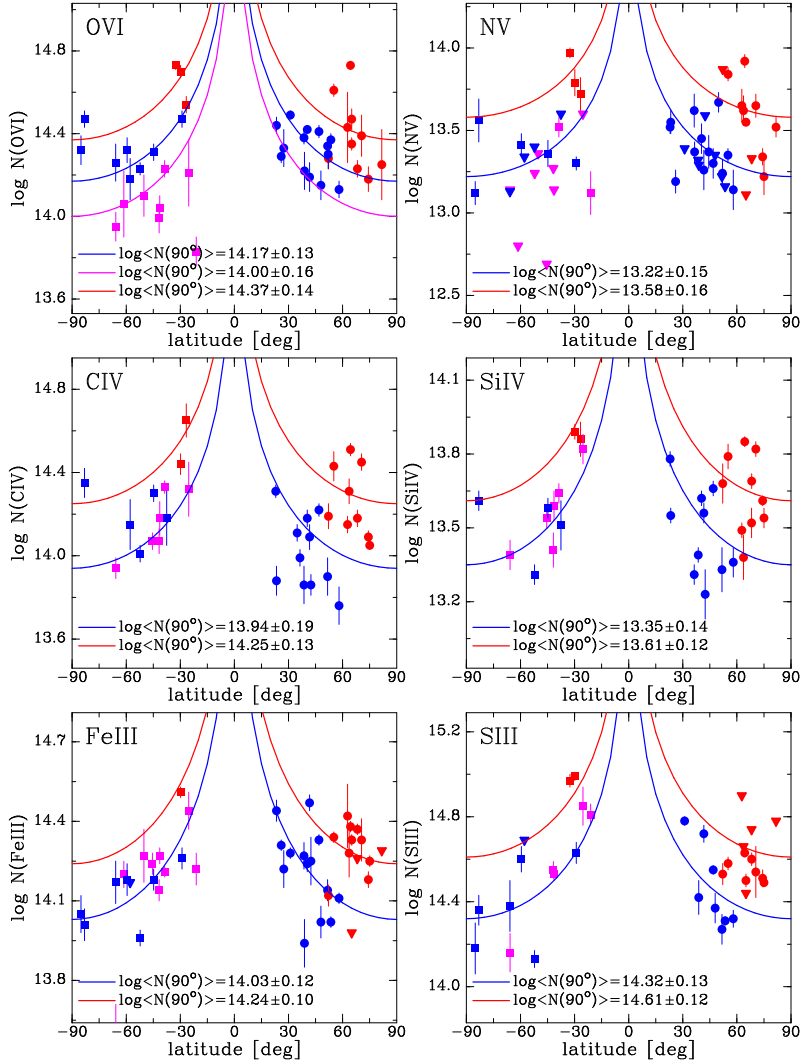}{0in}{0}{300}{400}{40}{0}
\figurenum{8}
\caption{%
Scatter plot of ionic column densities against Galactic latitude. Symbol types
and colors are the same as in Figure~\Fskycollongtot. In each plot we also give
a curve representing the \Nisinb{90\deg} distribution of column densities that
would result from a plane-parallel layer, with the values of the corresponding
polar column densities given in the lower left corner of each panel.
}\end{figure}

\begin{figure} 
\plotfiddle{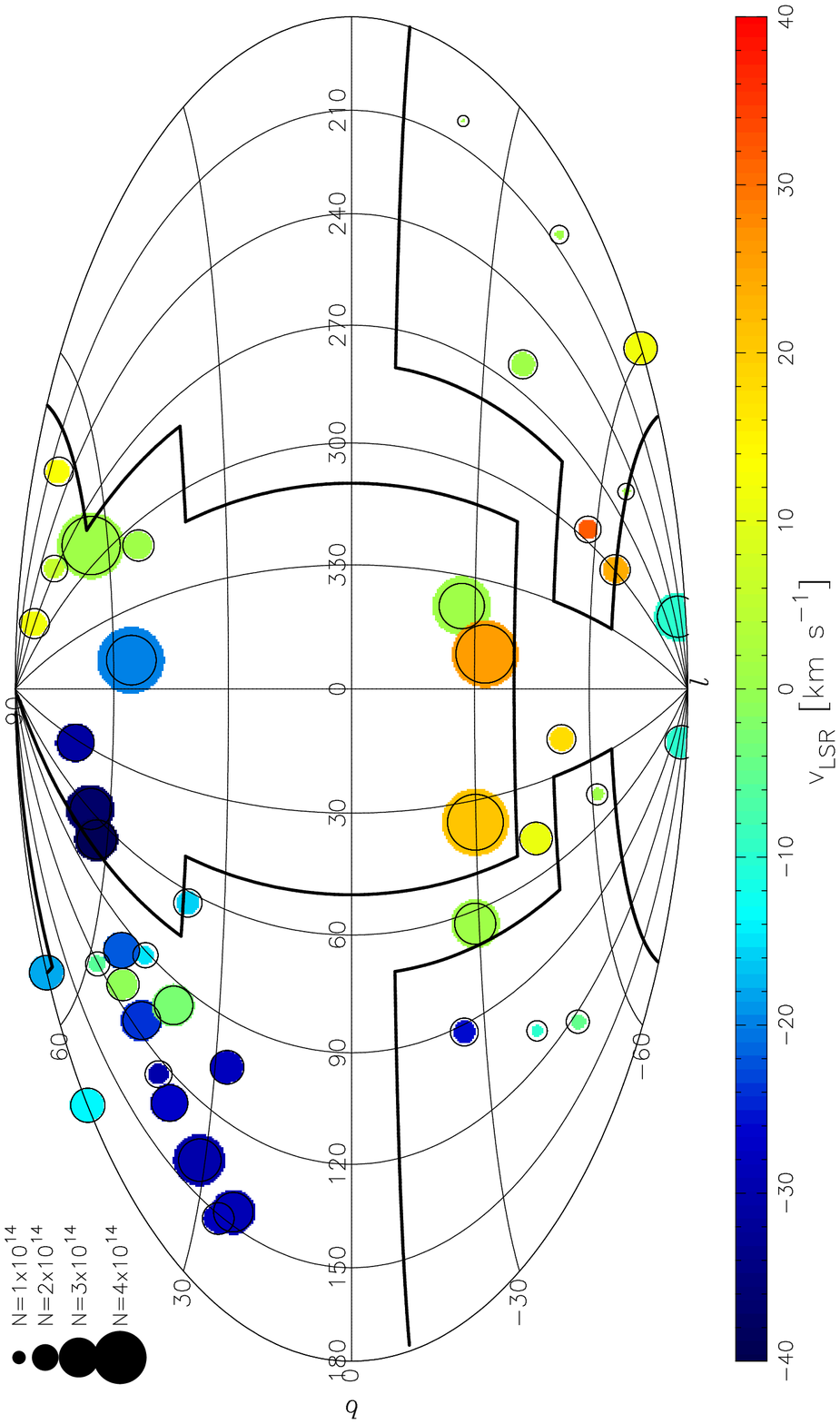}{0in}{270}{280}{450}{0}{0}
\figurenum{9}
\caption{%
All-sky map of the \OVI\ column density and centroid velocity, in aitoff
projection, centered on the Galactic Center. Colors correspond to the centroid
velocity of the Galactic \OVI\ absorption, following the scale on the bottom.
Symbol diameters are proportional to the value of $N$(\OVI), following the
legend in the upper left corner. The thick lines outline the regions with an
excess or deficit in the \OVI\ column density.
}\end{figure}

\par We further looked at the difference between absorption at negative
velocities and absorption at positive velocities, although comparison suffers
from having just 30 to 40 sightlines for which this can be done. We made this
comparison in many ways, looking at northern vs southern sightlines, sightlines
inside and outside the region of the \OVI\ excess, etc, but we found only one
(small) possible systematic difference. This is that the column densities at
negative velocities appear to be marginally larger than those at positive
velocities, by 0.10, 0.03, 0.22 and 0.26 dex for \OVI, \NV, \CIV\ and \SiIV,
respectively. In the region of the sky with excess total column density, these
differences are 0.07, 0.16, 0.25 and 0.20 dex, while in parts of sky away from
the excess or deficit region the differences are 0.19, 0.05, 0.23 and 0.31 dex.
However, in the region with the \OVI\ deficit, the differences are $-$0.15,
$-$0.17, 0.12 and 0.15 dex. Since the dispersions in the averages are about
0.20--0.25 dex, these differences are generally only about 0.5 times the spread,
which according to the $t$-test is barely significant (probability of it
happening by chance 2\%). However, all for four ions the column density at
negative velocities appears to be slightly larger than that at positive
velocities, so the effect is probably real.


\subsection{Scale heights}
\par The issue of the scaleheights of the different high-ionization species was
addressed by Savage \& Wakker (2009). They combined column density measurements
from a number of sources. Data from the \Copernicus\ and \FUSE\ satellites was
used to derive \OVI\ column densities toward stellar and extragalactic targets
(using measurements by Jenkins 1978b, Zsarg\'o et al.\ 2003, Wakker et al.\ 2003
and Bowen et al.\ 2008). \HST\ spectrographs provided \AlIII, \SiIV\ and \CIV\
spectra for a number of extragalactic targets, while \IUE\ (the {\it
International Ultraviolet Explorer}) was used to measure these ions in stellar
spectra and toward Magellanic Cloud stars (Savage et al.\ 2001b). The total
sample included 109 stars and 30 extragalactic objects, all having reliable
data.
\par Using AGNs at $-$90\deg$<$$b$$<$45\deg\ (i.e., omitting the region with
\OVI\ excess), Savage \& Wakker (2009) derived scaleheights of 2.6$\pm$0.6~kpc
for \OVI, 3.6$^{+1.0}_{-0.8}$~kpc for \CIV, and 3.2$^{+1.0}_{-0.6}$ for \SiIV.
They argued that the slightly smaller \OVI\ scaleheight compared to \CIV\ and
\SiIV\ is consistent with predictions of hybrid coronal gas models in which
these ions are produced by collisional ionization in cooling transition
temperature gas, but with photoionization from the extragalactic background
radiation increasing the amount of \CIV\ and \SiIV\ at high $z$. They also found
that \CIV\ and \SiIV\ are relatively tightly correlated, with an average
logarithmic ratio \SiIV/\CIV= $-$0.60$\pm$0.11 in the Thick Disk.
\par With the new set of measurements for \NV, we also tried to apply the
methods of Savage \& Wakker (2009) to derive a scaleheight for \NV, using data
from the literature for measurement of \NV\ toward disk stars (as listed by
Savage \& Wakker (2009) and complemented by values from Lehner et al.\ (2011).
However, a fair comparison between the different scaleheights requires using the
same set of stars and the same selection criteria for all ions. This results in
a sample of \NV\ measurements that is too small to give a reliable answer. With
somewhat less restrictive criteria, the derived scaleheight for \NV\ is similar
to that of \OVI\ and \CIV. A proper assessment of the \NV\ scaleheight will
require more data, however.
\par The fact that both the linewidths and the scaleheights of \CIV\ and \SiIV\
are comparable and that their centroids are very well aligned is strong evidence
in favor of the hypothesis that these ions originate in the same parcels of gas.
Comparing to \OVI\ we find that the scaleheight of \OVI\ is slightly smaller,
while the extent of the \OVI\ absorption tends to be slightly wider, but the
centroids of the \OVI\ and \CIV/\SiIV\ are similar. The most likely explanation
for these similarities and differences is that the \OVI\ is associated with the
same structures in which \CIV\ and \SiIV\ originate, but near the plane the
\OVI-containing gas is slightly more turbulent, while as one goes up, the
fraction of \OVI-containing gas decreases.

\subsection{Line ratio histograms}

\par Figure~\Fratiohist\ shows the distributions of all ratios between the four
highly-ionized species, as well as two ratios involving \FeIII, for the
absorption components where the derived ratio is flagged as being reliable. The
panels on the left side give the distributions of the logarithms of the ratios,
the panels on the right show the ratios linearly. The labels give the average
ratio and its dispersion, excluding from the calculation some outliers, which
are discussed below.

\begin{figure} 
\plotfiddle{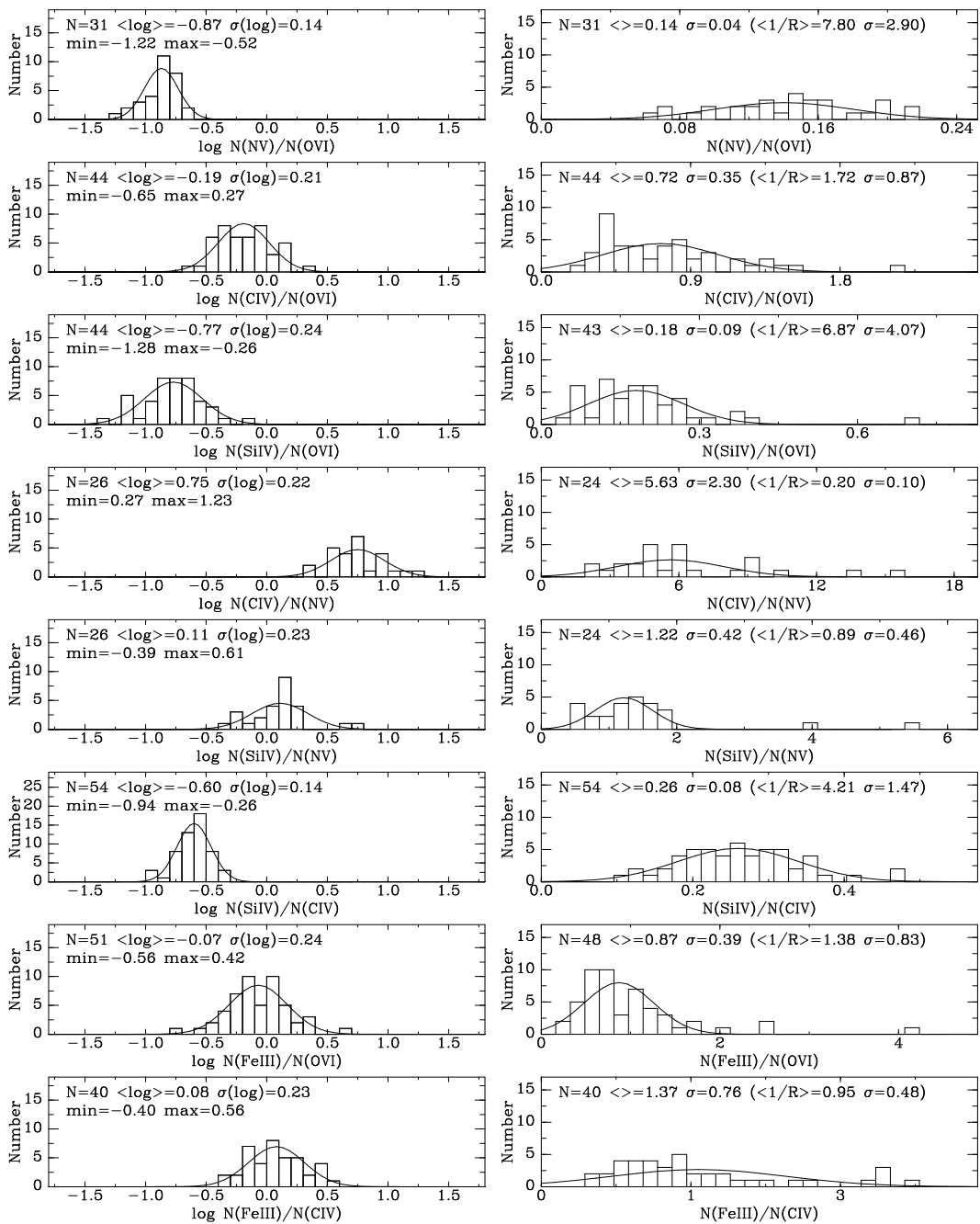}{0in}{0}{400}{500}{30}{0}
\figurenum{10}
\caption{%
Histograms of the distribution of the logarithmic (left panels) and linear 
ratios (right panels) between \OVI, \NV, \CIV, \SiIV\ and \FeIII. The number of
measurements ($N$), average ($<$log$>$) and dispersion ($\sigma$(log)) of each
ratio distribution are given in the top left corner. The second line gives the
range of values. Gaussians with the listed parameters are overlaid. To make
comparisons with other papers easier, in the panels on the right, the average
linear ratio and its dispersion are given, as are the average and dispersion of
the distribution of the inverse ratio. E.g., the average ratio of \NV/\OVI\ is
0.14, while the average ratio of \OVI/\NV\ is 7.80.
}\end{figure}

\par For the most part these distributions are rather narrow. In particular, the
histograms of log\,(\NV/\OVI) and log\,(\SiIV/\CIV) have a dispersion of only
\VALUE 0.14~dex ($\pm$40\%), with a spread between the highest and lowest ratio
of just 0.6~dex, or a factor $\sim$4. The dispersions in the distributions of
log\,(\CIV/\OVI), log\,(\SiIV/\OVI), log\,(\CIV/\NV), and log\,(\SiIV/\NV) are
\VALUE $\sim$0.23~dex (70\%), significantly larger than those in log\,(\NV/\OVI)
and log\,(\SiIV/\CIV). This again supports the notion that \OVI\ and \NV\
originate together, as do \SiIV\ and \CIV, but although the two pairs of ions
live in the same structure, they are not fully co-spatial. The median error in
each individual ratio is about 0.10~dex. Thus, in the case of \NV/\OVI\ and
\SiIV/\CIV, the observed dispersions in the distributions may reflect the
observational uncertainty more than an intrinsic spread. For the other ratios
there clearly are variations across the sky.
\par Previous determinations of these line ratios were presented by Sembach \&
Savage (1992), who measured \NV, \CIV\ and \SiIV\ toward 12 high-$z$ stars. They
listed an average \CIV/\NV\ ratio of 4.6$\pm$2.7, where we find \VALUE
5.6$\pm$2.3, using 24 components. For \CIV/\SiIV\ they found 3.6$\pm$1.3,
compared to our value of \VALUE 4.2$\pm$1.5. Indebetouw \& Shull (2004b) also
presented average ionic ratios, giving log\,(\NV/\OVI)=$-$0.83$\pm$0.28 (34
sightlines) and log\,(\CIV/\OVI)=$-$0.28$\pm$0.18 (11 sightlines), compared to
our results of \VALUE $-$0.87$\pm$0.14 (31 components) and \VALUE
$-$0.19$\pm$0.21 (44 components) for these ratios. I.e., we find similar values
for \CIV/\OVI, but a much narrower distribution for \NV/\OVI, which in large
part can be explained as the result of the very different velocity ranges over
which the lines were integrated.

\section{Ionization mechanisms}
\par The comparison between two line ratios can be a great tool to diagnose the
physical conditions in the gas giving rise to the high-ionization absorption. In
earlier work the diagnostic value of ionic ratios was often pointed out (e.g.
Sembach \& Savage 1992; Spitzer 1996; Savage et al.\ 1997b), and the observed
values of e.g., \CIV/\OVI\ or \CIV/\NV\ were compared with model predictions.
However, the diagnostic power of the ratio-ratio diagram (first discussed by
Indebetouw \& Shull 2004a and Fox et al.\ 2004, 2005) was only fully appreciated
after \FUSE\ \OVI\ data became available to complement \STIS\ observations of
\CIV, \NV\ and \SiIV.
\par In this section we summarize the potential ionization mechanisms and
compare the resulting predictions for ionic ratios and ionic column densities to
the observations. Much of this has been done in one form or another in previous
papers. However, unlike previous authors, we describe more explicitly the actual
predictions in the different papers in which a given model is calculated, rather
using a simplified range of possible values for the ratios. Further, for some
processes (CIE, turbulent mixing) revised models were published fairly recently,
using updated atomic parameters. We also include three calculations that
traditionally have not been part of such an overview: Thick Disk supernovae
(Shelton 1998), static non-equilibrium radiative cooling (Gnat \& Sternberg
2007), and cooling flows (described below). In the case of the last of these, we
present this model in a journal paper for the first time, although the
predictions were summarized in a conference  paper (Shapiro \& Benjamin 1991).
Further, we include a new numerical model for turbulent mixing, which includes
non-equilibrium ionization (Kwak \& Shelton 2010).
\par This set of mechanisms does not exhaust the range of ideas published in the
literature. Indebetouw \& Shull (2004a) also list as possibilities evaporating
shells, stellar wind bubbles, supernova remnants, and photoionization by white
dwarfs. However, some of these appear unlikely because they predict ionic ratios
that are very different from the observed values, or because one would expect a
small covering factor or large variations in column density across the sky since
objects like stellar wind bubbles and white dwarfs have a small volume filling
factor.
\par Historically, different elemental abundances have been used to create the 
different models. This is because the standard reference values of the solar
abundances of most elements (but especially C, N and O) have varied since 1980,
as determinations improved or problems with the interpretation of solar
absorption lines were corrected. We therefore adjusted the published predictions
of ionic column densities to reflect the most recent values for the solar
abundances of C, N, O and Si (Asplund et al.\ 2009: log\,A(C, N, O, Si)=
$-$3.57, $-$4.17, $-$3.31 and $-$4.49, respectively). However, this adjustment
is not a simple scaling, as to first order the \OVI\ column density is
independent of metallicity, as the cooling is dominated by the oxygen ions.
Thus, a lower oxygen abundance results in less efficient cooling, causing a
larger region with transition temperature gas, giving a fixed \OVI\ column
density. The column densities of other ions then scale with the change in the
{\it ratio} of elemental abundances. A different situation prevails in the
turbulent mixing model, where the heating up of cold gas is important. In that
case the column densities of the highly-ionized atoms {\it do} scale directly
with the elemental abundances.
\par We compare the models to the data in Figures~\Fratiomodel a, b, c and d,
which plot four combinations of ratios. Figure~\Fratiomodel a shows \NV/\OVI\
against \SiIV/\CIV, i.e.\ the ratios for the pairs of ions that seem to
originate in the same regions of the ionized structures in the sightline.
Figure~\Fratiomodel b gives \NV/\OVI\ against \CIV/\OVI\, which represents a
more global comparison through the ionized structure, with the x-axis relevant
for the hottest gas and the values in the vertical direction influenced by the
relative amount of the two different regions in the structure. In
Figure~\Fratiomodel c we show \CIV/\OVI\ against \SiIV/CIV, which focuses on the
cooler regions. Finally, Figure~\Fratiomodel d presents \CIV/\NV\ against
\SiIV/\CIV, equivalent to Figure~\Fratiomodel c, but chosen by bearing in mind
that \COS\ spectrograph on \HST\ is producing many additional spectra of QSOs
that will include data for \NV, \CIV\ and \SiIV, but not for \OVI.


\subsection{Observed ionic ratios}
\par The filled datapoints in Figure~\Fratiomodel\ represent the ratios (or
limits, shown by triangles) in the individual components of the Galactic
absorption profile.
We draw a box around the datapoints which represents the $\pm$2$\sigma$ range
around the average.
\par To determine whether our measurements for the highly-ionized gas in the
Milky Way corona are representative, we also looked at the measurements in other
environments: the Galactic disk, high-velocity clouds, and high-redshift damped
Lyman-alpha (DLA) absorbers. We discuss each of these below.
\par For the Galactic disk we used the measurements of highly-ionized atoms
toward 38 stars by Lehner et al.\ (2011), shown by open stars in
Fig.~\Fratiomodel. In most cases \OVI, \CIV\ and \SiIV\ can be measured, while a
value or upper limit is given for \NV\ in 22 sightlines. Lehner et al.\ (2011)
present a table of the integrated column densities, as well as one for component
column densities. However, for the latter the different ions are not explicitly
matched up, and a few times the fitted component structure differs for e.g.\
\CIV\ versus \OVI. In many sightlines the stellar continuum is difficult to
define, especially for \NV\ and \OVI. Toward many stars the \NV\ line is also
relatively noisy. In other sightlines the \CIV\ and/or \SiIV\ lines are
extremely deep and show evidence for saturation, making the derived column
densities suspect, and indeed in every one of those cases the \SiIV/\CIV\ ratio
is found to lie about 0.3~dex outside the range found at high latitude toward
AGNs. In a number of cases the apparent interstellar \OVI\ profile strongly
differs from the \CIV/\SiIV\ profiles, which may indicate the presence of
stellar absorption lines. Reassessing the Lehner et al.\ (2011) sample, we
decided to keep only the 12 sightlines where the stellar continuum is relatively
clean and toward which the interstellar absorption lines are not saturated.
These are the stars given number 1, 2, 6, 10, 11, 12, 13, 14, 27, 33, 35 and 38
in the Lehner et al.\ (2011) list. With this restriction of only including
measurements for stars with unsaturated and unambiguous absorption, the ratios
for these disk sightlines fall completely within the box determined from the
extragalactic targets (with the exception of \NV/\OVI\ toward HD\,99857, which
is 1.5$\sigma$ outside the box), indicating that the physical conditions in the
transition temperature gas do not change much with height above the Galactic
plane.
\par Another kind of cloud in which highly-ionized gas is often detected are the
high-velocity clouds (HVCs). Fox et al.\ (2004, 2005, 2010) and Collins et al.\
(2004, 2005) analyzed these lines toward the two largest HVCs (complex~C and the
Magellanic Stream). Sembach et al.\ (2003) presented measurements of \OVI\ in 84
HVCs seen in sightlines toward extra-galactic targets, while Lehner \& Howk
(2010) find highly-ionized high-velocity clouds toward several galactic stars.
Each of these papers shows a small subset of the total available data. In order
to look at the ionic ratios in the HVCs, we therefore measured \OVI, \NV, \CIV\
and \SiIV\ in each of the sightlines used in this paper. The full results will
be presented in a future paper on highly-ionized gas in HVCs. We find that
detecting \NV\ in HVCs is rare, but out of about 25 limits that can be formed
for \NV/\OVI\ only three fall outside the observed range for Milky Way coronal
gas. Similarly, of the about 20 HVCs in which \CIV\ and \OVI\ are both seen, the
ratio falls in the observed range for the Milky Way coronal gas. In the case of
\SiIV/\CIV\ this is true for all but two of the $\sim$25 possible ratios. The
two exceptions are clouds with relatively narrow lines, where it appears likely
that photoionization plays a role.


\begin{figure}
\plotfiddle{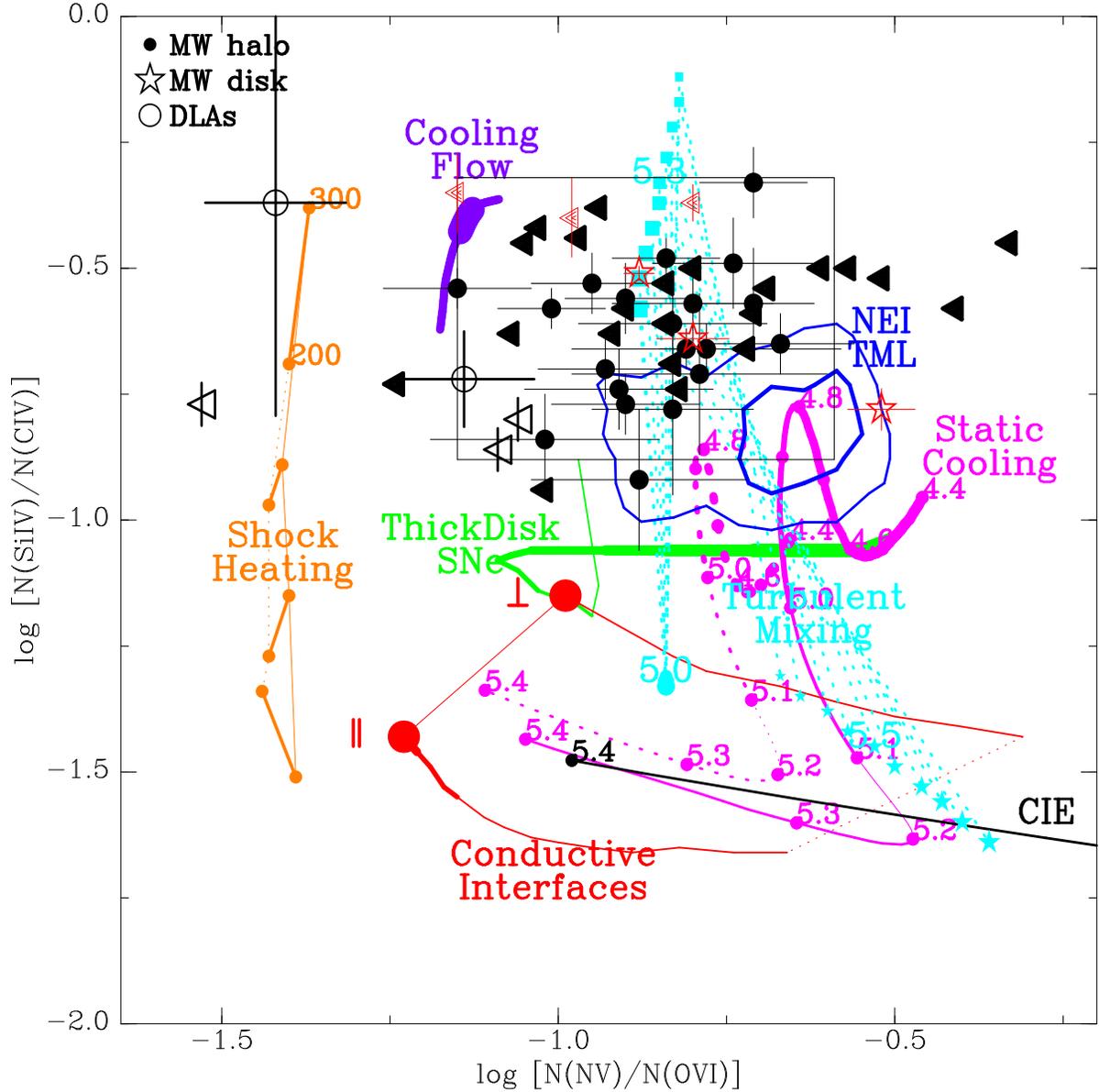}{0in}{0}{450}{450}{0}{0}
\figurenum{11a}
\caption{%
Theoretical ranges and observed values for pairs of ionic ratios.\ \ \ \
DATA: Filled black circles show ratios in individual components, as determined
in Section~\Smeas, black triangles show cases with one ratio measured, but the
other ratio being an upper or lower limit. A black box outlines the region given
by $<$$R$$>$$\pm$$\delta R$, where the average $<$$R$$>$ and spread $\delta R$
of the ratio distribution can be found in Figure~\Fratiohist. Open red stars and
red triangles represent the results of Lehner et al.\ (2011), limited to their
stars numbered 1, 2, 6, 8, 9, 10, 11, 12, 13, 14, 23, 27, 28, 32, 33, 35 and 38,
i.e., not including stars for which the lines are saturated, have very complex
component structure or difficult to determine stellar continua. Open black
circles and triangles represent measurements in DLAs at redshifts 2.0 to 3.1, as
presented by Fox et al.\ (2007). Note that this selection leaves just six
sightlines with measured values for \NV, while in two other sigthlines \OVI\ is
missing; thus fewer than 12 symbols show up in each individual ratio-ratio plot.
}\end{figure}

\bigskip
\bigskip
\bigskip
\bigskip
\bigskip
\bigskip
\bigskip
\bigskip
\bigskip
========================================
\bigskip
\bigskip

[FIG 11A CAPTION CONTINUED]

MODELS:
(a) Black line (label: CIE): ratios in collisional ionization equilibrium, as
calculated by Gnat \& Sternberg (2007), for temperatures ranging from
log\,$T$=5.4 to log\,$T$=4.4.
(b) Magenta lines (label: Static Cooling): non-equilibrium radiative cooling
models from Gnat \& Sternberg (2007) for temperatures ranging from log\,$T$=5.4
to log\,$T$=4.4, with a solid line for solar metallicity and a dotted line for
two times solar metallicity; the line gets thicker at lower temperatures, where
the gas spends more time; the gas traverses this trajectory in about
3\tdex4/$n_0$\,(\cmm3)~yr. In these models the gas is not flowing, and there is
no additional heat input or output.
(c) Orange lines (label: Shock Heating): Dopita \& Sutherland (1996) shock
models, with shock velocities of 200 and 300~\kms\ for the no-magnetic-field
case marked and connected by a thick line; The other three thick line segments
connect the 200 and 300~\kms\ case for higher magnetic parameters,
$B\,n^{-3/2}$=1, 2 and 4 $\mu$G\,cm$^{-3/2}$.
(d) Red lines (label: Conductive Interfaces): the Borkowski et al.\ (1990)
conductive interfaces predictions; A ``perpendicular'' and a ``parallel'' symbol
mark which of the two lines is for perpendicular or parallel magnetic field; the
line thickness increases linearly with the age of the interface, 0.1~Myr at the
thinnest point, 10~Myr at the other end, which is reached after 1--2~Myr. As a
result, the line for this model appears as a quickly-traversed thin red line
ending in a large dot, where the ratios stabilize.
(e) Green line (label: Thick Disk SN): the Shelton (1998) Thick Disk supernovae
model (SNe occuring at heights up to 2~kpc), giving the evolution of the ratios
with time (from 1 to 16~Myr in age), with the line thickness proportional to the
amount of time spent at each point.
(f) Light blue lines (label: Turbulent Mixing): predictions of the Slavin et
al.\ (1993) semi-analytic turbulent mixing layer model; circles, squares and
stars are shown for entrainment velocities ranging from 25 to 100~\kms\ at
temperatures of log\,$T$=5.0, 5.3 and 5.5, respectively; the symbol size is
proportional to the entrainment velocity.
(g) Solid blue contours (label: NEI TML): distribution of ratios in the Kwak \&
Shelton (2010) turbulent mixing layer models, corresponding to 256 column
densities derived by integrating through an interface at 1~Myr intervals for
ages between 20~Myr and 80~Myr.
(h) Dark purple line (label; Cooling Flow): the predictions of the model by
Benjamin \& Shapiro that is described in the appendix. This model is applicable
to cooling hot gas flowing through an interface. The range of flow velocities is
20--30~\kms\ (thicker part of the line) and 12--42~\kms\ (thinner part). For
these flow velocities the absolute column densities are within 2 and 1 times the
observed dispersion around the observed average value.

\bigskip
\bigskip
========================================
\bigskip
\bigskip

\begin{figure}\plotfiddle{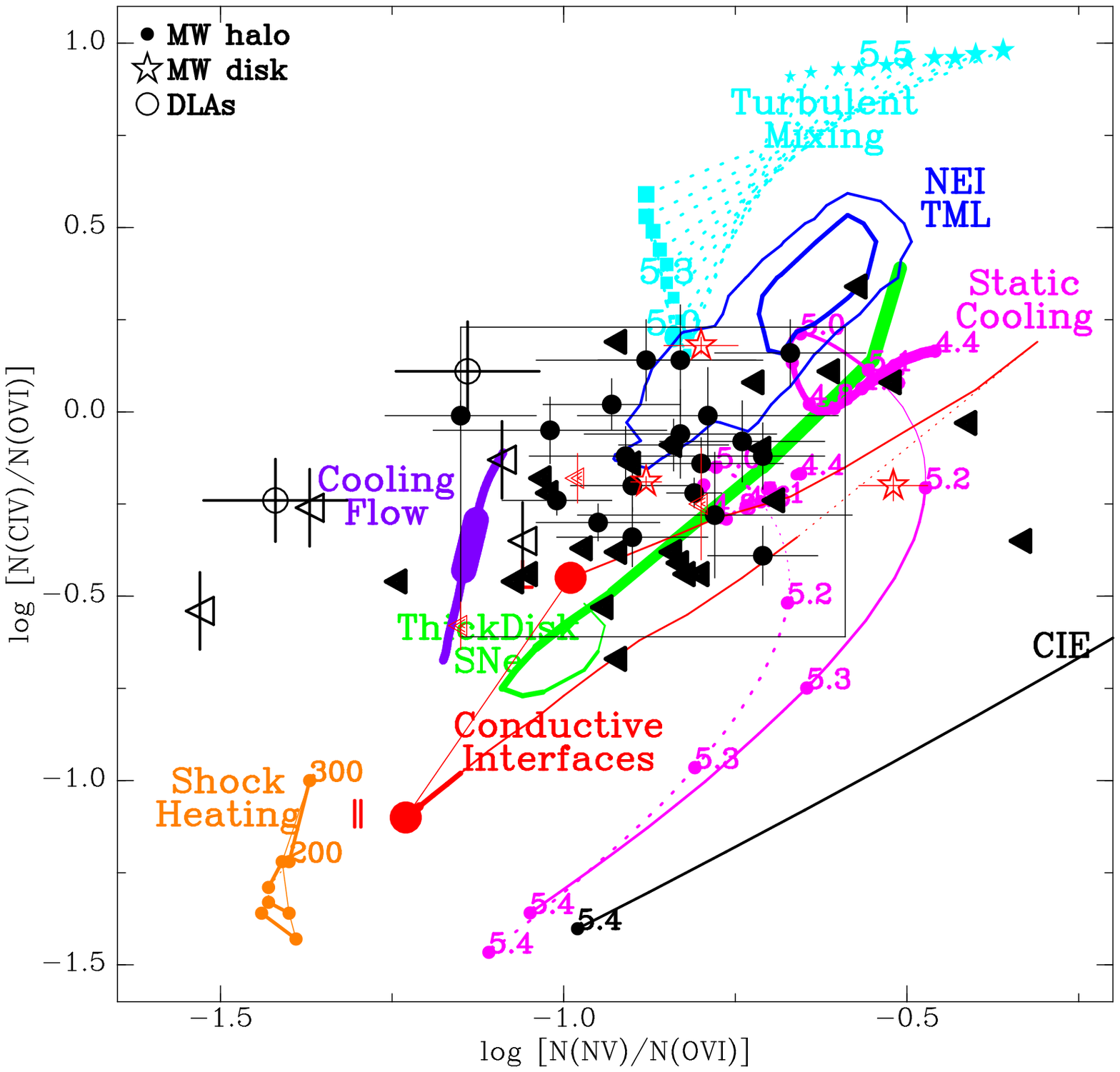}{0in}{0}{450}{450}{0}{20}\figurenum{11b}\caption{See caption for Fig.~\Fratiomodel a}\end{figure}
\begin{figure}\plotfiddle{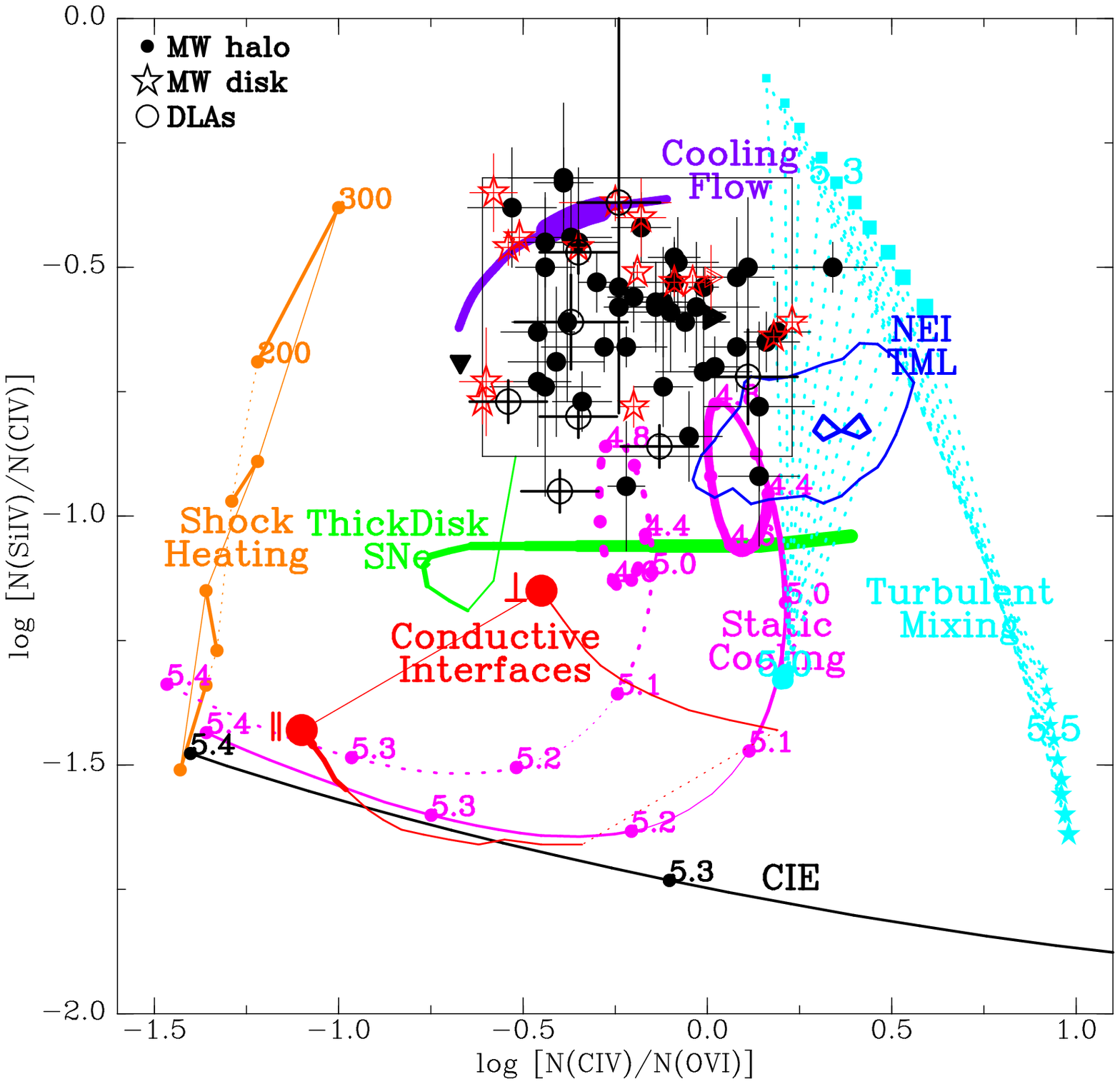}{0in}{0}{450}{450}{0}{20}\figurenum{11c}\caption{See caption for Fig.~\Fratiomodel a}\end{figure}
\begin{figure}\plotfiddle{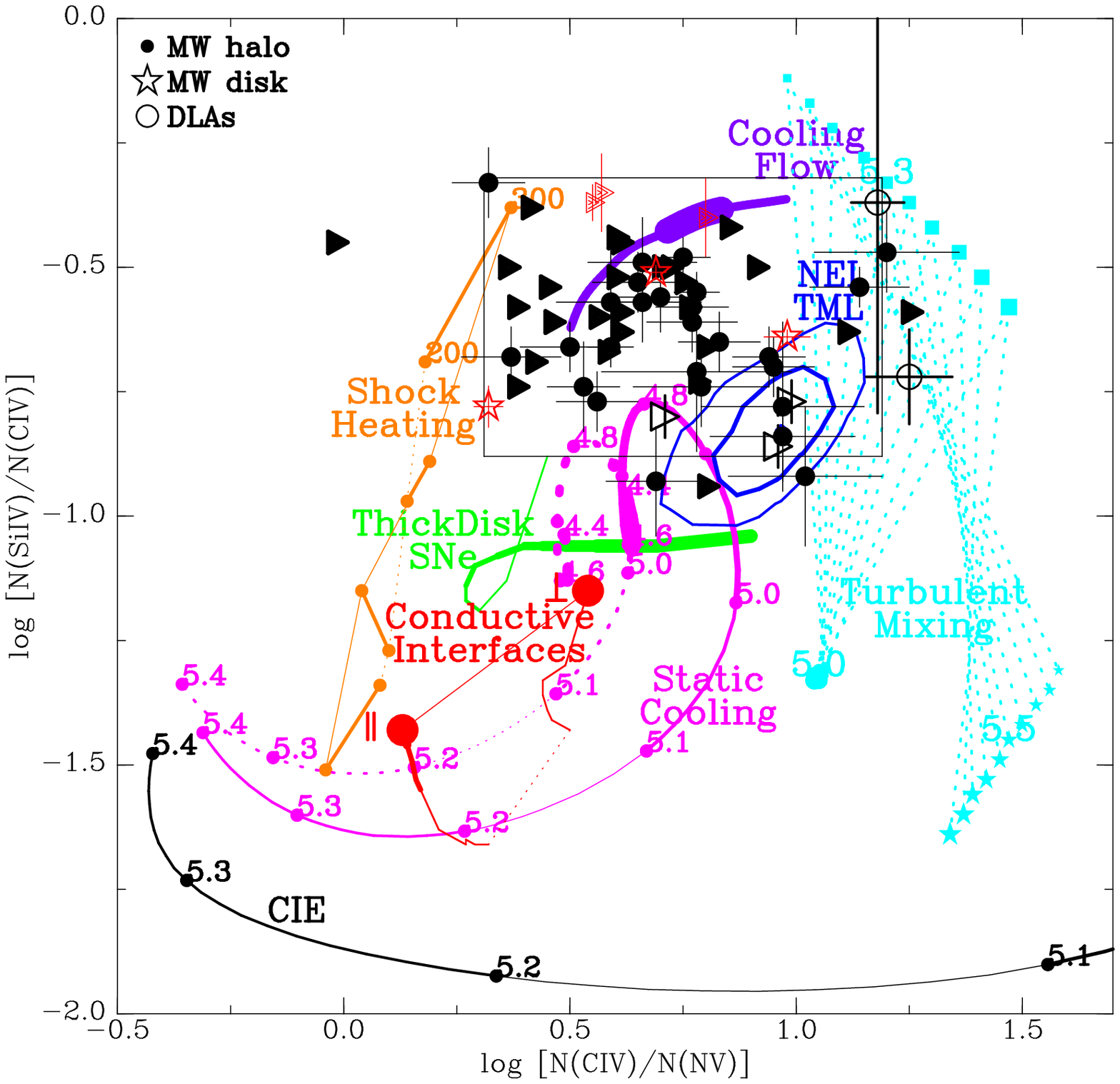}{0in}{0}{450}{450}{0}{20}\figurenum{11d}\caption{See caption for Fig.~\Fratiomodel a}\end{figure}

\par Open circles and triangles in Fig.~\Fratiomodel\ show the measurements of
Fox et al.\ (2007) associated with a number of Damped \Lya\ Absorbers (DLAs) at
redshifts 2.0 to 3.1. They found 12 absorption line systems in which at least
three of \OVI, \NV, \CIV\ and \SiIV\ could be measured. If we use these results
and derive ionic ratios, we find that the eight systems for which both
\SiIV/\CIV\ and \CIV/OVI\ can be calculated give ratios that fall in exactly the
same range as the Milky Way results. \NV, however, seems underabundant, with
only three detections and five upper limits. In all of these cases
log\,\NV/\OVI\ is $<$$-$1.1, which contrasts with the average value of $-$0.87
seen in the Milky Way. 
\par The metallicities found in the neutral portions of these DLAs (as
determined from S, Si or Zn absorption lines), are [Z/H]=$-$0.85, $-$0.91 and
$-$1.62 for the three DLAs in which \NV\ is detected, but much lower (between
$-$1.6 and $-$2.3) for the five cases with an \NV\ upper limit (Fox et al.\
2007). Thus, the non-detections are likely due to a lack of metals. In two of
the three DLAs with \NV\ and \OVI\ detections, the [N/O] ratio could also be
measured directly from \NI\ lines, resulting in [N/O]=$-$1.2 and $-$1.1 (Fox et
al.\ 2009). This follows the general trend that nitrogen is underabundant in
DLAs (e.g.\ Petitjean et al.\ 2008). Thus, the finding that the \NV/\OVI\ ratios
in DLAs are lower than in the Galactic Thick Disk can be explained as a
differential abundance effect for nitrogen, and is not indicative of any
difference in physical conditions.
\par We note that in nearby dwarf galaxies with metallicity [Z/H]$<$$-$1 times
solar, the N/O ratio is about 0.8~dex below its value in the Sun (and in the
Milky Way) (Skillman 1998). This effect can be explained as the result of a
difference in the relative amount of N produced by Type II supernovae (primary
N), as compared to N formed in intermediate-mass stars (secondary N). If the gas
sampled by the $z$=2--3 systems was blown out from galaxies at early times, all
N would be primary, and the N/O ratio would be subsolar, which can explain the
relatively low \NV/\OVI\ ratios.
\par Independent of the cause, however, it is clear that \NV/\OVI\ in the
high-redshift DLAs differs from that in the Milky Way. Thus, adapted numerical
models are necessary, requiring subsolar N/O. However, since most of the cooling
occurs through the ions of O and C, it is likely that the effects of using
subsolar N/O are minimal and that the predicted \NV\ column densities would
scale linearly with N/O.


\subsection{Collisional Ionization Equilibrium (CIE)}
\par The simplest calculation involves assuming a cloud that has fixed
temperature and density, in which every collisional ionization is balanced by a
recombination. Sutherland \& Dopita (1993) provided a table for the resulting
ion fractions, as did Gnat \& Sternberg (2007). The predictions in the latter
paper are about 0.3--0.4 dex higher for \NV, \CIV\ and \SiIV\ at temperatures
around log\,$T$=5.3, which is caused by changes in atomic parameters (see
references in Gnat \& Sternberg 2007). The \SiIV\ fraction differs even more
above log\,$T$=5.4. On the other hand, the predicted \OVI\ fraction is rather
similar, except above log\,$T$=6.0, where the Gnat \& Sternberg (2007) model
predicts less \OVI. All this results in predictions for the line ratios that can
differ by as much as 0.4 dex from the Sutherland \& Dopita (1993) values.
\par However, we do not expect the gas to be in CIE (and it is not observed to
be). First of all, \OVI\ will only exist at temperatures above about \dex5~K,
but then the \SiIV/\CIV\ ratio is $<$$<$$-$1.5, i.e.\ much smaller than what
is observed. Moreover, if we apply the CIE ionization fractions to a cloud with
log\,$N$(H)=19, then log\,$N$(\OVI) can only be $\sim$14.1 for temperatures
above log\,$T$=5.3, while log\,$N$(\CIV) is $\sim$13.8 for temperatures near
log\,$T$$\sim$4.8 and log\,$T$$\sim$5.2. log\,$N$(\SiIV) only reaches the
observed value of 13.2 if log\,$T$$<$5.0. I.e., for temperatures at which the
CIE model predicts the presence of \OVI, there is no \CIV\ and no \SiIV, while
at lower temperatures a sufficient amount of $N$(\CIV) and $N$(\SiIV) is only
present in the absence of \OVI.


\subsection{Static non-equilibrium ionization radiative cooling}
\par Edgar \& Chevalier (1986) were the first to predict the ionization
properties of gas starting at high temperature (\dex6~K) and cooling down
radiatively. In such models, the gas starts out in CIE at a high temperature
(5\tdex6~K in their case), and then is allowed to cool, either isochorically (at
constant density) or isobarically (at constant pressure). In these models there
are no external sources of heat or photoionization, the gas is not flowing, and
it is assumed that the column density is low enough that the gas is optically
thin. As most of the cooling is provided by metal-line emission, the metallicity
of the gas will determine the timescale of the cooling. At high metallicity, the
gas can cool, but highly-ionized atoms will recombine too slowly, leading to an
overabundance of ions such as \CIV, \NV\ and \OVI\ compared to gas in CIE at the
same temperature. That is, the ionization is frozen in for a while. The higher
the metallicity, the faster the cooling and the larger the discrepancy. 
\par The Edgar \& Chevalier (1986) paper gave predictions for both isochoric and
isobaric cooling. For the isobaric case the values of \SiIV/\CIV, \CIV/\OVI\ and
\CIV/\NV\ are 0.5--1.0~dex below the observed values, while in the isochoric
case these ratios are underpredicted by 0.2--0.4~dex. Although this does not
match the observations, the model was influential since it showed that it is
possible to get column densities and ionic ratios that are very different from
those in CIE.
\par Gnat \& Sternberg (2007) presented a more detailed set of predicted
ionization fractions for a variety of static non-equilibrium models.  They
provided tables for 10 cases: isochoric and isobaric cooling at metallicities
$Z$=0.001, 0.01, 0.1, 1 and 2 times solar. For metallicities $Z$$<$0.5 solar,
the difference with CIE predictions is minimal. Similarly, the difference
between the isochoric and isobaric cases is one of degree. In particular, for
$Z$=0.1, the isobaric and isochoric predictions differs by less than 0.1 dex,
while the $Z$=2 isobaric case is almost identical to the $Z$=1 isochoric case.
\par Gnat \& Sternberg (2007) also discussed when it would be appropriate to use
the isobaric or the isochoric case. They took a spherical cloud which starts out
cooling isobarically, and thus contracting as the temperature decreases. This
increases the column density. If it becomes larger than $N_{tr}$=1.2\tdex{19}
$\left({T\over10^5}\right)^{2.04}$~K the cooling become isochoric. With the
ionization fractions predicted by the isobaric cooling model, we can derive the
implied total hydrogen column density for an \OVI\ column density of
log\,$N$(\OVI)=14.1. This value is below $N_{tr}$ for temperatures
log\,$T$$>$5.3, implying that at lower temperatures the appropriate cooling
model to use is the isochoric one.
\par To assess whether a non-equilibrium ionization radiative cooling model is
realistic, one should not only look at ionic ratios, but also take into account
the predicted total ionic column densities, the cooling time and the integrity
of a cloud as it cools and moves. The column densities are within 0.2~dex of the
observed averages for temperatures between log\,$T$=4.3 and 5.1, if the cloud
has solar metallicity or higher and total hydrogen column density
log\,$N$(H)$\sim$19. For $Z$=0.1, however, the cooling is so slow that
recombination can keep up and the column densities of \OVI, \NV\ and \CIV\ only
barely match the observations at log\,$T$=5.2, long before any \SiIV\ is formed.
The cooling time to go from $T$=3\tdex5~K to \dex4~K is about
$t$=3\tdex4/$n_0$\,(\cmm3)~yr for a solar metallicity cloud (where $n_0$ is the
density of H$^+$; Gnat, private communication). Given log\,$N$(H)$\sim$19 and a
scaleheight of 3~kpc, the H$^+$ density at 1~kpc is about 7\tdex{-4}\,\cmm3,
implying a cooling time of 40~Myr. If the gas were moving at a rate of 10~\kms,
it would move over 0.4~kpc during that time, which is likely significant
relative to its size, but not so large that the cloud would completely disperse.
Thus, we conclude that it is possible to construct a realistic situation in
which the non-equilibrium ionization radiative cooling model might be applicable.
\par In Figure~\Fratiomodel, the magenta lines show the isochoric
non-equilibrium radiative cooling model for the temperature range of
log\,$T$=5.4 to 4.4. The solid curve is for solar metallicity, the dotted curve
for two times solar metallicity. Clearly, this model can match some of the
observed ratios, specifically \CIV/\NV\ and \CIV/\OVI. However, it underpredicts
\SiIV/\CIV\ and overpredicts \NV/\OVI. In Figure~\Fratiomodel d the pattern of
datapoints suggestively lies on a curve with the same shape as the
solar-metallicity curve, but with higher \SiIV/\CIV\ ratio. Thus, given the many
simplifying assumptions, the predictions of this model are close enough to the
observed values that we conclude that non-equilibrium radiative cooling is
likely to be an important part of the processes that produce the highly-ionized
atoms.


\subsection{Radiatively cooling gas flows}
\par The model described in the previous subsection does not take into account
the gas dynamics. Doing so substantially changes the predictions for the
relative ionic column densities, even with the same prescription for
non-equilibrium cooling. Shapiro \& Benjamin (1991, 1993) and Benjamin \&
Shapiro (unpublished) studied this in more detail. Their model follows the 
thermal and ionization state of gas in a one-dimensional planar, steady-state
flow, including the effects of radiative transfer of photons generated by the hot
phase of the gas. This results in a substantial boost for the column densities of 
CIV\ and especially \SiIV. Since this model has not been fully described in
detail in the literature, we present a summary in an appendix. We note that the
physical origin of the flow remains unspecified in this model -- it is a
description of the physical structure that results after a cooling front
develops in a flow, be it an outflowing wind of hot gas, or an inflow.
\par Here we compare the predictions of the cooling flow model to the
observations. This shows that the predicted column densities of all four ions
(\OVI, \NV, \CIV, \SiIV) are within 1 or 2 times the observed spread from their
average value (see Section~\Scoldens\ and Figure~\Fcoldens) for flow velocities
between 20--30 and 12--42~\kms, respectively. The resulting predictions for the
ionic ratios are shown by the thick magenta line in Fig.~\Fratiomodel. This
reveals that this model correctly predicts the observed range of the \CIV/\OVI,
\CIV/\NV\ and \SiIV/\CIV\ ratios. However, in general the model predicts slightly
too high values for $N$(\OVI) and slightly too low values for $N$(\NV) (see
appendix), leading to slightly lower \NV/\OVI\ ratios than observed (by
$\sim$0.4~dex). This is true both for the average value and for most of the
column densities seen in individual components where all four ions are detected.
Unlike what is the case for all other models, this model predicts good values
for {\it both} the observed column densities {\it and} the observed column
density ratios, suggesting that a cooling flow model including
self-photoionization represents the physical conditions in the Galactic lower
halo fairly well. It is likely that only relatively minor adaptations may be
needed to make the predictions for \OVI\ and \NV\ fit the observations even
better.

\subsection{Shock Ionization}
\par Shock fronts passing through the gas can create transition temperature gas.
A grid of low-density shock models was presented by Dopita \& Sutherland (1996),
for shock velocities of 200, 300, 400 and 500~\kms, magnetic parameter
$B\,n^{-3/2}$ = 0, 1, 2 and 4 $\mu$G\,cm$^{-3/2}$, and solar abundances. In the
Milky Way corona, shock velocities of 400~\kms\ are not expected, so we compare
data and model only for 200 and 300~\kms\ shocks. In these models the ionization
of the ions is determined taking into account non-equilibrium conditions. In
these models the logarithm of the \OVI\ column density is predicted to be
14.32--14.35, while log\,$N$(\NV) spans the range 12.85--12.95. The predictions
for log\,$N$(\CIV) and log\,$N$(\SiIV) are much more varied, ranging from
12.8--13.0 and 11.3--12.8, respectively, with higher column densities occurring
for higher shock velocities and stronger magnetic fields. Thus, shock models
predict \OVI\ and \NV\ column densities that are a factor 1.8 and 2.4 larger
than the observed average, while the column densities of \CIV\ and \SiIV\ are
underpredicted by an order of magnitude. Thus, in the ratio-ratio plots shown in
Figure~\Fratiomodel, the predictions for the shock models (shown as orange
lines) lie far away from the observed values. Therefore, shocks appear to be
unlikely explanation for the observed highly-ionized atoms in the Galactic
corona.


\subsection{Conductive Interfaces}
\par If a reservoir of hot gas comes into contact with a cool cloud, electron
collisions may conduct heat between the two media, and transition temperature
ions can be produced in the conduction front as the cooler gas evaporates and
the hotter gas condenses. This process was modeled by Borkowski et al.\ (1990)
and Gnat et al.\ (2009). In the first of these two papers predicted column
densities are shown for several ions as function of time and as function of the
angle between the magnetic field and the conduction front. The magnetic field is
important because the thermal conductivity of the plasma is far greater along
the magnetic field lines than across them.
\par In the Borkowski et al.\ (1990) model the \CIV\ column density stabilizes
at more or less constant values (less than 0.3~dex variation) after about
\dex4~yr. This is the case for \NV\ after \dex5~yr and for \OVI\ after about
2\tdex5~yr. At earlier times the evolution in the column densities (and thus the
ionic ratios) is rapid. Thus, in a ratio-ratio plot, the range of ratios that
this model can accommodate varies with the age of the youngest interface that is
included. Fox et al.\ (2004, 2005) chose \dex5~yr as the age of their youngest
interface. However, between \dex5 and 2\tdex5~yr the \OVI\ column density has
not yet stabilized, and $N$(\OVI)\ increases by 0.27~dex. On the other hand,
$N$(\NV) and $N$(\SiIV) are constant and $N$(\CIV) decreases by 0.07~dex. Thus,
the \CIV/\OVI\ ratio decreases by 0.34~dex as the interface ages from \dex5 to
2\tdex5~yr. It should be noted that these interfaces mature relatively quickly.
The ratios traverse the track in a few \dex5~yr, but then they stay constant for
the next several \dex6~yr. Thus, for over 90\% of the lifetime of the interface
this model predicts a unique value of the ionic ratios.
\par Gnat et al.\ (2010) recently revisited the conductive interface model,
exploring a greater range in the parameter space characterizing the interface.
They present column densities for mature interfaces in which the surrounding
medium has a temperature between \dex6 and \dex7~K, and in which the pressure
ranges from 0.1 to 50~K\,\cmm3. They further include photoionization by the
extragalactic radiation field, though not magnetic fields. However, unlike
Borkowski et al.\ (1990) they do not tabulate the time evolution of the column
densities. Comparing their predicted ionic ratios to those of Borkowski et al.\
(1990), we find that if the pressure $P/k$ is $>$5\,K\,\cmm3, all of the 
predicted ratios cluster within 0.2~dex of those found by Borkowski et al.\
(1990) models for the case of a perpendicular magnetic field. Thus, expanding
the parameter space does not substantially change the nature of the predictions,
and we therefore do not explicitly show the results of the Gnat et al.\ (2010)
predictions in Figure~\Fratiomodel. 
\par In Figure~\Fratiomodel\ we show with red lines the range of ratios for
times $>$2\tdex5~yr, with one line for the case of a perpendicular magnetic
field and one line for a parallel field. We connect the start and end points of
these lines to form a box that encloses the full range of possible predictions.
The line thickness increases with time. This shows that after about 1 to 2~Myr
the ionic ratios stabilize at a fixed value. Thus, unless one considers a very
young interface, this model predicts just a single set of ionic ratios. For the
case of the perpendicular magnetic field, the predicted final ratio for
\NV/\OVI\ and \CIV/\OVI\ lies toward the lower end of the observed values, while
\CIV/\NV\ is close to the average. However, \SiIV/\CIV\ is underpredicted by
about 0.5~dex, and all predictions for the case of a parallel magnetic field
differ by about 0.5~dex from the observations. The fact that Gnat et al.\ (2010)
added some photoionization to their models, but produce the same predictions for
the line ratios implies that adding photoionization will not help to increase
the relative amount of \SiIV.
\par An additional problem with these models is that each interface produces a
maximum column density of about log\,$N$(\OVI)=12.8, log\,$N$(NV)=12.1,
log\,$N$(\CIV)=12.5 and log\,$N$(\SiIV)=11.1). This is in the case with a
magnetic field that is perpendicular to the conduction front, and for an initial
temperature of the hot phase of 7.5\tdex5~K. For a parallel field these values
are 5--10 times smaller, but increasing the temperature of the hot phase to
\dex6~K increases them by a factor $\sim$2. Thus, the Borkowski et al.\ (1990)
model can be used to predict the column density {\it ratios}, but it cannot
explain the observed total column densities. That requires assuming that a
sightline intersects many interfaces. Assuming an average angle of 45\deg\
between the magnetic field and the interface (reducing the average column
density by a factor 2), but also assuming that \dex6~K is more typical for the
hot gas than 7.5\tdex5~K (increasing the average column density by a factor 2),
between five to fifty interfaces would be required to build up the observed
total column densities. As each interface has a thickness on the order of 10~pc,
the interfaces would have to occupy a total pathlength on the order of 100~pc,
which is easily accommodated within the several kpc scaleheight of the
highly-ionized atoms.
\par To summarize: (1) the conductive interface model strongly underpredicts
$N$(\SiIV); (2) the time spent as a mature interface with constant ionic ratios
is a factor 10 longer than the time spent along the tracks, (3) for a mature
interface the predicted \NV/OVI\ and \CIV/\OVI\ ratios are relatively low, and
(4) it is difficult to build up the total column densities that are seen, with
each interface only contributing a small fraction of the total. We thus conclude
that conductive interface physics is not the most important process producing
the highly-ionized atoms in the Galactic corona. We note that this conclusion is
consistent with the Bowen et al.\ (2008) result that the \OVI| absorbers in the
disk exhibit a broad range of column densities, with the more strongly absorbing
clouds having a lower space density than the weakly absorbing clouds. The latter
conclusion follows from the fact that while $N$(\OVI) correlates with distance,
the dispersion in $N$(\OVI) does not decrease as the detected clouds get more
distant.


\subsection{Thick Disk supernovae}
\par Another model that we consider here is that of Shelton (1998), who analyzed
the effect that old type Ia supernova remnants above the Galactic plane might
have on the state of the Thick Disk gas. She predicted the time-evolution of the
column densities of \OVI, \NV\ and \CIV\ (though not of \SiIV) inside a
supernova bubble for an explosion with energy 0.5\tdex{51}\,erg expanding into a
medium with ambient density 0.01\,\cmm3\ and ambient non-thermal pressure
1800\,K\,\cmm3. We were separately provided the predictions for the \SiIV\ ion,
so that we could properly compare this model to the observations.
\par In this model, the \OVI\ column density decreases from an early value of
log\,$N$(\OVI)=14.2 at 0.1~Myr to about 13.6 at 0.5~Myr, then slowly rises to
13.7 at 10~Myr, falls to 13.2 at 15~Myr and then it disappears. The other ions
behave similarly, plateauing at $\sim$12.8, 13.2, and 12.0 dex for \NV, \CIV\
and \SiIV, respectively. \NV/\OVI\ is about $-$1.0 up to an age of \dex7~yr,
then decreases to $-$0.5 dex at 1.5\tdex7~yr.  On the other hand, in the early
stages, \SiIV/\CIV\ first dips from $\sim$$-$1.5 to $\sim$$-$0.7 dex, then
reaches a value of $\sim$$-$1.1 dex after 2\tdex6~yr, where it remains
throughout.
\par As can be seen from the green line Figure~\Fratiomodel b, this model
predicts values for \NV/\OVI\ that span the range of observed values. However,
if the supernovae have random ages, there should be more ratios near $-$1.1
(ages $<$10~Myr) than at higher values (ratio slowly increasing until an age of
16~Myr). Further, this model underpredicts \SiIV/\CIV\ by about 0.5~dex and the
absolute values of the column densities during the plateau phase are about 0.4,
0.5, 0.6 and 1.2~dex too low for \OVI, \NV, \CIV\ and \SiIV, respectively. The
latter problem can be accommodated if between 2 and 3 remnants are intersected
in each sightline. Since Shelton (1998) derives that the sky coverage fraction
due to Thick Disk supernovae is 30--60\%, this implies that the predicted column
density is a factor $\sim$3 to 5 too low to account for the observed column
densities. 
\par In summary, this model predicts the correct range of \NV/\OVI\ ratios, but
the \SiIV/\CIV\ ratio is underpredicted by a 0.5~dex and it predicts a lopsided
distribution of \NV/\OVI\ ratios. Further, the implied supernovae rate is a
factor 3--5 too high, Thus, halo supernovae may contribute to the observed
quantities of highly-ionized atoms, but they cannot be the whole story.


\subsection{Turbulent Mixing Layers}
\par Conductive heat transfer is not the only process that can take place at the
interface between hot and cold gas. Relative sideways motions can cause
Kelvin-Helmholtz instabilities that mix the two, creating gas at transition
temperatures. This model was proposed by Begelman \& Fabian (1990), and
developed by Slavin et al.\ (1993), Esquivel et al.\ (2006) and Kwak \& Shelton
(2010). These papers present predictions for ionic column densities.
\par Slavin et al.\ (1993) presented predictions for entrainment velocities
between 25 and 100~\kms, at temperatures of log\,$T$= 5.0, 5.3 and 5.5, and for
two abundance cases -- solar abundances and gas in which half of the C, N, and O
(and 90\% of the Si) are on dust grains, for a total of 12 different
predictions. In these models the mechanical aspects of the mixing were treated
analytically (based on a simplified picture of how the mixing occurs), while the
ionic column densities were then predicted using non-equilibrium ionization
balance calculations including self-photoionization. We show the solar-abundance
results of Slavin et al.\ (1993) in Figure~\Fratiomodel\ (light blue points and
lines).
Since they showed that the change of ionic ratios with entrainment
velocity is almost linear, for a given temperature we interpolate the ratios at
25 and 100~\kms\ to form a series of points in the ratio-ratio diagram,
increasing in size with entrainment velocity. Points with log\,$T$=5.0 are shown
by closed circles, points at log\,$T$=5.3 by closed squares and points at
log\,$T$=5.5 by closed stars. High \SiIV/\CIV\ occurs at log\,$T$=5.3, while
high \NV/\OVI\ occurs at log\,$T$=5.5. We then use dotted lines to connect the
values for the three different temperatures at the same entrainment velocity.
This defines a region in the ratio-ratio diagnostic plot for the predictions of
the Slavin et al.\ (1993) turbulent-mixing model. Changing the abundance pattern
(i.e.\ assuming differential depletion onto dust) shifts these points somewhat,
but by less than 0.2~dex for \OVI, \NV\ and \CIV, although \SiIV\ is strongly
decreased.
\par In the Slavin et al.\ (1993) model, the \OVI\ column density in each
interface was only about \dex{11}--\dex{12}\,\cmm2, implying the need for more
than 100 interfaces to build up the total observed column density of
\dex{13}--\dex{14}\,\cmm2. de Avillez \& Breitschwerdt (2005) show that this may
indeed be possible. They took a hydrodynamical model of the ISM and calculated
the density of \OVI\ as implied by radiative cooling (assuming local CIE in each
gridcell). The \OVI\ originates in the interiors of regions created by
supernovae, as well as the result of local cooling promoted by turbulent mixing.
They found that on average each sightline out to 1~kpc contains many structures
that contribute a few \dex{12}\,\cmm2\ of \OVI, and that the total is about
\dex{14}\,\cmm2. Although the assumption of local CIE is probably false (given
that this predicts the wrong ionic ratios), this shows it could in principle be
possible to build up the total observed \OVI\ column density from many different
small structures in the sightline.
\par The study by Esquivel et al.\ (2006) addressed the effects of magnetic
fields, which mainly result in a reduction of the spread in the predicted ratios
for a given flow velocity and density choice. Since Esquivel et al.\ (2006) did
not give a table of predictions, and since Figures~8 and 9 in Esquivel et al.\
(2006) show that their results are similar to those of Slavin et al.\ (1993), we
do not include this model in Figure~\Fratiomodel.
\par Kwak \& Shelton (2010) revisited the turbulent mixing model, making a
two-dimensional hydrodynamical calculation, while following the ionization using
non-equilibrium physics (although they still used the CIE value for the cooling
function at each temperature). Their base model (``A'') uses a 100$\times$300~pc
box, with hot gas density and temperature \dex{-4}\,\cmm3\ and \dex6~K and cool
gas density and temperature 0.1\,\cmm3\ and \dex3~K, and an entrainment velocity
of 100~\kms. In their paper, they show the time evolution of the column
densities of \OVI, \NV\ and \CIV, and tabulate the time and sightline averages
for these three ions (which are log\,$N$(\OVI)=12.8, log\,$N$(\NV)=12.2,
log\,$N$(\CIV)=13.1), as well as the ionic ratios. Although Kwak \& Shelton
(2010) did not publish their predictions for \SiIV, they included it in their
calculations, and we obtained the \SiIV\ values from them.
\par For each time step, their model produces a two-dimensional map, with 256
directions along which a column density can be derived for each ion. We obtained
each of these values, once every Myr for an 80 Myr period. Since it takes about
20~Myr for the effects of initial conditions to disappear, we used the results
between 20 and 80~Myr to create a contour plot showing how often a given ratio
occurs. The resulting contours are shown in blue in Figure~\Fratiomodel.
Clearly, this model gets the ratios almost correct, with on average \NV/\OVI\
about 0.3~dex too high (observed average $-$0.87, model average $-$0.56),
\CIV/\OVI\ about 0.6~dex too high ($-$0.21 observed average, 0.39 predicted
average), \CIV/\NV\ about 0.2~dex too high (0.75 vs 0.94) and \SiIV/\CIV\ about
0.2~dex too low ($-$0.60 vs $-$0.83). Some of these differences are similar to
the mismatches for other models, but for the TML model none of the differences
is larger than 0.6~dex, unlike what is the case for any other model. The curve
showing the evolution of the ratios during static non-equilibrium cooling
crosses the turbulent mixing contours, suggesting that in the turbulent mixing
model the cooling is not highly affected by the motions of the gas. In this
model, the \OVI\ column density through each $\sim$100~pc thick layer is about
log\,$N$(\OVI)=12.8, so that about 10--20 layers are needed to make up the
observed total. These would take up about 1~kpc of the sightlines, which is a
substantial, but possible, percentage for gas with a scaleheight of 3~kpc.

\par We also compared the predictions of the Kwak \& Shelton (2010) model~A to
those of their models~D and F. They already showed that their results are not
very sensitive to the parameters (spatial resolution, grid size, entrainment
velocity, initial perturbation amplitude, temperature of the hot gas), but that
there are some differences. To be precise, Model~D (entrainment velocity
50~\kms\ instead of 100~\kms) results in relatively more \SiIV\ and \CIV,
improving the correspondence with the observed values for \SiIV/\CIV, but it
makes the predictions for \CIV/\OVI\ and \CIV/\NV\ slightly worse. The same
effect, but stronger, happens in Model~F (temperature of the hot gas 3\tdex6~K
instead of \dex6~K). Their Model~A appears to give the best fit. We conclude
that turbulent mixing layers may play an important role in explaining the
highly-ionized ions in the Galactic Corona.


\section{Discussion}
\par We showed in the Sect.~\Sextent\ that the highly-ionized gas in the
Galactic corona usually is seen at velocities 30 to 100~\kms\ more negative or
more positive than can easily be explained by differential galactic rotation. In
Sects.~\Swidth, \Srotate, \Scoldens, and \Ssky\ we analyzed the widths of the
absorption lines, the signature of Galactic rotation, the column densities and
the relation between these quantities and location on the sky. We found that the
widths (FWHM) of individual absorption lines range from $\sim$25~\kms\ to
$\sim$120~\kms. On average the widths of \OVI\ and \NV\ absorption lines are
similar, as are those of \CIV\ and \SiIV, while the average of the latter is
slightly, but significantly, narrower (42~\kms\ vs 60~\kms). The line width is
always much larger than the maximum contribution from thermal motions
($\sim$20~\kms), so most of the linewidth must be non-thermal. Therefore, the
four ions cannot be completely co-spatial. On the other hand, the widths of
\OVI/\NV\ vs \CIV/\SiIV\ correlate very well. That is, in sightlines where \OVI\
and \NV\ are wide, \CIV\ and \SiIV\ are also wide. Furthermore, there appears to
be no difference in the centroids of the absorption lines of these ions. Thus,
although they do not originate in the same parcels of gas, they must originate
in a single structure, in which non-thermal motions are larger in the region
where the hotter gas occurs.
\par There appears to be no relation between galactic latitude and the extent of
the absorption, which implies that the velocities of individual gas parcels in
the sightline are randomly oriented. If, for instance, the velocities were
mostly vertical, projection effects would lead to a relation between the extent
of the absorption and latitude, with wider lines at higher latitude. The large
spread of velocities in each direction make the effect of galactic rotation
difficult to discern, but we can show that it is responsible for shifting the
centroids of the profiles by up to 30~\kms\ from zero. We looked for evidence of
a vertical gradient in the rotation velocity, by comparing the expected
centroids to the observed values. Taking into account that the observed
scaleheight of the gas is about 3~kpc, we find that a co-rotating corona best
explains the centroids. We further find that although the distribution of the
highly-ionized gas is by no means uniform, the variations across the sky are
less than an order of magnitude, with no directions toward which these ions are
not seen. Much of the variation can be explained as a latitude effect, with a
larger pathlength through the layer at low latitudes.  However, over most of
outer galaxy below the Galactic plane there is a slight (0.2~dex, factor 1.6)
deficit in $N$(\OVI), while in the inner Galaxy and over the Galactic North Pole
there is a slight excess (0.2--0.4~dex, factor $\sim$2) in the column density of
all ions. A description in terms of a plane-parallel layer in which the column
density changes as \Nsinb\ works fairly well. The dispersion around \Nsinb\ is
then only about 0.15--0.20~dex (a factor $\sim$1.5). Thus, the highly-ionized
atoms appear to occur in a somewhat irregular thick layer with scaleheight
3~kpc, which rotates along with the Milky Way, and in which gas parcels have 
random velocities that are on the order of $\pm$60~\kms.
\par For a given absorption component and ion, the distribution of column
densities is relatively narrow. The range in column densities is no more than a
factor 10, with a dispersion of a factor of about 2, i.e.\ for 70\% of the
sightlines the observed ionic column density falls within a factor 2 of the
average. This is compatible with a situation in which the transition temperature
gas originates in just a few (3--5) individual structures in each sightline, in
each of which the column density is on the order of 1/3--1/2 the average
observed value. The distribution of column density ratios is even narrower, with
the ratios \NV/\OVI\ and \SiIV/\CIV\ having a dispersion of only 0.14~dex (a
factor 1.3 either way). This suggests that in each individual structure the same
process is responsible for producing the transition temperature gas everywhere.
The fact that the distribution of the \CIV/\OVI\ and \CIV/\NV\ ratios is
slightly wider 0.23~dex (a factor 1.7) again indicates that although these ions
all come from a single structure, the ions that should occur at somewhat higher
temperature are not fully co-spatial with the ions that are more prevalent at
lower temperatures.
\par From the ratio-ratio plots shown in Figure~\Fratiomodel, we conclude that
the physical process(es) that generate the highly-ionized gas in the Milky Way's
corona must include non-equilibrium ionization and involve radiative cooling. In
fact, the gas must be very far from collisional ionization equilibrium, since in
CIE \OVI\ cannot coexist with \SiIV, and there is no single temperature that
simultaneously explains the observed \NV/\OVI\ and \CIV/\NV\ ratios. Further,
for temperatures near \dex5~K all ionic ratios differ by at least a factor 10
from the observed values. Predictions based on shock heating strongly
overpredict the amount of \OVI\ relative to \NV\ and \CIV, and even the amount
of \NV\ relative to \CIV. A model in which the transfer of heat occurs purely
through electron conduction strongly underpredicts the \SiIV/\CIV\ ratio (by a
factor 5--10, and underpredicts \NV/\OVI\ and \CIV/\OVI, although it gets
\NV/\OVI\ correct, but only when the magnetic field is perpendicular to the
interface. For a non-perpendicular magnetic field the predictions of the
conductive interface model are rather different from the observations. This
model {\it does} seem to work reasonably well if one only considers the
\NV/\OVI\ vs \CIV/\OVI\ ratio-ratio plot. However, compared to the square box
used by Fox et al.\ (2004, 2005), a more detailed look at the actual predictions
shows that they fall in a strip parallel to but below the observations (see
Figure~\Fratiomodel b).
\par We explored four different models that include non-equilibrium radiative
cooling, including static cooling, the flow of gas through a cooling front,
Thick Disk supernovae and turbulent mixing. The original turbulent mixing model
of Slavin et al.\ (1993) predicts much larger ranges for the ratios than is
observed, as well as a factor 1--2 more \OVI\ than the newer numerical
simulations of Kwak \& Shelton (2010). Except for the static case, in these
models the \OVI, \NV, \CIV\ and \SiIV\ do not coexist in the same gas parcels,
but they are present in a related way in different parts of the same structure,
as is indeed indicated by the observations. Since the predicted column densities
are based on an integral through these structures, the predictions account for
this effect.
\par The Thick Disk supernovae model underpredicts \SiIV/\CIV\ by about 0.5~dex,
and predicts a distribution of ratios for \NV/\OVI\ and \CIV/\OVI\ that is much
more strongly weighted toward lower values than what is observed. For the
remaining three models, the predicted range of ratios differs by up to about
0.3~dex from the observations, with static radiative cooling and turbulent
mixing predicting a value for \NV/\OVI\, \NV/\CIV\ that is too high by about
0.25~dex and \CIV/\OVI\ too high by 0.5~dex. At the same time, \SiIV/\CIV\ is
too low by about 0.2~dex. That is, for a given amount of \OVI, these models (on
average) predict a factor 1.8 too much \NV, a factor 3 too much \CIV\ and a
factor 1.7 too much \SiIV. The cooling flow model (with self-photoionization
included) predicts the correct range of ratios for \CIV/\OVI\ and for
\SiIV/\CIV, but it underpredicts the amount of \OVI\ relative to \CIV\ and \NV.
Thus, none of the non-equilibrium ionization radiative cooling models can quite
get all four ions correct at the same time. However, compared to models having
equilibrium ionization and no radiative cooling, the former do a much better
job.


%


\section{Summary}
\par We have analyzed the absorption from highly-ionized atoms through the
complete Milky Way Thick Disk and halo as seen in 58 sightlines toward
extragalactic targets. These ions are \OVI, \NV, \CIV\ and \SiIV\ and they
sample gas at temperatures near \dex5~K, which originates where gas near \dex6~K
cools or interacts with \dex4~K gas. For each sightline we determined over which
range of velocities the absorption is likely to be associated with the lower
Milky Way corona. We then measured the centroid velocity, linewidth, and column
density for each ion. In many sightlines multiple components can be
distinguished, which were separately measured. From these measurements we find
that:
\par (1) The centroid velocities of \OVI, \NV, \CIV\ and \SiIV\ are very well
aligned, with a dispersion in the differences of $<$8~\kms.
\par (2) The highly-ionized absorption extends to velocities between 30 and
80~\kms\ more negative or more positive than the range that can easily be
explained by differential galactic rotation. For sightlines in the region near
the Galactic north pole there is an excess of absorption at velocities more
negative than $-$60~\kms. This excess does not correlate with the
intermediate-velocity clouds seen in \HI\ 21-cm emission.
\par (3) We determined linewidths by calculating the second moment of the
apparent column density profiles, converting to the full width at half maximum
(FWHM). The resulting widths of the \OVI\ and \NV\ lines are on average 60~\kms\
and correlate very well (correlation coefficient $\rho$=0.97). The average
widths of \CIV\ and \SiIV\ are 42~\kms\ (and $\rho$=0.94). The other
correlations (e.g.\ between the widths of \OVI\ and \CIV) are slightly less
strong ($\rho$$\sim$0.92). The difference between the widths of \OVI/\NV\ and
\CIV/\SiIV\ is shown to be significant. Since the thermal width of these lines
is only $\sim$15~\kms\ for gas at a temperature of \dex5~K, most of the width
must be non-thermal.
\par (4) We find that differential galactic rotation leaves its imprint on the
velocities of the absorption lines. We predict the locations of the centroids
from a simple model in which the gas co-rotates as one goes up above the disk,
and in which the density falls off exponentially with a scaleheight of 3~kpc.
Comparing these predictions to the observed centroids we can see the effect of
galactic rotation, and we find that the average difference between the data and
the prediction from the simple model is 5$\pm$16~\kms.
\par (5) The distributions of the integrated column densities of the
highly-ionized atoms is relatively narrow, with average and dispersion \VALUE
14.31$\pm$0.20 (\OVI), 13.48$\pm$0.23 (\NV), 14.17$\pm$0.21 (\CIV) ad
13.57$\pm$0.18 (\SiIV). For individual components these values are
14.06$\pm$0.27 (\OVI), 13.30$\pm$0.30 (\NV), 13.83$\pm$0.34 (\CIV) and
13.24$\pm$0.35 (\SiIV). Note that in the case of \NV\ upper limits were omitted
for this calculation.
\par (6) Over much of the sky the column densities generally increase with
decreasing galactic latitude, but with a dispersion of about 0.17~dex (a factor
1.5) around the average. The equivalent polar column densities are \VALUE 14.15,
13.19, 13.89 and 13.32 for \OVI, \NV, \CIV\ and \SiIV, respectively.
\par (7) Assuming solar abundances and an ionization fraction of about 10\% the
polar column densities imply a total hydrogen column density associated with the
transition temperature gas of about \dex{18.8}\,\cmm2.
\par (8) In sightlines near the Galactic Center and toward the Galactic north
pole, the column densities of all ions are enhanced by about 0.3~dex. This
suggests enhanced activity at small galactocentric radii, and near the Galactic
north pole. We find marginal evidence that this excess is associated with
enhanced absorption at negative velocities.
\par (9) At intermediate latitudes in the southern sky, \OVI\ is depressed by
about 0.2~dex, but none of the other ions are, indicating a subtle difference in
the physical conditions.
\par (10) Determining the ratios of the column densities of the different ions,
we find that the distributions of \NV/\OVI\ and \SiIV/\CIV\ are rather narrow,
with a dispersion of just 0.14~dex ($\pm$40\%), and a spread between the
smallest and largest value of 0.6~dex (factor 4). The distributions of other
ratios (such as \CIV/\OVI) give a larger spread of 0.23~dex ($\pm$70\%).
Important average ratios and spreads are (using a linear scale)
\NV/OVI=0.14$\pm$0.04, \OVI/\NV=7.80$\pm$2.90, \CIV/\OVI=0.71$\pm$0.36,
\OVI/\CIV=1.83$\pm$1.15, \SiIV/\CIV=0.26$\pm$0.08, \CIV/\SiIV=4.19$\pm$1.47
\par (11) We discuss a number of ionization mechanisms, going back to the
original papers to extract the detailed predictions. The models we discuss
include collisional ionization equilibrium (CIE, from Sutherland \& Dopita 1993
and Gnat \& Sternberg 2007), static non-equilibrium ionization radiative cooling
(Gnat \& Sternberg 2007), radiatively cooling gas flows (Shapiro \& Benjamin
1991, but fully described for the first time in the appendix), shock ionization
(Dopita \& Sutherland 1996), conductive interfaces (Borkowski et al.\ 1990 and
Gnat et al.\ 2010), Thick Disk supernovae (Shelton 1998) and turbulent mixing
layers (Slavin et al.\ 1993 and Kwak \& Shelton 2010). We compare the
predictions of ionic ratios for each model with the data in four different
ratio-ratio plots, which include \NV/OVI, \CIV/\OVI\ or \CIV/\NV\ vs \SiIV/\CIV\
and \NV/\OVI\ vs \CIV/\OVI. These were chosen so that for future observations at
least one diagram can be used if data are available for only three of the ions.
The historically most-used \NV/\OVI\ vs \CIV/\OVI\ diagram is the least
discriminative of the four.
\par (12) We find that single-phase CIE models fail to explain the observations.
\par (13) The model for static non-equilibrium ionization radiative cooling
shows that if the gas cools below $T$$\sim$\dex{4.9}~K the predicted ionic
ratios are close to the observed values, with \SiIV/\CIV\ underpredicted by
about 0.3~dex and \NV/\OVI\ overpredicted by the same amount.
\par (14) In radiatively cooling gas flows, there is a range of flow velocities
at which the amount of cooling gas is similar to the observed amount. The best
match for all four ions simultaneously requires a flow velocity on the order of
25~\kms. This model predicts that the different ions originate in different
parts of the flow, with the pairs \NV/\OVI\ and \CIV/\SiIV\ living more closely
together with each other than they do with the other ions. The model correctly
predicts the range of ratios for \CIV/\NV\ and \SiIV/\CIV, although it
overpredicts \OVI/\NV.
\par (15) Shock ionization models predict far too much \OVI\ relative to the
amount of \NV\ and \CIV, as well as too much \NV\ relative to \CIV. I.e., the
predicted ionization state of the gas is too high.
\par (16) Conductive interface models were favored by Sembach et al.\ (2003) and
Fox et al.\ (2004, 2005) to explain the transition temperature ions seen in
high-velocity clouds. However, Fox et al.\ (2004, 2005) used the full ranges for
\NV/\OVI\ and \CIV/\OVI\ predicted by Borkowski et al.\ (1990) for interfaces
older than \dex5~yr. Here, we instead follow the evolution of these ratios with
time and find that the delineated region is really a strip that lies below the
observations. Furthermore, it takes only 1--2~Myr for the ratios to stabilize at
a fixed value, compared to an interface lifetime that is $>$10~Myr. The other
three diagrams show that the model strongly underpredicts \SiIV/\CIV, while the
stabilized values for \NV/\OVI\ and \CIV/\OVI\ also tend to be too low.
Furthermore, the model predicts total column densities in a single layer of only
$\sim$\dex{12.5} for \OVI, so that $>$50 interfaces are required in each
sightline. In light of these considerations, we conclude that the absorption
from highly-ionized atoms in the Galactic corona probably does not originate in
conductive interfaces.
\par (17) The effects of Thick Disk supernovae on the ionization state of the
tenuous Galactic coronal gas was explored by Shelton (1998). In this model, the
\NV/\OVI\ and \CIV/\OVI\ ratios evolve slowly at first, but then increase almost
linearly over 16~Myr to cover the observed range. \SiIV/\CIV\ on the other hand
is basically constant over the lifetime of the remnant, but underpredicted by
about 0.5~dex. The integrated ionic column densities for \OVI, \NV\ and \CIV\
are only a factor 3--5 below the observed values. If a way could be found to
increase $N$(\SiIV) (e.g.\ including photoionization), this model might fit the
observations. However, the number of Thick Disk supernovae needed to produce the
integrated column densities is a factor of a few larger than the number of
supernovae observed.
\par (18) The original predictions for turbulent mixing layers by Slavin et al.\
(1993) cover a range in \SiIV/\CIV\ that is much larger than what is observed.
The predictions for the other ionic ratios also do not match the data well. On
the other hand, a more recent, numerical, model of turbulent mixing layers by
Kwak \& Shelton (2010) almost gets most of the ionic ratios correct, although it
appears to overpredict the amount of the ions with higher ionization potential
(\OVI\ and \NV).
\par (19) From the fact that the number of absorption components that can be
identified in each sightline is between 1 and 3, combined with the comparisons
between the data and predictions from different models, and taking into account
the results of our analysis of velocities and column densities, we conclude that
the transition temperature gas in the Milky Way corona originates in several (1
to 5) structures in each sightline. These are more common closer to the plane,
and have a scaleheight of $\sim$3~kpc. Each structure has a randomly-oriented
velocity, with a dispersion between different structures of about 75~\kms. In
these structures radiative cooling takes place, fast enough for the ionization
to be frozen in. Whether the structures originate in a cooling flow, or as the
result of the mixing of cool and hot gas remains unclear.
\par (20) Comparing the observed ionic ratios in Thick Disk gas to those seen in
Disk gas and high-velocity clouds, we find that all of these fall in the same
regions in the ratio-ratio plots, suggesting a similar origin for the transition
temperature gas. In high-redshift Damped \Lya\ Absorbers the \OVI, \CIV\ and
\SiIV\ ions also exist in the same proportions, but \NV\ appears underabundant,
which is consistent with the subsolar N/O ratios measured in these systems.

\bigskip
B.P.W.\ thanks the Space Telescope Science Institute for financial support
through archival grant HST-R-10984.01-A. R.A.B. would like to acknowledge the
support of NASA Astrophysical Theory grant NNX10AI70G to the University of
Wisconsin-Whitewater. R.A.B.\ also would like to thank Paul Shapiro for
suggesting the calculations described in the Appendix. We thank Orly Gnat for
providing a useful plot of the timescales on which gas cools in her radiative
cooling models. We thank Robin Shelton for providing us with the \SiIV\
predictions for the Thick Disk supernova model. We thank Kyujin Kwak and Robin
Shelton for giving us the detailed Myr-by-Myr column density predictions for
their turbulent mixing model.


\newpage
{\bf Appendix}

\medskip\par{\it A cooling flow model}
\par In this appendix we describe a model to predict the column densities and
emissivities of highly-ionized atoms. A summary of this model was presented by
Shapiro \& Benjamin (1991, 1993) and Benjamin \& Shapiro (1993); the full model
is described in Benjamin (1994). This is the model labeled ``Cooling Flow'' in
Fig.~\Fratiomodel.
\par The model starts from the realization that a common element of the models
used to analyze the column densities of the highly-ionized atoms is that they
originate in gas that is not thermally stable. As a result, these ions are
necessarily associated with flows of heating or cooling material. Since the
heating and cooling timescales of this gas can be shorter than the ionization or
recombination timescales, it is necessary to account for non-equilibrium effects
when calculating the ionization. Moreover, it is important to consider the
effects of ionizing radiation, produced either externally or by the cooling gas
itself; this can substantially change the colum denities for certain ions under
the right circumstances.
\par Such models of a radiatively cooling flow can be considered a
generalization of post-shock flow models. However, unlike the case of radiative
shock models, the initial flow velocity and the initial gas temperature are not
necessarily coupled. The parameters for a flow model are (1) the initial
temperature, (2) the initial density, and (3) the flow's velocity, as well as
(4) the density compression evolution. In addition, one must specify an
elemental abundance set and a set of atomic data to calculate the
non-equilibrium ionization and cooling of the gas. In the model calculations
described below the abundances used were from Allen (1973); the atomic data were
completely described in Benjamin (1994), but come principally from Raymond \&
Smith (1977), Raymond (1987, priv.\ comm.) Shull \& van Steenberg (1985) and
Romanik (1988). There would be a (small) correction associated with adjusting
the results to the more recent Asplund et al.\ (2009) abundances, which changes
\CIV/\OVI\ by 0.04~dex and \SiIV/\CIV\ by $-$0.08~dex. We include these two
changes in all the results shown below. The absolute changes in the abundances
are a bit larger ($-$0.13~dex for oxygen, $-$0.09~dex for carbon, $-$0.13~dex
for nitrogen and +0.01~dex for silicon), but (as argued below), a change in the
abundance of oxygen will not significantly affect the \OVI\ column density. For
the other elements it is only the change relative to the change in oxygen
abundance that matters.

\begin{figure}\plotfiddle{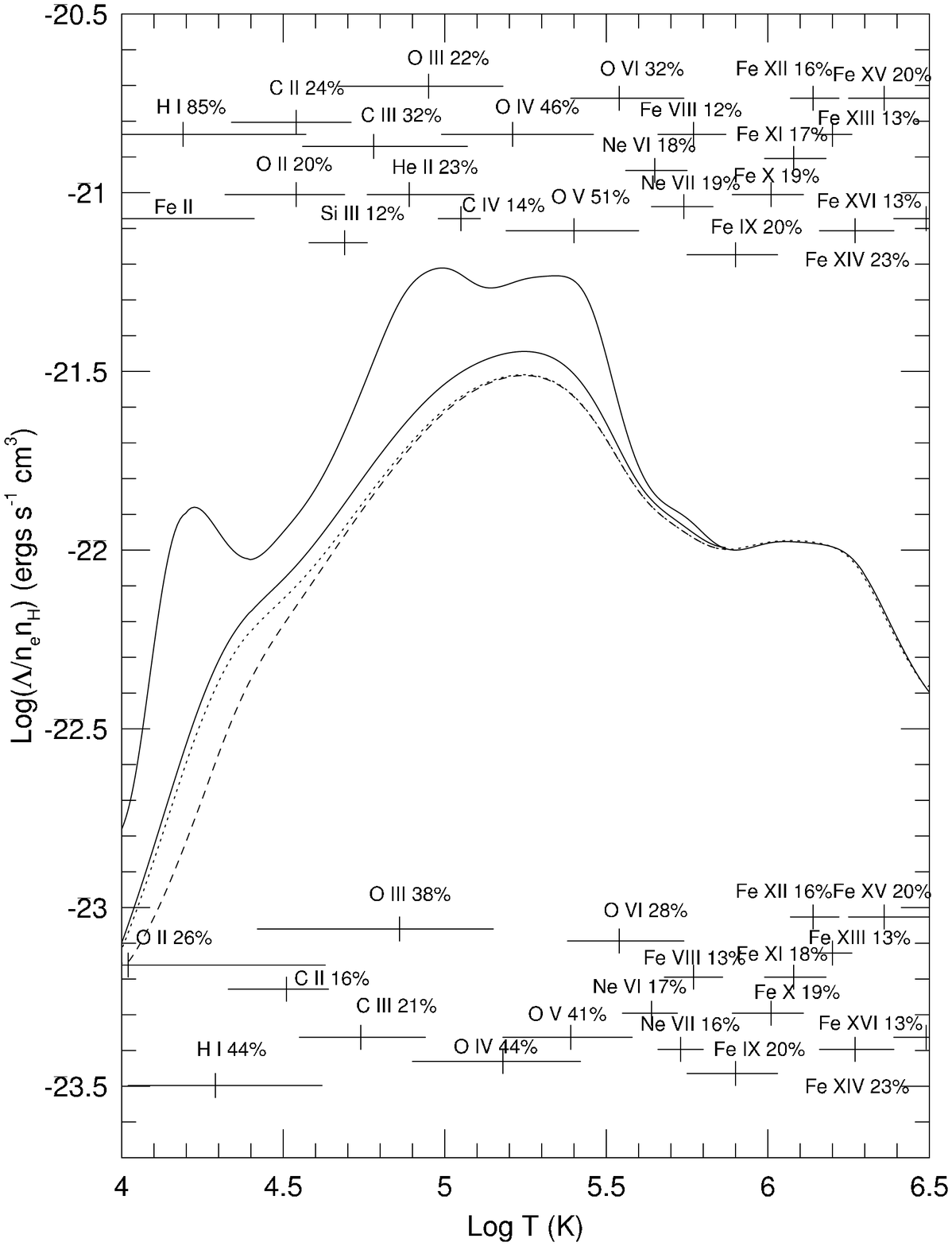}{0in}{0}{350}{350}{0}{0}\figurenum{A1}
\caption{Radiative cooling rate of a solar metallicity gas
($\log[\Lambda/n_e\,n_H]$~(erg\,\cmm3\,s$^{-1})$) versus $\log T$(K). Three
cases are shown: cooling in collisional equilibrium ionization (upper solid
curve), isobarically with non-equilibrium ionization evolution (lower solid
curve), and isochorically with non-equilibrium ionization evolution (dotted
curve), in the absence of the effects of ionizing radiation. The dashed line
shows the change in the isochoric cooling curve when the effects of radiation
are accounted for assuming an initial flow velocity of $v_0$=30~\kms. These
curves are very similar in shape and level to the ones used by others (see e.g.\
Sutherland and Dopita 1993 and Gnat \& Sternberg 2007). The horizontal bars at
the top of the graph indicate the temperature regime in which the labeled ion
contributes more than 10\% of the total cooling rate for gas cooling in
ionization equilibrium. The temperature at which this ion makes its maximum
fractional contribution to the cooling rate is denoted by a vertical tick mark
and the percentage is noted. The lower bars show the same quantities but for
non-equilibrium, isobarically cooling gas.}
\end{figure}

\medskip\par{\it Cooling curve}
\par The effects of the non-equilibrium ionization and cooling are encapsulated
in the cooling curve. These are given in Fig.~\FAcoolcurve\ for equilibrium,
isobaric and isochoric cooling gas, along with the dominant coolants at
different temperature ranges. The instantaneous cooling time for a solar
metallicity gas in ionization equilibrium with initial temperature
$T_6$=$T$/\dex6~K is $t_c$=3\tdex7 $n_{-2}^{-1}\ T_6\ {{\cal L}_{-23}}^{-1}$~yr,
where $n_{-2}$=$n_{\rm H}$/(\dex{-2}\,\cmm3) and the normalized cooling rate is
${\cal L}_{-23} = \Lambda/(n_{\rm e} n_{\rm H})/(10^{-23}$
erg\,\cmm3\,s$^{-1}$). This complicated function can be approximated by
functional forms that are good to within 50\% (for the equilibrium case) or even
30\% (for the isobaric and isochoric cases) by the following formulae:
\begin{equation}
{\cal L}_{-23}(T)= \left\{
\begin{array}{cl}
13.4~T_6^{-0.74}                                         & \ 10^5\,    {\rm K}\ \simlt\ T\ \simlt\ 2\times10^7\,{\rm K}\ ({\rm equilibrium}) \\
36.3 \exp \left[-3.4({\rm log}_{10}(T)-5.23)^{2} \right] & \ 10^{4.4}\,{\rm K}\ \simlt\ T\ \simlt\        10^6\,{\rm K}\ ({\rm isobaric}) \\
31.2 \exp \left[-3.4({\rm log}_{10}(T)-5.23)^{2} \right] & \ 10^{4.4}\,{\rm K}\ \simlt\ T\ \simlt\        10^6\,{\rm K}\ ({\rm isochoric}) 
\; ,
\end{array} \right\}
\end{equation}
When the gas cools to a temperature $T$$\sim$\dex{4.4}~K, the dominant source of
cooling is collisional excitation of H \Lya, which is extremely sensitive to the
ionization fraction of hydrogen, and thus depends sensitively upon the assumed
value of the incident ionizing flux.

\medskip\par{\it Structure of the flow}
\par The initial conditions of the one-dimensional model are marked by a flow
boundary. On one side of this boundary the gas has some high initial temperature
$T_0$ ($>$\dex6~K), and on the other side the gas has started to cool due to
radiative losses. Hot material flows through the boundary, and moves away from
it. This sets up a steady-state temperature and ionization-level gradient along
the direction of the flow. The rate of mass ($M$) flowing through per unit area
($A$) equals $\dot{M}/A=\bar{m} n_{{\rm H},0} v_0$, where $\bar{m}$ is the mean
mass per hydrogen atom (which depends on the metallicity), $n_{{\rm H},0}$ is
the initial hydrogen particle density, and $v_0$ is the flow velocity. This
velocity is defined in the (unique) reference frame where the particle flux
remains constant.
\par As the gas cools, the loss of thermal energy reduces its pressure, and thus
its density should increase. However, the actual pressure profile along the flow
is also affected by the constant addition of new material, by geometrical
constraints (e.g.\ expansion) and/or by magnetic fields. For simplicity, we only
consider two bounding cases for the density evolution of the gas as it travels
along the flow: isobaric (constant pressure) and isochoric (constant volume and
thus constant density). In the isobaric case the temperature drops along the
flow, but it is assumed that the volume increases proportionally, such that the
pressure is constant. This occurs, for example, in a non-magnetized steady-state
environment, where the sound crossing time is shorter than cooling time. Most of
the other models that are being considered elsewhere in this paper (turbulent
mixing, conductive interfaces, radiative shocks) are also characterized by
pressure equilibrium. In the isochoric case, the temperature drops, but the gas
does not compress. This occurs, for example, if magnetic pressure exceeds the
thermal pressure in the gas, or if the flow is unconstrained by external
pressure and allowed to diverge. For either of these cases (isobaric or
isochoric), fixing the pressure or density profile along the flow fixes the
temperature profile as function of the time elapsed since the material passed
the boundary.
\par In order to compare the model predictions to observations, we derive the
column densities of the different ions along sightlines along the flow
direction, i.e., perpendicular to the boundary. The column density of any given
ion is equal to the integral of the ion's density times the length of the
segment in the flow where the ion is present. Given a flow velocity $v_0$, and
an amount of time $dt$ spent at some given temperature, the segment's length is
$v_0\,dt$. Each ion $i$ has an elemental abundance, $A_i$. Its ionization
fraction after some amount of time, $t$, since passing through the boundary is
$f_i(t)$, as derived from the radiative cooling calculation. Thus, the ion's
volume density at time $t$ since crossing the boundary is $f_i(t)\,A_i\,n_{{\rm
H},0}$. This implies a column density along the flow direction equal to
$$N_i = \int_0^t\,f_i(t')\,A_i~v_0~d(n_{{\rm H},0}t') = v_0 \mu_i,$$
where this equation defines $\mu_i$ as a constant of the flow for each ion.
Values for $\mu_i$ for a number of ions can be found in Table~A1.
\par The fact that the column densities are proportional to the flow velocity,
but do not depend on the density or ionization structure can be understood as
follows. Start with a zone in the flow at a given temperature. If the density is
high, the cooling time is short, and thus the thickness of the zone is small. At
lower densities, the cooling time is long and thus the zone is thicker. The
thickness of zone is thus proportional to the product of the density, the flow
velocity and the time spent at a given temperature. Since the flow velocity is
assumed to be a constant of the flow, the time evolution of a fluid element only
depends on the product ($n_{{\rm H},0}t)$. This is referred to as the fluence,
and it may be interpreted indirectly as a scaled spatial coordinate. The
resulting scaling property allows us to run a case for a fixed density, and then
rescale our results for an arbitrary density. We note that other models
(radiative shocks, mixing layers, conductive interfaces) also exhibit scaling
relations.
\par The same argument shows that the column density predictions will not change
much with metallicity. Lowering the metal abundance increases the cooling time
by the same factor, since the principal coolants are ions of oxygen. This means
that the predictions for oxygen ions are nearly independent of oxygen
abundances. To first order, the column densities for the other ions will also
stay constant. However, a proper handling of this issue would require
calculations using a number of different assumed abundances.

\medskip\par{\it Photoionization}
\par A crucial (and unique) aspect of the model described here is that it takes
into account the ionizing radiation that is produced by the hot, upstream gas.
Here we discuss a few general implications of including photoionization.
\par For a given cooling history, the quantity $\phi_{\nu} = F_{\nu}/(n_{{\rm
H},0}v_0)$ is a constant of the flow, where $F_{\nu}$ is the photon flux emitted
by the cooling gas at a given frequency and $(n_{{\rm H},0} v_0)$ is the particle
flux. The effectiveness of photoionization depends on the photoionization
parameter, $U_i=n_{\gamma,i}/n_{\rm H}$, where $n_{\gamma,i}=(F_i/c)$ is the
density of photons capable of ionizing some ion $i$. Therefore, we can express
the ionization parameter at a given point in the cooling flow as $U_i=F_i/(c
n_{\rm H})=(\Phi_i/X)(v_0/c)$, where $X=n_{\rm H}/n_{{\rm H},0}$ is the factor
by which the flow has compressed, and $\Phi_i$ is the integral of $\phi_\nu$
for photons that can ionize ion $i$.
\par For example, for a flow velocity of $v_0$=60~\kms, the ionization parameter
in the photoionized part of an isochoric flow, where the hydrogen ionizing
photons per H atom is $\Phi_{\rm HI}$=4.3 and the gas compression is $X=1$, will
be $\log U_{\rm HI}$=$-$3.1. In contrast, for isobarically cooling gas, where
$\Phi_{\rm HI}$=7.0 and $X=112$, the corresponding ionization parameter is $\log
U_{\rm HI}$=$-$4.9. This low ionization parameter in isobarically cooling gas
implies that there is no photoionization boost in this case. Despite the
increased number of ionizing photons produced per atom in the cooling gas (due
to $p dV$ work), the compression lowers the effectiveness of photoionization.
\par The width of the photoionization zone, where the radiation from the hot
section is reprocessed, can be estimated by equating the number flux of ionizing
photons produced by the hot flow to the number of recombinations per hydrogen
atom per second in the photoionized zone. Using the scaling relationships above
this yields $L \sim v_0 \Phi_{\rm HI}/(n_{{\rm H},0} \alpha_{\rm H} X^2)$, where
$\alpha_{\rm H}$ is the Case A radiative recombination coefficient of hydrogen.
Our numerically calculated lengths agree well with this analytical estimate,
yielding $L$$\sim$$4/n_{-2}$~pc for the isobaric case and
$L$$\sim$$800/n_{-2}$~pc for the isochoric case with $v_0$=30~\kms. The
corresponding flow times are $t$=3.3\tdex5/$n_{-2}$ years for the isobaric case
and 2.5\tdex7/$n_{-2}$ years for the isochoric case.
\par The inclusion of photoionization produces a steeper than linear
relationship between the column density and $v_0$, as demonstrated in 
Fig.~\FAcoldens\. In this case, the column density can be written as $N_i = v_0
\mu_i b_i(v_0)$, where $b_i(v_0)$ is a photoionization boost for ion $i$, which
increases with flow velocity.

\medskip\par{\it Column density predictions}
\par The cumulative column density as a function of (normalized) position is
shown in Fig.~\FAzone\ for an isochoric flow with $v_0$=30~\kms, both with and
without photoionization. Figure~\FAfractions\ shows the temperature evolution
and the ionization fractions as function of position. In this example, the
spatial thickness of the zone containing the highly ionized gas is
$\sim$100\,$n_{-2}$~pc. When the effects of photoionization are added in, the
pathlength occupied by \CIV\ and particularly \SiIV\ is extended by a factor of
2 to 8.

\begin{figure}\plotfiddle{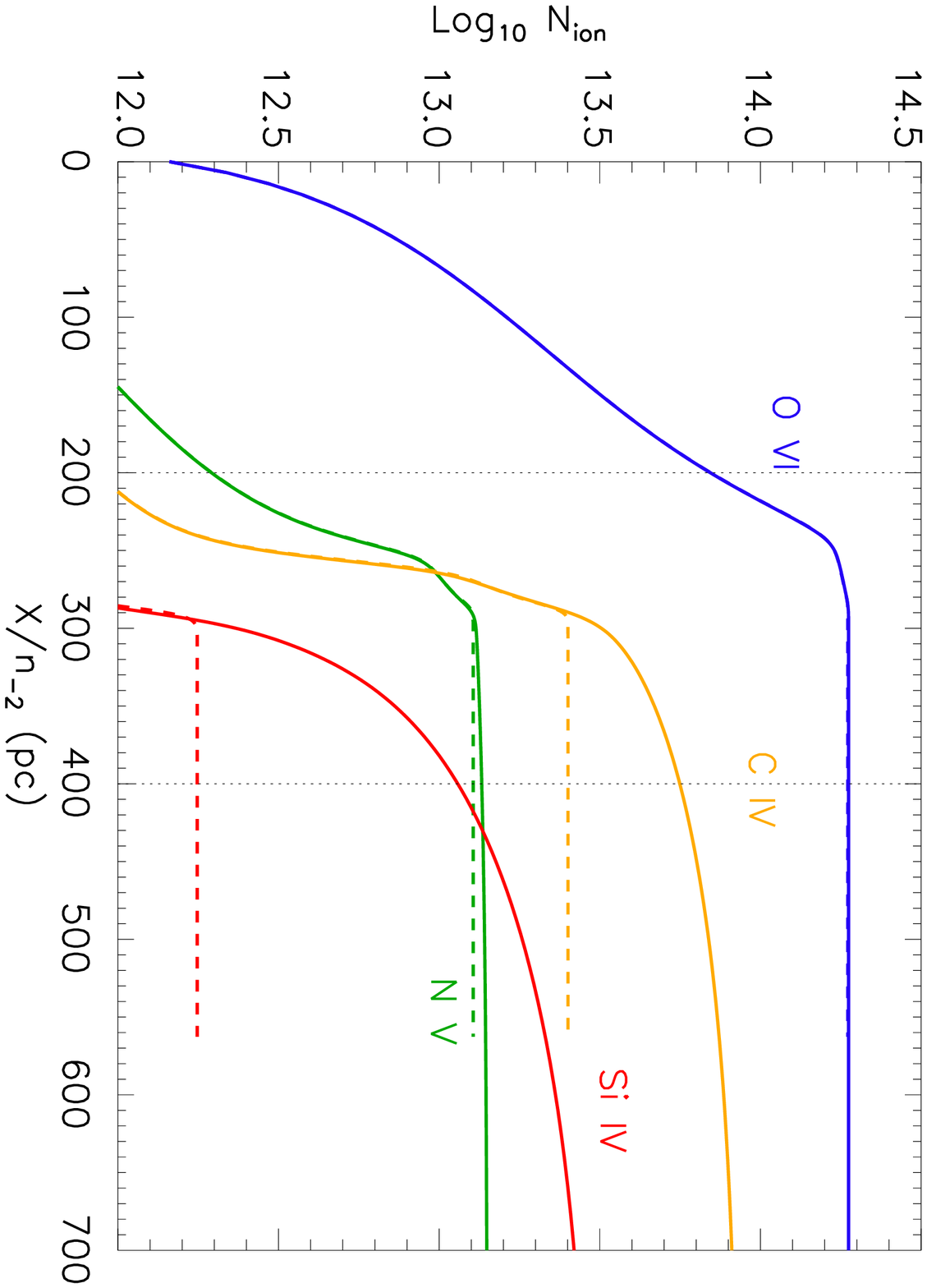}{0in}{90}{320}{350}{0}{0}\figurenum{A2}
\caption{Cumulative column density of highly ionized gas as a function of
(normalized) path length through an isochoric radiative cooling flow with
$v_0$=30~\kms. The solid and dotted lines show the column density with and
without the inclusion of photoionization from the radiation field produced by
the cooling gas.}
\end{figure}

\begin{figure}\plotfiddle{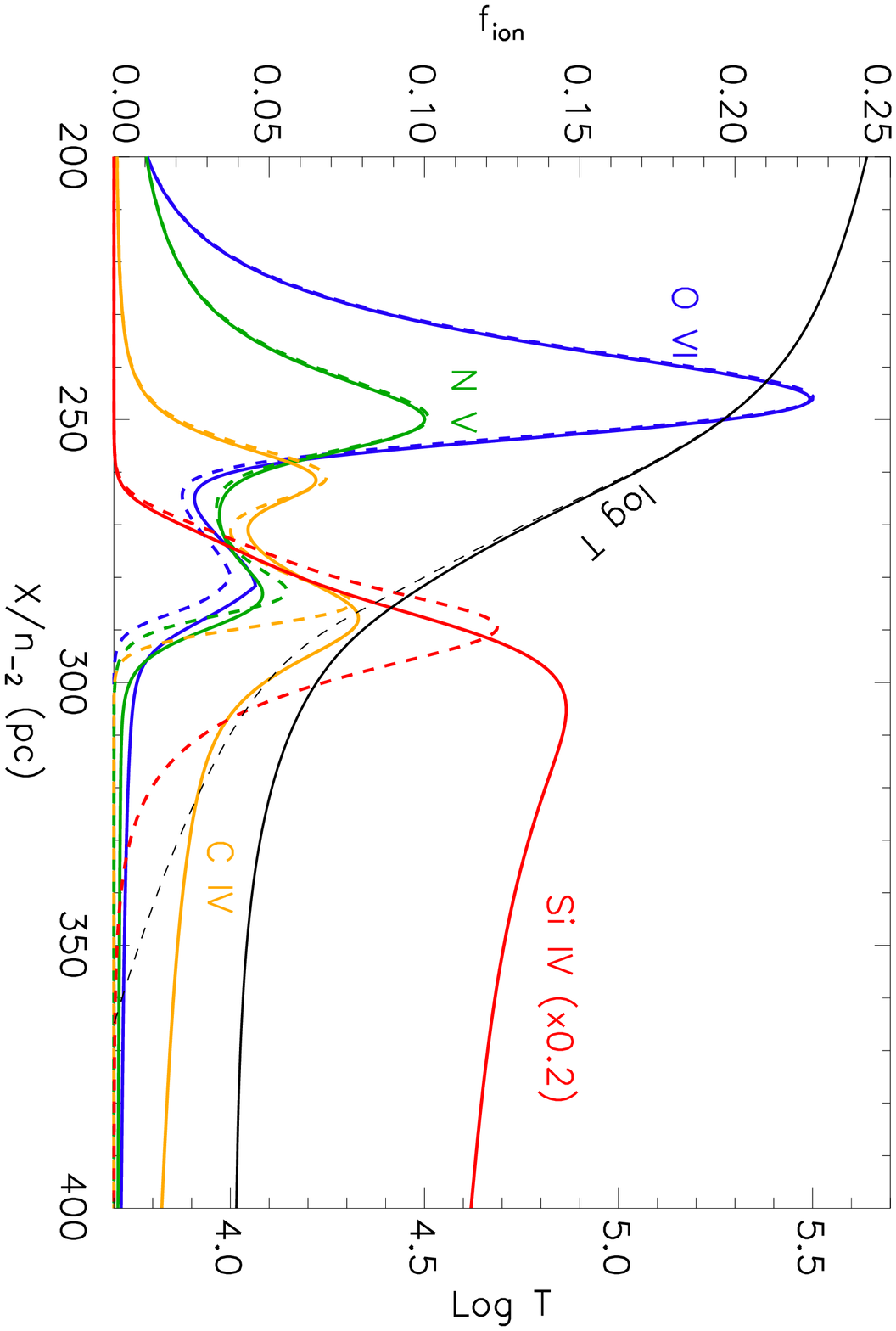}{0in}{90}{320}{350}{0}{0}\figurenum{A3}
\caption{Ionization fractions (left axis) and temperature (right axis) as a
function of normalized position for isochorically cooling gas with
$v_0$=30~\kms\ for cases that include (solid) and do not include (dashed) the
effects of photoionization. For clarity, the ionization fraction of \SiIV\ has
been multiplied by 0.2. The thickness of the zones scales linearly with $v_0$,
although the zones where \CIV\ and \SiIV\ occur become more extended due to the
enhanced photoionization.}
\end{figure}

\begin{figure}\plotfiddle{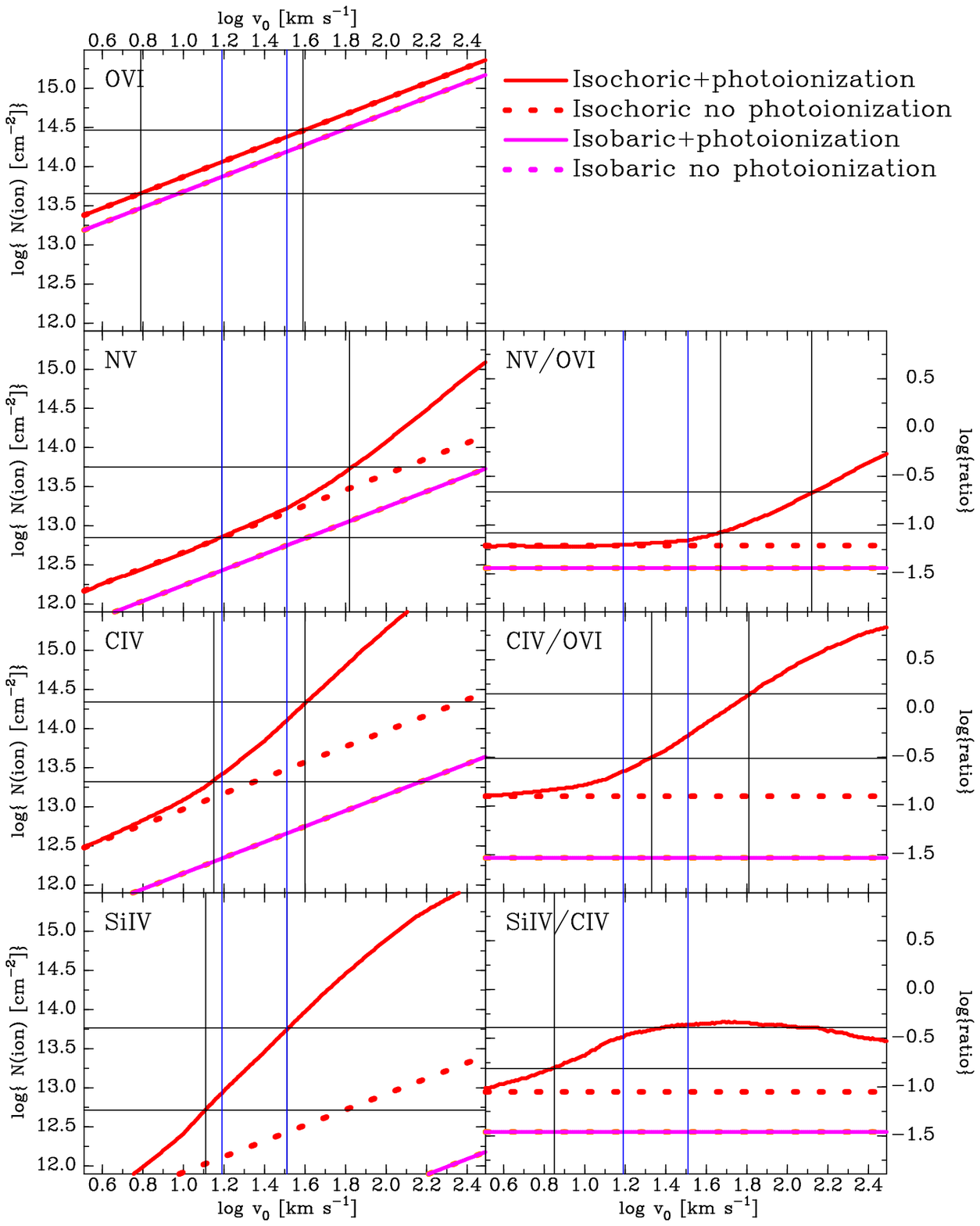}{0in}{0}{390}{400}{0}{0}\figurenum{A4}
\caption{Left panels: column densities as a function of the logarithm of the
flow velocity, $v_0$, for cases with isobaric cooling (magenta line) and for
isochoric cooling with and without photoionization (red solid and red dashed
line). Note that the column densities are linearly proportional to flow velocity
for low velocities. Note the significant photoionization boost for \NV, \CIV\
and \SiIV, setting in at log\,$v_0$$\sim$1.4, 1.0 and 0.7, respectively. Right
panels: selected column densities ratios as a function of the logarithm of the
flow velocity. In all panels the observed 1.5$\sigma$ range around the average
ionic column density or ratio observed in the Galactic corona is shown with
horizontal lines. The averages and dispersions can be found in Sects.~\Scoldens\
and \Sratiohist. In all panels vertical lines shows the range of velocities,
$v_0$ for which the predicted column density or ratio falls within the observed
range. The solid blue vertical lines show the velocity range within which all
four column density predictions fit the observed range (15--32~\kms).}
\end{figure}

\par Figure~\FAcoldens\ shows the predicted column densities as function of the
flow velocity, $v_0$. There are three useful insights that one can take from
this figure. First, when comparing models to observations, {\it column density
ratios are not necessarily independent of total column densities}. Models which
successfully produce the observed ratios of column densities might do so in
increments so large as to be inconsistent with the estimated column per
component or so small as to require hundreds of interfaces along the line sight
(as is e.g. the case for conductive interface layers). Second, for any
reasonable set of parameters, the \OVI\ and \NV\ ions originate only in the
thermal zone and are basically independent of the radiation field. This is
because their ionization potentials are much higher than the typical photon
energy (I.P., for going from \OV\ to \OVI\ 113.9~eV, for going from \NIV\ to
\NV\ 77.47~eV). Third, if there is resistance to density compression (or a
higher radiation field than that produced in the cooling gas alone),
photoionization can become important in contributing to the column densities of
\SiIV\ (I.P.\ 33.49~eV), and \CIV\ (I.P.\ 47.89~eV).
\par Figure~\FAcoldens\ shows that the observed ranges of column densities
through the Galactic corona for \CIV\ and \SiIV\ both are consistent with flow
velocities between 15 and 35~\kms. The range for \NV\ (15 to 65~\kms) overlaps
with this, but in some directions higher flow velocities would be needed to
produce the observed \NV. On the other hand, such flows would predict too much
\OVI. Nevertheless, since this cooling flow is modeled as a simplified
1-dimensional flow, it is encouraging that there is a range of flow velocities
for which all four column densities are similar to the observed values.
\par It is relatively easy to understand why the thermal and ionization
structure of the cooling gas is independent of density or metallicity. However,
if photoionization is important one might expect the volume density to influence
the column densities. However, the photon flux from the flow is linearly
proportional to the particle flux. This means that cooling flows produce a fixed
number of ionizing photons per hydrogen atom, independent of the density. Thus,
the column density predictions do not depend on the absolute value of the volume
density (although they do depend on the density structure).

\medskip\par{\it Emissivity predictions}
\par In general, the density of the flow can be determined by comparing a column
density ($\propto n_{H}$) with an emission line (or broadband) intensity
($\propto n_{\rm H}^{2}$). With a model, any absorption/emission pair can be
chosen. In the original Galactic Fountain model of Shapiro \& Field (1976), for
example, the combination of matching a model of soft X-ray emission and \OVI\
column densities yielded a density large enough that it indicated outflowing gas
would cool and fall back before reaching an equilibrium scaleheight, suggesting
that hot gas in the disk might establish a convective-radiative cycle. The least
model dependent results come from comparing the emission and absorption from a
single ion. Martin \& Bowyer (1990), for example, used the absorption/emission
from \CIV. However, the chief uncertainty in this approach is the possibility
that a significant fraction of \CIV\ could be produced in a photoionized region
and therefore absorbing but not emitting. The body of \OVI\ absorption and
emission measurements now available from FUSE alleviates this problem, since it
is expected that \OVI\ absorption and emission arise in the same spatial region.
\par Since the intensity of any emission line is proportional to the particle
flux, $I_j=(n_{\rm H,0}\,v_0) b(v_0) \iota_j$, and the column densities are
proportional to the flow velocity only, $N_i = v_0 a_i(v_0) \mu_i$, one can use
the comparison of the two quantities to solve for the initial density of the
flow, $n_{\rm H,0} = (I_{j,obs}/N_{i,obs}) (\mu_i/\iota_j) (a(v_0)/b(v_0))$,
where $I_{j,obs}$ and $N_{i,obs}$ are the observed emission line intensity and
column density from a region of interest. The constant of proportionality
between the particle flux and emission line intensity, $\iota_j$, is given in
Table~A1 for selected lines. The photoionization boost term for the intensities
is $b(v_0)$. For infrared and optical lines, this correction can be quite
significant. However, for ultraviolet and x-ray resonance lines, this correction
is negligible, as the temperature in the photo-ionized zone is too low to excite
these higher energy transitions. For example, the \CIVd\ doublet is enhanced by
less than 10\% in the photo-ionized isochoric flows for example.
\par Combining the results for the emission and absorption, and scaling these
numbers to reasonable values, $N_{i,14} = N_i$/\dex{14}\,\cmm2\ and
$I_{j,1000}=I_j$/(1000 photons~s$^{-1}$\,\cmm2\,sr$^{-1}$) (line units=LU),
we can solve for the initial flow density, 
\begin{equation}
n_{\rm H,0}=(10^{-3}~{\rm cm^{-2}})\left\{
\begin{array}{cl}
(0.34/3.31)~a_{{\rm CIV}}(v_0)~I_{{\rm CIV},1000}/N_{{\rm CIV},14} & ({\rm isobaric/isochoric}) \\
(1.62/3.91)~a_{{\rm OVI}}(v_0)~I_{{\rm OVI},1000}/N_{{\rm OVI},14} & ({\rm isobaric/isochoric}) 
\; 
\end{array} \right\}\; ,
\end{equation}
\noindent where the intensity lines are the combined intensities of the
resonance doublets of \OVI\ and \CIV. In this above equation, we neglect the
(small) photoionization boost, $b_i(v_0)$ to the UV emission lines, and $a_{\rm
OVI}(v_0)$$\sim$1. Note that since the isochoric case does not compress it must
have a higher initial density to produce the same emergent intensity as an
isobaric flow.
\par Observations of I(\OVI) fall in the range 3000 to 7000 line units (Dixon \&
Sankrit 2008), while the 1$\sigma$ range of N(\OVI) is 1.3--3.0\tdex{14}\,\cmm2\
(this work). Unfortunately, there are not enough simultaneous measurements to
judge whether these two quantities are correlated. Moreover, the intensity must
be corrected for the extinction along the line of sight, and both quantities
must be estimated for the same physical path length, c.f.\ Lei et al (2009) and
Shelton et al (2010). Just using the values above to bracket the possible range
of densities, we find that for isochoric flows, $n_{\rm
H,0}$=0.3-1.5\tdex{-2}\,\cmm3. Using a mean \OVI\ temperature of $\log T$=5.3
gives a thermal pressure in the \OVI\ zone of $p/k$=600-3000\,\cmm3~K. For both
isobaric and isochoric cases, the radiative cooling models predict I(\OVI) to be
$\sim$6 times greater than I(\CIV) (in line units). There are very few
directions toward which both lines have been estimated Shelton et al.\ (2010)
combined SPEAR data for \CIV\ with \OVI\ from FUSE, finding a ratio $\sim$1.
Using just SPEAR data, Welsh et al.\ (2007) find ratios $<$1 to $\sim$3 in
directions toward the North Galactic Cap. These ratios indicate a larger \CIV\
abundance or higher \OVI\ abundances in thermal (non-photoionized) zones than
expected from cooling gas. However, until more and especially more detailed
observations are done, it will remain unclear how serious this discrepancy is.


\input tableA1.tex


\newpage
\def\ref {\par\noindent}
\def\REF {\par\noindent}
\noindent {\bf References}
\ref Allen C.W., 1973, Astrophysical Quantities (London: The Athlone Press)
\ref Asplund M., Grevesse N., Sauval A.J., Scott P., 2009, ARA\&A, 47, 481
\ref Begelman M.C., Fabian A.C., 1990, NNRAS, 244, 26P
\ref Benjamin R.A., Shapiro P.R., 1993, in ``UV and X-ray Spectroscopy of Laboratory and Astrophysical Plasmas'', eds.\ Silver E.H. \& Kahn S.M., (Cambridge: Cambridge University Press), p280
\ref Benjamin R.A., 1994, Ph.D.\ thesis University of Texas
\ref Borkowski K.J., Balbus S.A., Fristrom C.C., 1990, ApJ, 355, 501
\ref Bregman, J.N. 1980, ApJ, 236, 577
\ref Bunner A.N., Coleman P.L., Kraushaar W.L., McCammon D., Williamson F.O., 1973, ApJ, 179, 781
\ref Bowen D., et al.\ 2008, ApJS, 176, 59
\ref Collins J.A., Shull J.M., Giroux M.K., 2004, ApJ, 605, 216
\ref Collins J.A., Shull J.M., Giroux M.K., 2005, ApJ, 623, 195
\ref Cowie L.L., Jenkins E.B., Songaila A., York D.C., 1979, ApJ, 232, 467
\ref Cowie L.L., Taylor W., York D.G., 1981, ApJ, 248, 528
\ref de Avillez M.A., Breitschwerdt D., 2005, ApJ, 634, L65
\ref Dixon W.V.D., Sankrit R., 2008, ApJ, 686, 1162
\ref Dopita M.A., Sutherland R.S., 1996, ApJS, 102, 161
\ref Edgar R.J.. Chevalier R.A., 1986, ApJ, 310, L27
\ref Esquivel A., Benjamin R.A., Lazarian A., Cho J., Leitner S.N., 2006, ApJ, 648, 1043
\ref Everett J.E., Zweibel E.G., Benjamin R.A., McCammon D., Rocks L., Gallagher J.S., 2008, ApJ, 674, 258
\ref Fox A.J., Savage B.D., Sembach K.R., Fabian D., Richter P., Meyer D.M., Lauroesch J., Howk J.C., 2003, ApJ, 582, 793 
\ref Fox A.J., Savage B.D., Wakker B.P., Richter P., Sembach K.R., Tripp T.M., 2004, ApJ, 602, 738 
\ref Fox A.J., Wakker B.P., Savage B.D., Tripp T.M., Sembach K.R., Bland-Hawthorn J., 2005, ApJ, 630, 332 
\ref Fox A.J., Savage B.D., Wakker B.P., 2006, ApJS, 165, 279 
\ref Fox A.J., Petitjean P., Ledoux C., Srianand R., 2007, A\&A, 465, 171
\ref Fox A.J., Prochaska J.X., Ledoux C., Petitjean P., Wolfe A.M., Srianand R., 2009, A\&A, 503, 731
\ref Fox A.J., Wakker B.P., Smoker J.V., Richter P., 2010, ApJ, 718, 1046
\ref Fraternali F., Oosterloo T.A., Sancisi R., 2004, A\&A, 424, 485
\ref Gaensler B.M., Madsen G., Chatterjee S., Mao S.A., 2008, PASA, 25, 184
\ref Ganguly R., Sembach K.R., Tripp T.M., Savage B.D., 2005, ApJS, 157, 253 
\ref Gnat O., Sternberg A., 2007, ApJS, 168, 213 
\ref Gnat O., Sternberg A., 2009, ApJ, 693, 1514 
\ref Gnat O., Sternberg A., McKee, 2010, ApJ, 718, 1315 
\ref Heald G.H., Rand R.J., Benjamin R.A., Bershady M.A., 2006, ApJ, 647, 1018
\ref Heald G.H., Rand R.J., Benjamin R.A., Bershady M.A., 2007, ApJ, 663, 933
\ref Henley D.B., Shelton R.L., Kwak K., Joung M.R., Mac Low M.-M., 2010, ApJ, 723, 935
\ref Houck, J.R., Bregman, J.N. 1990, ApJ, 352, 506
\ref Howk J.C., Savage B.D., Sembach K.R., Hooper C.G., 2002, ApJ, 572, 264 
\ref Howk J.C., Sembach K.R., Savage B.D., 2003, ApJ, 586, 249 
\ref Hulsbosch A.N.M., Wakker B.P. 1988, A\&AS, 75, 191
\ref Indebetouw R., Shull J.M., 2004a, ApJS, 605. 205
\ref Indebetouw R., Shull J.M., 2004b, ApJ, 607, 309 
\ref Jenkins E.B., Meloy D.A., 1974, ApJ, 193, L121 
\ref Jenkins E.B., 1978a, ApJ, 219, 845 
\ref Jenkins E.B., 1978b, ApJ, 220, 107 
\ref Jenkins E.B., et al., 1998, ApJ, 492, 147 
\ref Jenkins E.B., Oegerle W.R., Gry C., Vallerga J., Sembach K.R., Shelton R.L., Ferlet R., Vidal-Madjar A., York D.G., Linsky J.L., Roth K.C., Dupree A.K., Edelstein J., 2000, ApJ, 538, L81 
\ref Kalberla P.M.W., Burton W.B., Hartmann D., Arnal E.M., Bajaja E., Morras R., P\"oppel W.G.L., 2005, A\&A, 440, 775
\ref Kahn, F.D. 1981, in Investigating the Universe: Papers presented to Zdenek Kopal on the occasion of his retirement, September 1981 (Dordrecht, Reidel), 1
\ref Kim-Quijano, J. et al. 2007, STIS Instrument Handbook, Version 8.0, (Baltimore: STScI).
\ref Keeney B.A., Danforth C.W., Stocke J.T., Penton S.V., Shull J.M., Sembach K.R., 2006, ApJ, 646, 951 
\ref Kwak K., Shelton R.S., 2010, ApJ, 719, 523
\ref Knauth D.C., Howk J.C., Sembach K.R., Lauroesch J.T., Meyer D.M., 2003, ApJ, 592, 946 
\ref Kuntz K.D., Danly L. 1996, ApJ, 457, 703
\ref Lehner N., Jenkins E.B., Gry C., Moos H.W., Chayer P., Lacour S., 2003, ApJ, 595, 858 
\ref Lehner N., Howk J.C., 2010, ApJ, 709, L138
\ref Lehner N., Zech W.F., Howk J.C., Savage B.D., 2011, ApJ, 727, 461
\ref Lei S., Shelton R.L., Henley D.B., 2009, ApJ, 699, 1891
\ref Martin C., Bowyer S., 1990, ApJ 350, 242
\ref M\"unch, G. 1952, PASP, 64, 312
\ref M\"unch, G., Zirin, H., 1961, ApJ, 133, 11
\ref Oegerle W.R., Tripp T.M., Sembach K.R., Jenkins E.B., Bowen D.V., Cowie L.L., Green R.F., Kruk J.W., Savage B.D., Shull J.M., York D.G., 2000, ApJ, 538, L23 
\ref Oegerle W.R., Jenkins E.B., Shelton R.L. Bowen D.V., Chayer P., 2005, ApJ, 622, 377 
\ref Oosterloo T.A., Fraternali F., Sancisi R., 2007, AJ, 134, 1019
\ref Petitjean P., Ledoux C., Srianand R., 2008, A\&A, 480, 349
\ref Raymond J.C., Smith, B.W., 1977, ApJS 35, 419
\ref Richter P., 2006, Reviews in Modern Astronomy, 19, 31
\ref Rogerson J.B., York D.G., Drake J.F., Jenkins E.B., Morton D.C., Spitzer L., 1973, Ap, 181, L110
\ref Romanik C.J., 1988, ApJ, 330, 1022
\ref Savage B.D., de Boer K.S., 1979, ApJ, 230, L77
\ref Savage B.D., de Boer K.S., 1981, ApJ 243, 460
\ref Savage, B.D., Massa, D. 1987, ApJ, 314, 380
\ref Savage B.D., Edgar R., Diplas A., 1990, ApJ, 361, 107
\ref Savage B.D., Sembach K.R., 1991, ApJ, 379, 245 
\ref Savage B.D., Sembach K.R., 1994, ApJ, 434, 145 
\ref Savage B.D., Sembach K.R., Lu L., 1995, ApJ, 449, 145 
\ref Savage B.D., Sembach K.R., Tripp T.M., 1997a, ApJ, 480, 216 
\ref Savage B.D., Sembach K.R., Lu L., 1997b, AJ, 113, 2158 
\ref Savage B.D., Sembach K.R., Howk J.C., 2001a, ApJ, 547, 907 
\ref Savage B.D., Meade M., Sembach K.R., 2001b, ApJS, 136, 631 
\ref Savage B.D., Sembach K.R., Wakker B.P., et al., 2003, ApJS, 146, 125
\ref Savage B.D., Wakker B.P., Fox A.J., Sembach K.R., 2005, ApJ, 619, 863 
\ref Savage B.D., Lehner N., 2006, ApJS, 162, 134 
\ref Savage B.D., Wakker B.P., 2009, ApJ 702, 1472
\ref Sembach K.R., Savage B.D., Massa D., 1990, ApJ, 355, 114 
\ref Sembach K.R., Savage B.D., Massa D., 1991, ApJ, 372, 81 
\ref Sembach K.R., Savage B.D., 1992, ApJS, 83, 147 
\ref Sembach K.R., Savage B.D., Cardelli J.A., 1994a, ApJ, 420, 183 
\ref Sembach K.R., Savage B.D., Jenkins E.B., 1994b, ApJ, 421, 585 
\ref Sembach K.R., Savage B.D., 1994, ApJ, 431, 201 
\ref Sembach K.R., 1994, ApJ, 434, 244 
\ref Sembach K.R., Savage B.D., Lu L., 1995, ApJ, 439, 672 
\ref Sembach K.R., Savage B.D., Hurwitz M., 1999, ApJ, 524, 98 
\ref Sembach K.R., Howk J.C. Savage B.D., Shull J.M., Oegerle W.R., 2001, ApJ, 561, 573 
\ref Sembach K.R., Wakker B.P., Savage B.D., Richter P., Meade M., Shull J.M., Jenkins E.B., Sonneborn G., Moos H.W., 2003, ApJS, 146, 165
\ref Shapiro P.R., Field G.B., 1976, ApJ, 205, 762
\ref Shapiro P.R., Benjamin R., 1991, PASP, 103, 923
\ref Shapiro P.R., Benjamin R.A., 1993, in ``Star-Forming Galaxies and their Interstellar Media'', ed.\ Franco J.J., (Cambridge: Cambridge University Press), p273
\ref Shelton R., 1998, ApJ, 504, 785
\ref Shelton R.L., Henley D.B., Dixon W.V., 2010, ApJ, 722, 302
\ref Shull J.M., van Steenberg M.E., 1985, ApJ 298, 268
\ref Skillman E.D., 1998, in ``VIII Canary Islands Winter School of Astrophysics: Stellar Astrophysics for the Local Group'', eds.\ A.\ Aparicio, A.\ Herrero, F.\ S\'anchez
\ref Slavin J.D., Shull J.M., Begelman M.C., 1993, ApJ, 407, 83
\ref Sutherland R.S., Dopita M.A., 1993, ApJS, 88, 253
\ref Spitzer L., 1956, ApJ, 124, 20
\ref Spitzer L., 1990, ARA\&A, 28, 71
\ref Spitzer L., 1996, ApJ, 458, L29
\ref Sterling N.C., Savage B.D., Richter P.R., Fabian D., Sembach K.R., 2002, ApJ, 567, 354 
\ref Tripp T.M., Sembach K.R., Savage B.D., 1993, ApJ, 415, 652 
\ref Wakker B.P., 2001, ApJS, 146, 463
\ref Wakker B.P., 2004, in ``High-velocity clouds'', Astrophysics and Space Science Library, vol.\ 312, Kluwer-Dordrecht, p25
\ref Wakker B.P., 2006, ApJS, 163, 282
\ref Wakker B.P., Howk J.C., Savage B.D., van Woerden H., Tufte S.R., Schwarz U.J., Benjamin R., Reynolds R.J., Peletier R.F., Kalberla P.M.W., 1999, Nature, 400, 388
\ref Wakker B.P., Savage B.D., Sembach K.R., et al., 2003, ApJS, 146, 1
\ref Wakker B.P., York D.G., Howk J.C., Barentine J.C., Wilhelm R., Peletier R.F., van Woerden H., Beers T.C., Ivezi\'c Z, Richter P., Schwarz U.J., 2007, ApJL, 670, L113
\ref Wakker B.P., Savage B.D., 2009, ApJS, 182, 378
\ref Welsh B.Y., Edelstein J., Korpela E.J., Kregenow J., Sirk M., Min K.-W., Park J.W., Ryu K., Jin H., Yuk I.-S, Park J.-H., 2007, A\&A 472, 509
\ref Williamson F.O., Sanders W.T., Kraushaar W.K., McCammon D., Borken R., Bunner A.N., 1974, ApJ, 193, L133
\ref York D.G., 1974, 193, L127 
\ref York D.G., 1977, ApJ, 213, 43
\ref Zsarg\'o J., Sembach K.R., Howk J.C., Savage B.D., 2003, ApJ, 586, 1019 
\end{document}

%% file: table1.tex
\begin{deluxetable}{lrrllccccccl}
\tabletypesize{\scriptsize}
\tabcolsep=3pt
\tablecaption{Observational data}
\tablehead{%
\ch{Object} &\ch{Lon}   &\ch{Lat}   &\ch{Status$^a$} &\ch{Instrument$^b$} &\ch{Resol$^c$} &\ch{T$_{\rm exp}$} &\ch{S/N}  &\ch{S/N} &\ch{S/N}  &\ch{S/N}   &\ch{Obs. ID.} \\
            &\ch{[deg]} &\ch{[deg]} &                &                    &\ch{[\kms]}    &\ch{[ks]}          &\ch{\OVI} &\ch{\NV} &\ch{\CIV} &\ch{\SiIV} &\ch{ }        \\
\ch{(1)} &\ch{(2)} &\ch{(3)} &\ch{(4)} &\ch{(5)} &\ch{(6)} &\ch{(7)} &\ch{(8)} &\ch{(9)} &\ch{(10)} &\ch{(11)} &\ch{(12)} }
\tablecolumns{12}
\tablenum{1}
\tablewidth{0pt}
\startdata
3C232                & 194.17 &  52.32 & Skip     & FUSE            & 20 &  10.8 &  1 &    &    &    & P10714 \\
                     &        &        &          & STIS$-$1222       & 30 &  11.2 &    & 12 &    &    & O6700 \\
                     &        &        &          & STIS$-$1400       & 30 &  11.2 &    &    &    & 11 & O6700 \\
                     &        &        &          & STIS$-$1567       & 30 &  22.3 &    &    & 13 &    & O6700 \\
3C249.1              & 130.39 &  38.55 & Ratios   & FUSE            & 20 & 260.5 & 16 &    &    &    & D11701 P10716 S60109 U10275 \\
                     &        &        &          & STIS$-$E140M      &  8 & 137.6 &    &  7 &  6 &  8 & O6E10 O6E12 O6E13 \\
3C273.0              & 289.95 &  64.36 & Ratios   & FUSE            & 20 &  43.3 & 27 &    &    &    & P10135 \\
                     &        &        &          & STIS$-$E140M      &  8 &  37.3 &    & 28 & 21 & 43 & O5D30 \\
3C351.0              &  90.08 &  36.38 & Ratios   & STIS$-$E140M      &  8 & 156.3 &    &  9 &  9 &  9 & O5790 \\
ESO141$-$G55           & 338.18 & $-$26.71 & Ratios   & FUSE            & 20 &  40.9 & 16 &    &    &    & I90401 \\
                     &        &        &          & GHRS$-$1241       & 19 &  15.6 &    & 16 &    &    & Z3E70 \\
                     &        &        &          & GHRS$-$1247       & 19 &   9.8 &    & 16 &    &    & Z3I70 \\
                     &        &        &          & GHRS$-$1400       & 16 &   9.8 &    &    &    & 13 & Z3I70 \\
                     &        &        &          & GHRS$-$1538       & 14 &  11.1 &    &    & 12 &    & Z3I70 \\
ESO438$-$G09           & 277.55 &  29.36 & Skip     & STIS$-$1222       & 30 &  11.1 &    & 10 &    &    & O5EW0 \\
Fairall9             & 295.07 & $-$57.83 & Meas.    & FUSE            & 20 &  34.8 &  6 &    &    &    & P10106 \\
                     &        &        &          & GHRS$-$1238       & 19 &  14.4 &    & 26 &    &    & Z3E70 \\
                     &        &        &          & GHRS$-$1250       & 18 &   8.1 &    & 26 &    &    & Z26O0 \\
                     &        &        &          & GHRS$-$1258       & 18 &   6.9 &    &    &    &    & Z3E70 \\
                     &        &        &          & GHRS$-$1539       & 14 &   8.1 &    &    &  6 &    & Z26O0 \\
H1821+643            &  94.00 &  27.42 & Skip     & FUSE            & 20 & 278.4 & 28 &    &    &    & C09502 P10164 \\
                     &        &        &          & STIS$-$E140M      &  8 & 101.9 &    & 13 & 16 & 19 & O5E70 \\
HE0226$-$4110          & 253.94 & $-$65.77 & Ratios   & FUSE            & 20 & 204.8 & 24 &    &    &    & D02701 P10191 P20713 \\
                     &        &        &          & STIS$-$E140M      &  8 &  87.5 &    &  8 &  8 & 11 & O6E10 O6E11 \\
HE0340$-$2703          & 222.68 & $-$52.12 & Meas.    & STIS$-$1222       & 30 &   4.9 &    & 11 &    &    & O8EI0 \\
HE1029$-$1401          & 259.33 &  36.52 & Meas.    & STIS$-$1222       & 30 &   4.1 &    & 33 &    &    & O4EC0 \\
                     &        &        &          & STIS$-$1272       & 30 &   3.4 &    &    &    &    & O4EC0 \\
HE1228+0131          & 291.26 &  63.66 & Ratios   & FUSE            & 20 &   4.0 &  5 &    &    &    & P10190 \\
                     &        &        &          & STIS$-$E140M      &  8 &  54.5 &    &  5 &  6 &  7 & O56A0 \\
HS0624+6907          & 145.71 &  23.35 & Ratios   & FUSE            & 20 & 132.3 & 11 &    &    &    & P10710 S60112 U10210 \\
                     &        &        &          & STIS$-$E140M      &  8 & 123.9 &    &  7 &  9 & 14 & O6E10 O6E11 \\
HS1543+5921          &  92.40 &  46.36 & Skip     & FUSE            & 20 &   8.5 &  2 &    &    &    & P10805 \\
                     &        &        &          & STIS$-$1222       & 30 &  25.2 &    & 13 &    &    & O8MR0 \\
                     &        &        &          & STIS$-$1272       & 30 &  54.0 &    &    &    &    & O8MR0 \\
                     &        &        &          & STIS$-$1321       & 30 &  49.5 &    &    &    &    & O8MR0 \\
MCG+10$-$16$-$111        & 144.21 &  55.08 & Meas.    & STIS$-$1222       & 30 &  19.5 &    & 30 &    &    & O5EW0 \\
MRC2251$-$178          &  46.20 & $-$61.33 & Meas.    & FUSE            & 20 &  50.2 & 12 &    &    &    & P11110 \\
                     &        &        &          & STIS$-$1222       & 30 &   6.0 &    & 27 &    &    & O4EC0 \\
                     &        &        &          & STIS$-$1272       & 30 &   4.6 &    &    &    &    & O4EC0 \\
Mrk33                & 156.20 &  52.80 & Skip     & FUSE            & 20 &  23.7 &  6 &    &    &    & C04805 \\
                     &        &        &          & STIS$-$1222       & 30 &  15.8 &    &  9 &    &    & O5CA0 \\
Mrk110               & 165.01 &  44.36 & Meas.    & FUSE            & 20 &  11.5 &  1 &    &    &    & P10713 \\
                     &        &        &          & STIS$-$1222       & 30 &   4.5 &    & 10 &    &    & O4N30 O4N35 \\
Mrk205               & 125.45 &  41.67 & Ratios   & FUSE            & 20 & 231.9 & 17 &    &    &    & D05401 Q10602 S60108 U10311 \\
                     &        &        &          & STIS$-$E140M      &  8 & 124.0 &    &  8 &  6 &  9 & O62Q0 \\
Mrk279               & 115.04 &  46.86 & Ratios   & FUSE            & 20 & 235.3 & 44 &    &    &    & C09002 D15401 F32501 P10803 \\
                     &        &        &          & STIS$-$E140M      &  8 & 107.4 &    & 30 & 21 & 25 & O6JM0 O8K10 \\
Mrk290               &  91.49 &  47.95 & Ratios   & FUSE            & 20 & 112.9 & 19 &    &    &    & D07601 E08401 P10729 \\
                     &        &        &          & GHRS$-$1250       & 18 &   7.1 &    & 25 &    &    & Z3KH0 \\
Mrk335               & 108.76 & $-$41.42 & Ratios   & FUSE            & 20 &  99.2 & 28 &    &    &    & P10102 \\
                     &        &        &          & STIS$-$E140M      &  8 &  32.3 &    & 11 &  5 &  7 & O8N50 \\
Mrk421               & 179.83 &  65.03 & Meas.    & FUSE            & 20 &  82.0 & 30 &    &    &    & P10129 Z01001 \\
                     &        &        &          & GHRS$-$1239       & 19 &  15.7 &    & 24 &    &    & Z2IA0 \\
Mrk478               &  59.24 &  65.03 & Meas.    & FUSE            & 20 &  14.0 &  8 &    &    &    & P11109 \\
                     &        &        &          & STIS$-$1222       & 30 &   7.6 &    & 29 &    &    & O4EC1 \\
                     &        &        &          & STIS$-$1272       & 30 &   6.3 &    &    &    &    & O4EC1 \\
Mrk501               &  63.60 &  38.86 & Meas.    & FUSE            & 20 &  30.2 &  9 &    &    &    & C08101 P10733 \\
                     &        &        &          & GHRS$-$1239       & 19 &  31.3 &    & 11 &    &    & Z1A65 \\
Mrk509               &  35.97 & $-$29.86 & Ratios   & FUSE            & 20 &  90.4 & 30 &    &    &    & P10806 X01701 \\
                     &        &        &          & STIS$-$E140M      &  8 &  15.7 &    & 12 &  8 &  9 & O6AP0 \\
Mrk771               & 269.44 &  81.74 & Meas.    & FUSE            & 20 &   6.3 &  5 &    &    &    & P10723 \\
                     &        &        &          & STIS$-$1222       & 30 &   7.8 &    & 24 &    &    & O4EC0 O4N30 \\
                     &        &        &          & STIS$-$1272       & 30 &   5.2 &    &    &    &    & O4EC0 \\
Mrk817               & 100.30 &  53.48 & Meas.    & FUSE            & 20 & 186.8 & 43 &    &    &    & P10804 \\
                     &        &        &          & GHRS$-$1241       & 19 &  26.8 &    & 40 &    &    & Z3E70 \\
Mrk876               &  98.27 &  40.38 & Ratios   & FUSE            & 20 & 139.4 & 33 &    &    &    & D02802 P10731 U10366 \\
                     &        &        &          & STIS$-$E140M      &  6 &  58.4 &    & 11 &  8 & 16 & O8NN0 \\
Mrk926               &  64.09 & $-$58.76 & Skip     & FUSE            & 20 &  11.2 &  3 &    &    &    & P10740 \\
                     &        &        &          & STIS$-$1222       & 30 &   3.9 &    &  8 &    &    & O4EC1 \\
                     &        &        &          & STIS$-$1272       & 30 &   3.7 &    &    &    &    & O4EC1 \\
Mrk1044              & 179.69 & $-$60.48 & Skip     & FUSE            & 20 &  12.6 &  6 &    &    &    & D04101 \\
                     &        &        &          & STIS$-$1222       & 30 &   2.4 &    & 29 &    &    & O8K40 \\
                     &        &        &          & STIS$-$1272       & 30 &   4.8 &    &    &    &    & O8K40 \\
                     &        &        &          & STIS$-$1567       & 30 &   2.4 &    &    & 12 &    & O8K40 \\
Mrk1095              & 201.69 & $-$21.13 & Meas.    & FUSE            & 20 &  56.0 & 10 &    &    &    & P10112 \\
                     &        &        &          & GHRS$-$1241       & 19 &  21.0 &    & 20 &    &    & Z3E70 \\
Mrk1383              & 349.22 &  55.12 & Ratios   & FUSE            & 20 &  63.1 & 23 &    &    &    & P10148 P26701 \\
                     &        &        &          & STIS$-$E140M      &  8 &  38.3 &    &  9 &  8 &  8 & O8PG0 \\
Mrk1502              & 123.75 & $-$50.18 & Meas.    & FUSE            & 20 &  39.1 &  7 &    &    &    & P11101 \\
                     &        &        &          & GHRS$-$1239       & 19 &  49.5 &    & 10 &    &    & Z1A60 Z2IA0 \\
Mrk1513              &  63.67 & $-$29.07 & Meas.    & FUSE            & 20 &  64.3 & 15 &    &    &    & P10183 \\
                     &        &        &          & STIS$-$1222       & 30 &   7.3 &    & 23 &    &    & O4EC1 \\
                     &        &        &          & STIS$-$1272       & 30 &   6.2 &    &    &    &    & O4EC1 \\
NGC985               & 180.84 & $-$59.49 & Meas.    & FUSE            & 20 &  44.3 & 10 &    &    &    & P10109 \\
                     &        &        &          & STIS$-$1222       & 30 &   3.7 &    & 25 &    &    & O4EC1 \\
                     &        &        &          & STIS$-$1272       & 30 &   3.8 &    &    &    &    & O4EC1 \\
NGC1705              & 261.08 & $-$38.74 & Ratios   & FUSE            & 20 &  24.0 & 27 &    &    &    & A04601 \\
                     &        &        &          & STIS$-$E140M      &  8 &  34.2 &    & 13 &  8 & 13 & O58N0 \\
NGC3516              & 133.24 &  42.40 & Meas.    & FUSE            & 20 & 114.5 &  9 &    &    &    & G91701 P11104 P21101 \\
                     &        &        &          & STIS$-$E140M      &  8 &  13.7 &    &  2 & 20 &  3 & O57B0 \\
NGC3783              & 287.46 &  22.95 & Ratios   & FUSE            & 20 & 178.4 & 26 &    &    &    & B10701 E03101 P10133 \\
                     &        &        &          & STIS$-$E140M      &  8 & 187.4 &    & 34 & 35 & 28 & O57B0 O63M0 O63M1 O63M5 \\
NGC4051              & 148.88 &  70.09 & Skip     & FUSE            & 20 &  86.6 & 16 &    &    &    & B06202 C01901 \\
                     &        &        &          & STIS$-$E140M      &  8 &  20.6 &    &  7 &  8 &  6 & O5F00 \\
NGC4151              & 155.08 &  75.06 & Ratios   & FUSE            & 20 &  96.6 & 36 &    &    &    & C09201 P11105 P21102 \\
                     &        &        &          & STIS$-$E140M      &  8 &  55.4 &    & 21 & 17 & 22 & O5780 O5KT0 O5KT5 O61L0 O6JB0 \\
NGC4593              & 297.48 &  57.40 & Skip     & STIS$-$E140M      &  8 &  22.0 &    & 10 &  7 &  7 & O5L50 \\
NGC5548              &  31.96 &  70.50 & Ratios   & FUSE            & 20 &  56.1 &  9 &    &    &    & D15501 P10146 \\
                     &        &        &          & STIS$-$E140M      &  8 & 107.5 &    & 24 & 11 & 10 & O6JD0 O6KW0 \\
                     &        &        &          & STIS$-$E140M      &  8 &   4.8 &    & 24 & 11 & 10 & O4LL0 \\
NGC7469              &  83.10 & $-$45.47 & Ratios   & FUSE            & 20 &  44.0 & 16 &    &    &    & C09001 P10187 \\
                     &        &        &          & STIS$-$E140M      &  8 &  71.6 &    & 25 & 11 & 12 & O6BN0 O8N50 \\
PG0804+761           & 138.28 &  31.03 & Meas.    & FUSE            & 20 & 170.9 & 31 &    &    &    & P10119 S60110 \\
                     &        &        &          & STIS$-$1222       & 30 &   7.3 &    & 20 &    &    & O4EC0 O4N30 \\
                     &        &        &          & STIS$-$1272       & 30 &   4.2 &    &    &    &    & O4EC0 \\
PG0953+414           & 179.79 &  51.71 & Ratios   & FUSE            & 20 &  75.2 & 24 &    &    &    & P10122 \\
                     &        &        &          & STIS$-$E140M      &  8 &  49.0 &    &  9 &  9 & 10 & O4X00 \\
PG1001+291           & 200.08 &  53.21 & Skip     & FUSE            & 20 &  11.4 &  5 &    &    &    & P20731 \\
                     &        &        &          & STIS$-$E140M      &  8 &  96.8 &    &  7 &  5 &  8 & O6E10 O6E11 O6E12 \\
PG1004+130           & 225.12 &  49.12 & Skip     & FUSE            & 20 &  83.4 &  8 &    &    &    & A03501 \\
                     &        &        &          & STIS$-$1222       & 30 &   5.2 &    & 11 &    &    & O5EW0 \\
PG1049$-$005           & 252.28 &  49.88 & Meas.    & STIS$-$1222       & 30 &   1.5 &    &  6 &    &    & O4N30 \\
PG1116+215           & 223.36 &  68.21 & Ratios   & FUSE            & 20 &  77.0 & 25 &    &    &    & P10131 \\
                     &        &        &          & STIS$-$E140M      &  8 &  53.1 &    & 12 & 10 & 15 & O5A30 O5E70 \\
PG1211+143           & 267.55 &  74.32 & Ratios   & FUSE            & 20 &  52.2 & 17 &    &    &    & P10720 \\
                     &        &        &          & STIS$-$E140M      &  8 & 110.0 &    & 18 & 14 & 20 & O61Y0 \\
PG1216+069           & 281.07 &  68.14 & Ratios   & FUSE            & 20 &  12.4 &  4 &    &    &    & P10721 \\
                     &        &        &          & STIS$-$E140M      &  8 & 139.6 &    &  8 &  7 &  8 & O6E10 O6E13 \\
PG1259+593           & 120.56 &  58.05 & Ratios   & FUSE            & 20 & 528.1 & 38 &    &    &    & P10801 U10318 \\
                     &        &        &          & STIS$-$E140M      &  8 & 191.5 &    &  7 &  8 & 11 & O63G0 O63G1 \\
PG1302$-$102           & 308.59 &  52.16 & Ratios   & FUSE            & 20 & 144.7 & 15 &    &    &    & P10802 \\
                     &        &        &          & STIS$-$E140M      &  8 &  44.2 &    &  7 &  5 &  5 & O5BU0 O5BU6 \\
PG1341+258           &  28.71 &  78.15 & Skip     & STIS$-$1222       & 30 &   8.1 &    & 15 &    &    & O5EW0 \\
PG1351+640           & 111.89 &  52.02 & Meas.    & FUSE            & 20 & 119.2 & 17 &    &    &    & P10725 S60107 \\
                     &        &        &          & STIS$-$1222       & 30 &   8.5 &    & 10 &    &    & O4EC5 \\
                     &        &        &          & STIS$-$1272       & 30 &   6.2 &    &    &    &    & O4EC5 \\
PG1444+407           &  69.90 &  62.72 & Ratios   & FUSE            & 20 &  10.1 &  4 &    &    &    & P10727 \\
                     &        &        &          & STIS$-$E140M      &  8 &  97.2 &    &  8 &  8 &  8 & O6E10 \\
PG1718+481           &  74.38 &  34.83 & Meas.    & FUSE            & 20 &  59.5 &  2 &    &    &    & E01001 P21001 \\
                     &        &        &          & STIS$-$1567       & 30 &  11.3 &    &    & 22 &    & O4VA0 \\
PHL1811              &  47.47 & $-$44.81 & Ratios   & FUSE            & 20 &  72.7 & 17 &    &    &    & P10810 P20711 \\
                     &        &        &          & STIS$-$E140M      &  8 &  67.8 &    & 12 &  9 & 18 & O8D90 \\
PKS0312$-$77           & 293.44 & $-$37.55 & Meas.    & STIS$-$E140M      &  8 &  46.3 &    &  3 &  3 &  3 & O65T0 O65T1 \\
PKS0405$-$12           & 204.93 & $-$41.76 & Ratios   & FUSE            & 20 & 142.2 & 18 &    &    &    & B08701 D10301 \\
                     &        &        &          & STIS$-$E140M      &  8 &  54.4 &    &  7 &  7 &  9 & O55S0 \\
PKS2005$-$489          & 350.37 & $-$32.60 & Meas.    & FUSE            & 20 &  48.6 & 17 &    &    &    & C14903 P10738 \\
                     &        &        &          & STIS$-$1222       & 30 &   6.1 &    & 67 &    &    & O4EC0 \\
                     &        &        &          & STIS$-$1272       & 30 &   5.5 &    &    &    &    & O4EC0 \\
PKS2155$-$304          &  17.73 & $-$52.25 & Ratios   & FUSE            & 20 & 119.8 & 31 &    &    &    & P10807 \\
                     &        &        &          & STIS$-$E140M      &  8 &  57.0 &    & 11 & 11 & 15 & O5BY0 \\
RX J0100.4$-$5113      & 299.48 & $-$65.84 & Meas.    & FUSE            & 20 &  22.6 &  9 &    &    &    & D80603 E89702 \\
                     &        &        &          & STIS$-$1222       & 30 &   2.3 &    &  8 &    &    & O8P80 \\
                     &        &        &          & STIS$-$1272       & 30 &   1.2 &    &    &    &    & O8P80 \\
RX J1830.3+7312      & 104.04 &  27.40 & Meas.    & FUSE            & 20 &  23.8 &  6 &    &    &    & G02003 \\
                     &        &        &          & STIS$-$1222       & 19 &   5.8 &    & 29 &    &    & O5EW0 \\
Ton S180             & 139.00 & $-$85.07 & Meas.    & FUSE            & 20 &  26.7 & 14 &    &    &    & D02801 P10105 \\
                     &        &        &          & STIS$-$1222       & 30 &   4.1 &    & 27 &    &    & O4EC0 \\
                     &        &        &          & STIS$-$1272       & 30 &   3.5 &    &    &    &    & O4EC0 \\
Ton S210             & 224.97 & $-$83.16 & Ratios   & FUSE            & 20 &  52.5 & 20 &    &    &    & P10703 \\
                     &        &        &          & STIS$-$E140M      &  8 &   5.1 &    &  4 &  5 &  7 & O6L00 \\
UGC12163             &  92.14 & $-$25.34 & Ratios   & FUSE            & 20 &  49.9 &  8 &    &    &    & B06201 \\
                     &        &        &          & STIS$-$E140M      &  8 &  20.6 &    &  4 &  2 &  4 & O5IT0 \\
VIIZw118             & 151.36 &  25.99 & Meas.    & FUSE            & 20 & 194.4 & 18 &    &    &    & P10116 S60113 U10215 \\
                     &        &        &          & STIS$-$1222       & 30 &   9.5 &    & 16 &    &    & O4EC1 \\
\enddata
\tablecomments{
a: Status of the analysis --
``Ratios'' means that this sightline is used to analyze ionic ratios;
``Measured'' means that only one or two of the four ions were measured;
``Skip'' implies that the S/N of the data is sufficient, but the measurement is impossible or ambiguous;
b: \FUSE\ or \HST\ spectrograph, with in the case of \GHRS\ data the central wavelength of the grating setting added;
an added central wavelength for \STIS\ data means that the G140M grating was used, otherwise the spectrum was obtained using the E140M echelle;
c: approximate size of resolution element.
}
\end{deluxetable}

%% file: table2.tex
\def\rulei{-------------------------------------------------}
\def\ruler{----------------------------------}
\def\EM{Emission}
\def\PN{PN}
\def\UNK{Unid. Contam.}
\def\BADCONT{Complex continuum}
\def\NOFLUX{No flux}
\def\STELLAR{Stellar}
\begin{deluxetable}{lcccrrcrrcr}
\tablenum{\Tres} \tablewidth{0pt}
\tabcolsep=3pt
\tablecaption{Measurements$^a$}
\tabletypesize{\scriptsize}
\tablehead{%
\ch{Object} &\ch{vel. range}&\ch{ion} &\multicolumn{3}{c}{strong line}         &\multicolumn{3}{c}{weak line}           &\multicolumn{2}{c}{ratios}            \\
            &               &         &\multicolumn{3}{c}{\rulei}              &\multicolumn{3}{c}{\rulei}              &\multicolumn{2}{c}{\ruler}            \\
            &               &         &\ch{log\,$N$} &\ch{$<v>$}  &\ch{FWHM}   &\ch{log\,$N$} &\ch{$<v>$}  &\ch{FWHM}   &\ch{ions} &\ch{log\,ratio}            \\
            &\ch{\kms}]     &         &\ch{[\cmm2]}  &\ch{[\kms]} &\ch{[\kms]} &\ch{[\cmm2]}  &\ch{[\kms]} &\ch{[\kms]} &                                      \\
\ch{(1)} &\ch{(2)} &\ch{(3)} &\ch{(4)} &\ch{(5)} &\ch{(6)} &\ch{(7)} &\ch{(8)} &\ch{(9)} &\ch{(10)} &\ch{(11)} }
\tablecolumns{11}
\tablenum{2}
\tablewidth{0pt}
\startdata
3C249.1        & $-$100:70   & OVI   & 14.38\e0.03\e0.01 &         &        &[14.36\e0.05\e0.17]&         &        &           &             \\ 
               &           & NV    & $<$13.32          &         &        &[\Lya]             &         &        & NV/OVI    & $<$$-$1.06    \\ 
               &           & CIV   & 13.86\e0.04\e0.01 & $-$38\e2  & 68\e7  & 13.85\e0.08\e0.01 & $-$35\e2  & 75\e7  & CIV/OVI   & $-$0.52\e0.09 \\ 
               &           & SiIV  & 13.39\e0.03\e0.01 & $-$40\e1  & 58\e6  &[\Lya]             &         &        & SiIV/CIV  & $-$0.47\e0.09 \\ 
               &           & FeIII & 14.27\e0.05\e0.01 & $-$39\e3  & 72\e7  &                   &         &        & FeIII/CIV &  0.41\e0.10 \\ 
               &           & SIII  &[\Ht]              &         &        &                   &         &        &           &             \\ 
               & $-$100:$-$30  & OVI   & 14.10\e0.04\e0.02 &         &        &[14.18\e0.05\e0.19]&         &        &           &             \\ 
               &           & NV    & $<$13.05          &         &        &[\Lya]             &         &        & NV/OVI    & $<$$-$1.05    \\
\enddata
\tablecomments{
a: The columns contain the following.
Column~1: Target name.
Column~2: Velocity range of integration for either all Milky Way Disk/Halo absorption or for individual components.
Columns~4, 5, 6: Logarithmic column density, central velocity and second-moment width of the strong line of each doublet (\OVIa, \NVa, \CIVa, \SiIVa), {\it if} the measurement was deemed to
be reliable; note that the listed widths are corrected for instrumental broadening (FWHM$_{\rm corr}$=$\sqrt{{\rm FWHM}_{\rm obs}^2-{\rm Res}^2}$). Instrumental widths are 19~\kms\ for GHRS-G160M
spectra, 30~\kms\ for STIS-G140M spectra, 6.5~\kms\ for STIS-E140M spectra, and $\sim$20~\kms\ for FUSE spectra.
Values in square brackets are given for cases where we deemed it possible to make a measurement, but it is too unreliable to use in the analyses;
notes in square brackets indicate the reason for a contamination or the identification of the line blending with the Galactic high-ionization absorption.
Columns~7, 8, 9: Logarithmic column density, central velocity and second-moment width of the weak line of each doublet (\OVIb, \NVb, \CIVb, \SiIVb).
Columns~10, 11: log of the ratio of the column densities of an ion pair, as identified in Col.~10. Values are omitted for all cases where the ratio was deemed untrustworthy.
}
\end{deluxetable}

%% file: tableA1.tex
\begin{deluxetable}{cccccccccc}
\tablenum{A1}
\tablecaption{Cooling flow parameters (without photoionization effects)} 
\tablehead{
\colhead{Case}&\colhead{$X^a$}&\colhead{$\phi_{HI}^b$}&\multicolumn{5}{c}{$\mu_{i,6}^c$}                                            &\multicolumn{2}{c}{$\iota_{j,3}^d$}              \\
\colhead{ }   &\colhead{}     &\colhead{}             &\colhead{\SiIV}&\colhead{\CIV}&\colhead{\NV}&\colhead{\OVI}&\colhead{\NeVIII}&\colhead{\CIV\,1548\,\AA}&\colhead{\OVI\,1031\,\AA} }
\startdata
Isobaric      &        112    &           7.0         &      0.037    &      1.35    &      1.67   &      41.40   &       27.93     &        4.0            &         25.6            \\
Isochoric     &         1     &           4.3         &      0.59     &      8.39    &      4.26   &      62.33   &       23.87     &        2.5            &         15.9            \\
\enddata
\tablenotetext{a}{Density compression of gas, $n_{\rm H}/n_{\rm H,o}$ between $T_o$=\dex6 K and $T$$\sim$\dex4~K.}
\tablenotetext{b}{Hydrogen ionizing photons produced per H atom in flow.} 
\tablenotetext{c}{Column density normalized to initial flow velocity, $\mu_i=N_i/v_o$, in units of \dex6\,\cmm3~s when photoionization is neglected.}
\tablenotetext{d}{Emission line intensity of resonance doublet normalized to flow mass flux, $\iota_{j,-3}=I_j/(n_{\rm H,o} v_0)$ in units of \dex3~photons~ster$^{-1}$.}
\end{deluxetable}